\documentclass[12pt]{article}
\usepackage[english]{babel}
\usepackage{amsmath,amssymb,amstext,amsthm,booktabs,cite,color,simplewick,exscale,relsize,slashed,graphicx,amsfonts,upgreek}
\usepackage{multirow,dcolumn,bm,enumerate,url,hyperref}

\usepackage{url}
\usepackage{comment}
\usepackage{graphicx}
\usepackage{caption}
\usepackage{subcaption}
\usepackage{amssymb}
\usepackage{cleveref}
\usepackage{siunitx}

\date{}

\title{\bf Gravitational Waves From Dark Sectors, Oscillating Inflatons, and Mass Boosted Dark Matter}

\author{Amit Bhoonah$^\eth$, Joseph Bramante$^{\eth,\dagger}$, Simran Nerval$^{\eth}$, Ningqiang Song$^{\eth,\dagger}$\\
{\small $^\eth$ The Arthur B. McDonald Canadian Astroparticle Physics Research Institute and} \\ {\small Department of Physics, Engineering Physics, and Astronomy,} \\ {\small Queen's University, Kingston, Ontario, K7L 2S8, Canada}\\
{\small $^\dagger$Perimeter Institute for Theoretical Physics, Waterloo, Ontario, N2L 2Y5, Canada}}

\begin{document}
\maketitle

\begin{abstract}
Gravitational wave signatures from dynamical scalar field configurations provide a compelling observational window on the early universe. Here we identify intriguing connections between dark matter and scalars fields that emit gravitational waves, either through a first order phase transition or oscillating after inflation. To study gravitational waves from first order phase transitions, we investigate a simplified model consisting of a heavy scalar coupled to a vector and fermion field. We then compute gravitational wave spectra sourced by inflaton field configurations oscillating after E-Model and T-Model inflation. Some of these gravitational wave signatures can be uncovered by the future Big Bang Observatory, although in general we find that MHz-GHz frequency gravitational wave sensitivity will be critical for discovering the heaviest dark sectors. Intriguingly, we find that scalars undergoing phase transitions, along with E-Model and T-Model potentials, can impel a late-time dark matter mass boost and generate up to Planck mass dark matter. For phase transitions and oscillating inflatons, the largest dark matter mass boosts correspond to higher amplitude stochastic gravitational wave backgrounds.
\end{abstract}

\section{Introduction}\label{sec:intro}
Despite a wealth of cosmological data collected in the last two decades, almost nothing is certain about the dynamics of our universe at temperatures above an MeV, when nuclei were synthesized. Primordial fluctuations evident in the cosmic microwave background (CMB) and the large scale distribution of galaxies are both consistent with a period of inflation and subsequent matter or radiation dominated expansion \cite{Aghanim:2018eyx,Akrami:2018odb}. However, both of these primordial structures become apparent at temperatures below an MeV, and both are consistent with a spectrum of initial scalar perturbations. With only scalar primordial perturbations, it is not possible to pinpoint the energy density of the universe either during or after inflation. In contrast, primordial gravitational waves are eminently sensitive to the energy density of the universe at the time of their creation \cite{Lyth:1996im,Knox:2002pe,Seljak:2003pn,Bramante:2016yju}. It follows that the discovery of primordial gravitational waves, either as tensor perturbations to the CMB or as a stochastic gravitational wave background (SGWB) sourced by scalar field dynamics \cite{Bertone:2019irm,Witten:1984rs,Grojean:2006bp,Das:2009ue,Schwaller:2015tja,Croon:2018kqn,Breitbach:2018ddu,Croon:2018erz,Amin:2019qrx,Croon:2019rqu,Marfatia:2020bcs,Guo:2020grp,Fornal:2020esl,Gavrilik:2020mjy,Fornal:2020ngq}, could provide a fascinating new window on the dark history of our universe at energy densities above an MeV. In this work we will consider two separate sources of gravitational waves in the early universe: first order phase transitions and inflatons oscillating after inflation. 

Because they would provide a stunning new view of the early universe, it is natural to consider whether primordial gravitational waves can also inform us about the formation of dark matter, which is inextricably linked to the state of the universe at temperatures above an MeV. Although we have accumulated extensive gravitational evidence for dark matter, the couplings, mass, and mechanism by which dark matter was produced in the early universe remain a mystery.  While much work has scrutinized production of thermally equilibrated weakly interacting massive particle (WIMP) dark matter in a radiation dominated universe, for dark matter masses ranging up to 10 TeV \cite{Jungman:1995df}, simple physical mechanisms that result in the observed relic abundance of dark matter with a mass ranging up to the Planck mass are still being uncovered. A canonical example of superheavy dark matter produced out of thermal equilibrium is the WIMPzilla, gravitationally produced at the end of cosmic inflation \cite{Chung:1998zb,Kuzmin:1998kk}. However, this nonthermal process is limited to producing dark matter with a mass below $\sim 10^{14}$ GeV, since gravitational production is exponentially suppressed for masses exceeding the Hubble constant at the end of inflation, which itself has been limited to $H_e \lesssim 10^{14}$ GeV by the observed tensor-to-scalar ratio of primordial perturbations \cite{Akrami:2018odb}. On the other hand, dark matter heavier than $10^{14}$ GeV is particularly compelling, since it may be connected to Grand Unified Theories (GUT) at mass scales above $10^{14}$ GeV \cite{Raby:1997pb,Dimopoulos:1982gm,Dine:1982zb}, and could soon be discovered at underground experiments using multiscatter and other techniques \cite{Bramante:2018tos,Bramante:2018qbc,Coskuner:2018are,Davoudiasl:2018wxz,Bai:2019ogh,Bramante:2019yss}. A number of papers have considered how suppressed gravitational production of dark matter heavier than $10^{14}$ GeV can be supplemented by production through the decay of the inflaton or out of equilibrium ``freeze-in" processes during the radiation dominated epoch following inflation \cite{Harigaya:2014waa,Harigaya:2016vda,Kolb:2017jvz}. 

In this paper we consider a simple mechanism for gravitational production of dark matter with a mass well above $10^{14}$ GeV. The mechanism operates by coupling dark matter weakly to a scalar field, which undergoes a phase transition that boosts the mass of the dark matter after its initial abundance is set by gravitational processes at the end of inflation. In contrast, some previous work has considered how mass shifts and phase transitions can alter the relic abundance of thermally produced dark matter \cite{Hui:1998dc,Baker:2017zwx,Bramante:2017obj,Davoudiasl:2019xeb}. Here we find distinct features and compelling gravitational wave signatures associated with a scalar field that can boost the mass of dark matter with a nonthermal abundance fixed by the end of inflation. In particular, we find that parameters associated with a strong first order phase transition that source a substantial SGWB are associated with a large dark matter mass boost. We also find that a large dark matter mass associated with certain inflaton potentials also corresponds to a large source of gravitational waves generated by the inflaton oscillating in its potential.

The remainder of this paper is organized as follows. In Section \ref{sec:Yukawa}, we detail a simplified model consisting of a scalar with a quartic self-coupling generated at one-loop order through coupling to a vector and fermion field. First order phase transitions and associated gravitational wave signatures of this potential are explored in Section \ref{sec:phasetransition}. Stochastic gravitational waves sourced by E-Model and T-Model inflation are computed using lattice simulations in Section \ref{sec:inflat}. Section \ref{sec:massboost} details the cosmology of very heavy dark matter that receives a late-time mass boost from a scalar field. In Section \ref{sec:conc}, we conclude with a summary of observational prospects for gravitational wave signatures. In Appendices \ref{sec:thermappendix} and \ref{sec:RGEImprovement} we detail higher order corrections and running couplings for the thermal potential studied in Section \ref{sec:Yukawa}.

\section{Phase Transitions from a Simplified Model}
\label{sec:Yukawa}

In this section we introduce a simplified model that will be useful for investigating phase transitions and dark matter mass boosts in the early universe. Our model consists of a scalar field with a small tree-level quartic self-coupling, coupled to a fermion and gauge field. More specifically, for the parameter space we will find most interesting, the scalar field's quartic self-coupling is predominantly loop-induced, through coupling to fermion and gauge fields. To explore the salient features of such a model, we consider the case that the gauge group is Abelian - the non-Abelian generalization is relatively straightforward. 

Then in total the simplified dark sector consists of a scalar $\varphi$ charged under a $U_{D}\left(1\right)$ group, a vector $A_{\mu}$, and a fermion $\chi$. For the moment, we do not consider the couplings of this sector to the Standard Model. The Lagrangian can be written as a sum of tree level terms, one loop contributions, $\mathcal{L}_{1}$, and counter-terms which we omit for clarity,
\begin{equation}
\label{eq:Lagrangian}
\begin{split}
\mathcal{L} = -\frac{1}{4}F_{\mu\nu}F^{\mu\nu} + i\overline{\chi}\gamma^{\mu}\partial_{\mu}\chi + \frac{1}{2}D_{\mu}\varphi \left(D^{\mu}\varphi\right)^{\dagger} - V_{0}\left(\varphi\right) - iy \overline{\chi}\chi\varphi + \mathcal{L}_{1},
\end{split}
\end{equation}
where $D_{\mu}\varphi = \partial_{\mu}\varphi + igA_{\mu}\varphi$ and $V_{0}\left(\varphi\right)$ is the tree level scalar potential. Here we assume that the fermion and the vector boson acquire masses through spontaneous symmetry breaking when the scalar develops a non-zero vacuum expectation value. In future sections when it is cosmologically relevant, we will also add an ``initial'' mass term for the fermion of the form $-m_{fi} \overline{\chi}\chi$, where $\chi$ will be our candidate dark matter field.

Here we assume that the scalar has no self-coupling or couplings to other scalars. Our motivation for doing so is twofold. First, such scalars can arise naturally in theories containing a scalar field with a broken shift symmetry \cite{Coleman:1969sm,Callan:1969sn,Low:2014oga}. Second, such a scalar gains a large vacuum expectation value (vev) at the minimum of its potential. This second feature can be understood by considering a scalar potential with self coupling $\lambda$,
\begin{equation}
\begin{split}
V_{0}\left(\varphi\right) & =  \frac{\lambda}{4!}\varphi^{4} - \frac{m^{2}_{S}}{2}\varphi^{2}. 
\end{split}
\end{equation}
Assuming both $m^{2}_{S}$ and $\lambda$ are greater than zero, the minimum of this potential is given by
\begin{equation}
\left< \varphi \right> \sim \frac{m_{S}}{\sqrt{\lambda}}.
\end{equation}
Clearly, a large vacuum expectation value can be generated either through a large scalar mass or a small self-coupling. Of course if the scalar is coupled to other fields, its self-coupling cannot be arbitrarily small, as it will receive radiative corrections from coupling to other fields. In our setup, a small loop-induced effective quartic coupling $\lambda_{eff}$ will arise from the scalar field's couplings to a vector boson and a fermion. In the case that the effective self-coupling is generated at loop level by the fermion coupling, $\lambda_{eff} \propto y^4$. In this case, if $y$ is small, $\lambda_{eff}$ will be very small, resulting in a very large scalar vev. 

Through its Yukawa coupling to the scalar field, the fermion will obtain a mass. However, it is interesting to note that for the setup described above, the Yukawa mass for the fermion can actually increase as the Yukawa coupling shrinks, since in the absence of other couplings, the Yukawa term would scale as $m_{\chi} \sim y m_S/ \sqrt{\lambda} \sim m_S/y$. We explore this feature in more detail in Section \ref{sec:massboost}. One might also consider a fine-tuned scenario, where both the Yukawa coupling and final scalar vacuum expectation value are large. As we will see in Eq. \eqref{eq:potential}, the loop-level fermion and vector contributions to the scalar's self-coupling carry opposite signs and can be tuned to cancel, thereby generating an extremely small $\lambda_{eff}$. In this fine-tuned case, the size of the vacuum expectation value is then an interplay between the scalar mass, the size of the Yukawa and gauge couplings, and the degree of tuning between them. 

Spontaneous symmetry breaking in this model is realised through radiative corrections, similar to the Coleman Weinberg (CW) \cite{Coleman:1973jx} mechanism. An important difference to highlight is that while CW assumes the scalar is massless but self-interacting at tree level, we assume that the scalar is massive and its self-interaction parameterized by $\lambda_{eff}$ are generated at loop-level through coupling to other fields,
\begin{equation}
\label{eq:V}
    V\left(\varphi\right) =\frac{\lambda_{eff}}{4!}\varphi^{4} - \frac{m^{2}_{S}}{2}\varphi^{2}.
\end{equation}
A comment about the potential can be made here. Since $\lambda_{eff}$ is zero at tree level, the potential may at first appear tachyonic. Specifically, in Eq.~\eqref{eq:V}, if $m_{S}^{2} >$ 0 the tree level potential is unbounded from below. However, as we show after computing the one loop effective potential, quantum effects stabilise the potential. For the parameters we will consider hereafter, the potential is bounded from below and the physical mass of the scalar - given by the second derivative of the potential evaluated at its minimum - is positive, meaning that the system has no tachyonic degrees of freedom. Here as in the CW model the scalar $\varphi$ obtains a vacuum expectation value which depends on its coupling to other fields.

The method we will use for studying loop effects on spontaneous symmetry breaking is effective potentials. Here we take care to define what we mean by ``effective potentials" since the term can appear in a few different contexts in the literature. The first arises when one is mainly looking to study the low energy regime of a theory by integrating out heavy degrees of freedom whose effects are unimportant at energies far below the high energy cutoff of the theory. The second context, which is the one considered in this work, consists of integrating out quantum fluctuations of fields in the presence of classical sources \cite{Andreassen:2014eha}. While there are different ways of doing this, the most widely used technique for one loop computations is the background field method, where all the fields in the action are split into a classical background piece (c) and a quantum piece (q). In our case this means shifting the scalar, vector, and fermionic fields as
\begin{equation}
\begin{split}
    \varphi & \rightarrow \varphi_{c} + \varphi_{q} \\
    A^{\mu} & \rightarrow A^{\mu}_{q} \\
    \chi & \rightarrow \chi_{q}
\end{split}
\end{equation}
respectively. Here, as is in prior literature \cite{PhysRevD.10.3455}, we turn on a background field for the scalar. When we then integrate out the pieces quadratic in the quantum fields, we are left with an effective potential that depends only on the classical background field $\varphi_{c}$, which is a space-time independent constant (in general it need not be) and encodes all the one particle irreducible graphs of the original theory at one loop. We have been quite sparse in our review the background field method as there are already many helpful references on this topic in the literature. For examples of background field methods applied to potentials significant to cosmology like those considered here, we refer interested readers to \cite{Quiros:1999jp,Dine:1992wr,Patel:2011th} for a review. 

The one-loop effective potential, which we evaluate in the Landau gauge, contains a divergent piece and an ``effective quartic" interaction - effective because it is generated from loop-level corrections and so takes the form coupling multiplied by $\varphi^{4}$ - and logarithmic pieces. In more detail, we take the scalar in our theory to be complex and so there are actually two real scalar fields, $\varphi_{1}$ and $\varphi_{2}$. However, the effective potential in the Landau gauge only depends on $\varphi = \sqrt{\varphi^{2}_{1} + \varphi^{2}_{2}}$. This introduces factors of two in our loop calculations which we can absorb in normalizations for the scalar field. In the interest of clarity, we therefore use $\varphi$ throughout our calculations. If we had a tree level quartic coupling, we would also have another divergent piece from the $\lambda\varphi^{2}$ term that would require renormalization using the mass counter-term. However, in our current setup this mass renormalization occurs at second order in perturbation theory and will not be considered in this work. Here we use the following renormalization conditions: 
\begin{subequations}\label{eq:renormconditions}
\begin{align}
 V^{\prime\prime}\left(\varphi\right) \vert_{\varphi = 0} & =-  m_{S}^{2} \label{eq:CWRenormMass}\\
 V^{\prime\prime\prime\prime}\left(\varphi\right) \vert_{\varphi = M} & = \frac{9}{16\pi^2}\left(4y^4 - 3g^4\right) \label{eq:CWRenormQuartic} 
\end{align}
\end{subequations}
where primes denote derivatives with respect to $\varphi$. These are similar to the renormalization conditions in CW, but with the loop-generated quartic substituted for a tree-level quartic in~\eqref{eq:CWRenormQuartic}. 

With these conditions, the renormalized one-loop effective potential at scale $M = \left< \varphi \right> \equiv \nu$ is
\begin{equation}
V\left(\varphi\right) = -\frac{m^{2}_{S}}{2}\varphi^2 + \frac{17}{192\pi^2}\left(4y^{4} - 3g^{4}\right)\varphi^{4} - \frac{4y^{4} - 3g^{4}}{64\pi^2}\log\left[\frac{\varphi^{2}}{\nu^{2}}\right]\varphi^{4}.
\label{eq:potential}
\end{equation}
The vacuum expectation value at one loop is given by the zero of the first derivative of the potential at scale $\varphi = \nu$,
\begin{equation}\label{eq:vev}
\nu^{2}  = \frac{96\pi^{2}}{31}  \frac{m_{S}^{2}}{\left(4y^{4} - 3g^{4}\right)}, 
\end{equation}
where we emphasize that this expression is valid to one loop order at $\varphi= \nu$, at which field point logarithmic terms like those in Eq.~\eqref{eq:potential} are absent for the one loop expression.
Imposing that spontaneous symmetry breaking happens, $\nu^{2} > 0$, requires $4y^{4} > 3g^{4}$. Of course, this spontaneous symmetry breaking will also generate a mass for the fermion $m_{f} = y\nu$ and the vector field $m_{V} = g \nu$. There are two ways in which this would result in a large vev for $\varphi$ and corresponding large masses for the fermion and vector boson. First, if the Yukawa and gauge couplings are small, $g,y \ll 1$, this can result in a large vacuum expectation value for $\varphi$. In lower regions of Figure \ref{fig:mf}, we show parameter space where a large fermion mass results from small Yukawa and gauge couplings. Secondly, even if these couplings are order unity, $g,y \sim 1$, in the case that $4y^4 \approx 3 g^4$, the fermionic and vector contributions may cancel to some degree, again leading to heavy fermions and vector bosons but a comparatively light scalar. However, we note that if a large $\nu$ is generated via a cancellation between the gauge and Yukawa couplings, some additional consideration may be required. The results here have been computed at one loop level, and so determining cancellation between couplings to a certain precision would require a more precise two loop treatment. While a two loop calculation is beyond the scope of this work, we can gain some insights into the effect of higher order loops using parametric estimates. The leading two loop corrections to the effective potential from the gauge and Yukawa sectors are $\mathcal{O}\left(g^{6}\right)$ and $\mathcal{O}\left(y^{6}\right)$ respectively, and as long they are both perturbatively smaller than the leading terms $\mathcal{O}\left(g^{4}\right)$ and $\mathcal{O}\left(y^{4}\right)$, higher order effects should be subdominant. In our setup we anticipate that two loop corrections will not significantly affect our results, so long as $g/y \lesssim 1.07$, since for the strongest couplings we consider $g =1.07, y= 1$, the leading terms correcting the one loop $\lambda_{eff}$ result should scale like $\frac{g^2}{16 \pi^2} \sim \frac{y^2}{16 \pi^2} \sim 0.007$, compared to the larger one loop term $4y^{4} - 3g^{4} \approx 0.07$. In future work it might be interesting to explicitly calculate the effective potential at two loop order in perturbation theory.

A few additional remarks are in order. We define the physical mass of the scalar, $M_{S}$ as the second derivative of the potential at $\varphi = \nu$,
\begin{equation}
 M^{2}_{S}  = V^{\prime\prime}\left(\varphi\right) \vert_{\varphi = \nu} = - m^{2}_{S} + \frac{27}{32\pi^{2}}\left(4y^{4} - 3g^{4}\right)\nu^{2} 
 \label{eq:massdiscussion}
\end{equation}\\
It can be verified by substituting in the expression for $\nu$ from~\eqref{eq:vev} that the square of this scalar physical mass, $M^{2}_{S}$, is positive.  
For completeness, we also comment on gauge invariance for the scalar potential. A detailed discussion of gauge dependence in potentials like the one we consider can be found in Ref.~\cite{Andreassen:2014eha}. In our derivation, we have performed our calculations in Landau gauge where the gauge fixing parameter (usually called $\xi$ in the literature) does not appear. For our potential we anticipate that as expected (and proven for the CW model in \cite{PhysRevD.10.3455}), once couplings are appropriately fixed at renormalization scale $M$, our results are gauge invariant.

\section{First Order Phase Transitions \& Gravitational Waves}
\label{sec:phasetransition}
We now turn to the cosmological evolution of our simplified model in a radiation dominated background. When thermal effects are included, the effective potential~\eqref{eq:potential} acquires temperature dependent pieces
\begin{equation}\label{eq:temppotential}
\begin{split}
& V\left(\varphi, T\right) = -\frac{m^{2}_{S}}{2}\varphi^2 + \frac{17}{192\pi^2}\left(4y^{4} - 3g^{4}\right)\varphi^{4} - \frac{\left(4y^{4} - 3g^{4}\right)}{64\pi^2}\log\left[\frac{\varphi^{2}}{\nu^{2}}\right]\varphi^{4} \\
& + \frac{T^{4}}{2\pi^{2}}\int_{0}^{\infty} dx \ 3x^{2}\log\left[1-e^{-\sqrt{x^2 + \frac{g^{2}\varphi^{2}}{T^{2}}}}\right] - 4 x^{2}\log\left[1+e^{-\sqrt{x^2 + \frac{y^{2}\varphi^{2}}{T^{2}}}} \right]\,.
\end{split}
\end{equation}  
The integrals above have no known analytic forms and must be evaluated numerically. However, it is possible to expand this potential in the high limit, where $\frac{y\varphi}{T} \ll 1$ and $\frac{g\varphi}{T} \ll 1$. The effective potential then takes the form
\begin{equation}\label{eq:temppotentialhighT}
\begin{split}
V\left(\varphi, T\right) = D\left(T^{2} - T_{0}^{2}\right)\varphi^2 -E T \varphi^{3} + \frac{\lambda\left(T\right)}{4}\varphi^{4}\,,
\end{split}
\end{equation}  
where we define the following terms 
\begin{subequations}
\begin{align}
E & = \frac{3 g^{3}}{4\pi}\,, \\
D  & = \frac{\left(4y^{2} + 3g^{2}\right)}{24}\,, \\
T_{0} & = \sqrt{\frac{m_{S}^{2}}{2D}} =  \sqrt{\frac{12 m_{S}^{2}}{\left(4y^{2} + 3g^{2}\right)}} \,,\\ 
\lambda\left(T\right) = \frac{17}{48\pi^2}\left(4y^{4} - 3g^{4}\right) + \frac{1}{16\pi^{2}}&\left(4y^{4}\log\left[\frac{y^{2}\nu^{2}}{a_F T^{2}}\right] - 3g^{4}\log\left[\frac{g^{2}\nu^{2}}{a_B T^{2}}\right]\right)\,,
\label{eq:thermalparameters}
\end{align}
\end{subequations}
which can be compared to similar terms defined in \cite{Dine:1992wr}.
Here a few numerical constants have been added, $\log a_{F} \approx 1.14$ and $\log a_{B} \approx 3.91$. These constants are obtained through matching numerical solutions of Eq.~\eqref{eq:temppotential} to Eq.~\eqref{eq:temppotentialhighT}. The temperature dependent extrema of this potential, not including the one at zero, are given by
\begin{equation}\label{eq:vevT}
\nu_{\pm} (T) = \frac{3ET}{2\lambda} \pm \frac{1}{2\lambda }\sqrt{\left(9E^{2} - 8\lambda D\right)T^{2} + 8\lambda D T_{0}^{2}}\,.
\end{equation}
We note that the cubic term $\propto ET$ is responsible for creating a barrier between the two extrema, which is typical of a first order phase transition. 

Before detailing the phase transitions facilitated by this potential, let us first review the qualitative features of a first order phase transition. An analogy can be made to boiling water in a container. As the temperature inside the container is raised to the boiling point of water, the vapour phase is energetically favoured; the liquid phase becomes the false vacuum and the vapour phase becomes the true vacuum. A growing vapour bubble converts regions of liquid into vapour (in other words, converts regions of the false vacuum state into the true vacuum state), and soon the entire container is filled with vapour, the new true vacuum. For cosmological phase transitions where the universe is undergoing expansion and dilution, matters proceed in the opposite temperature direction. 

\begin{figure}
    \centering
    \includegraphics[width=.495\textwidth]	{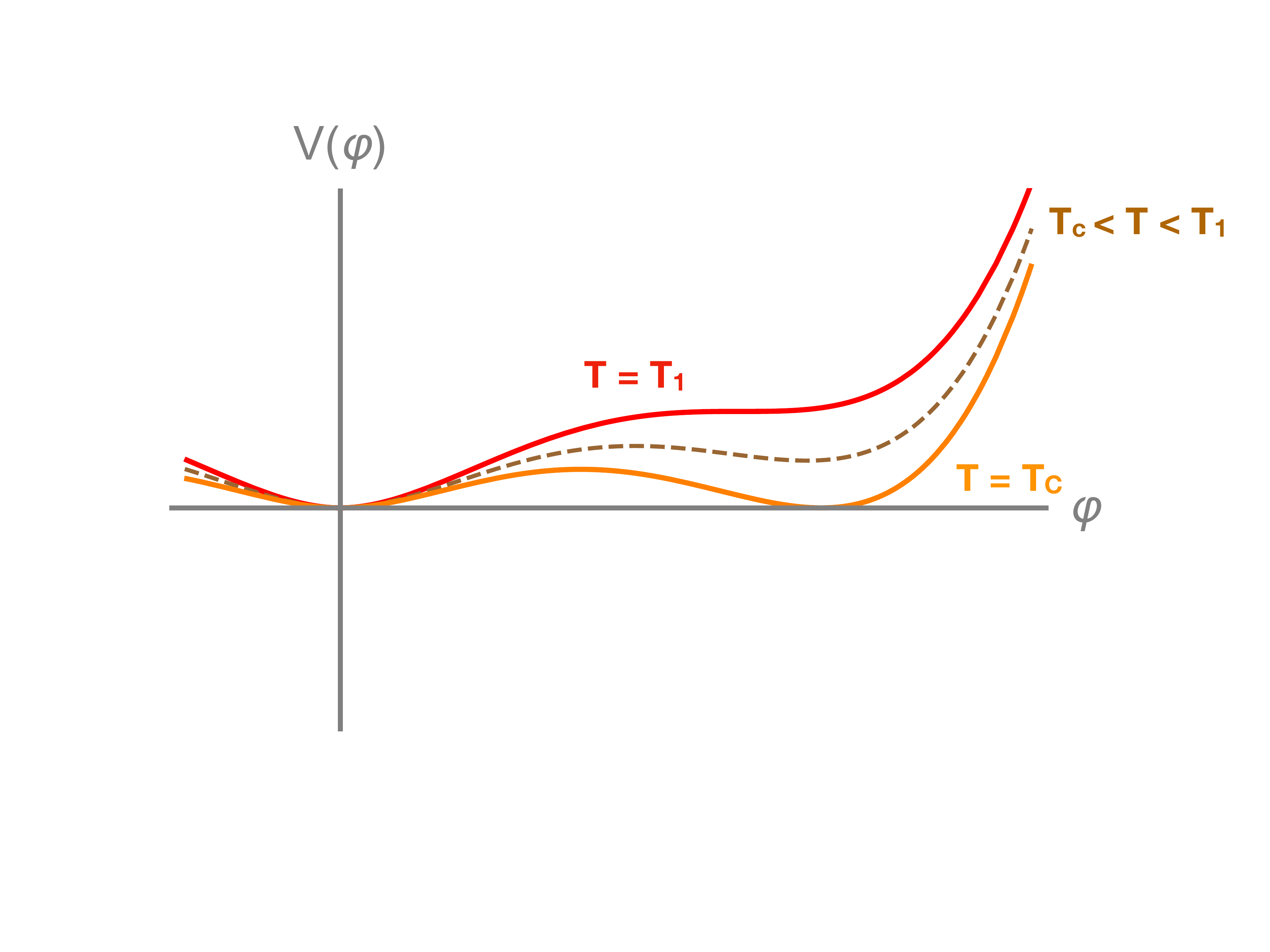}
    \includegraphics[width=.495\textwidth]	{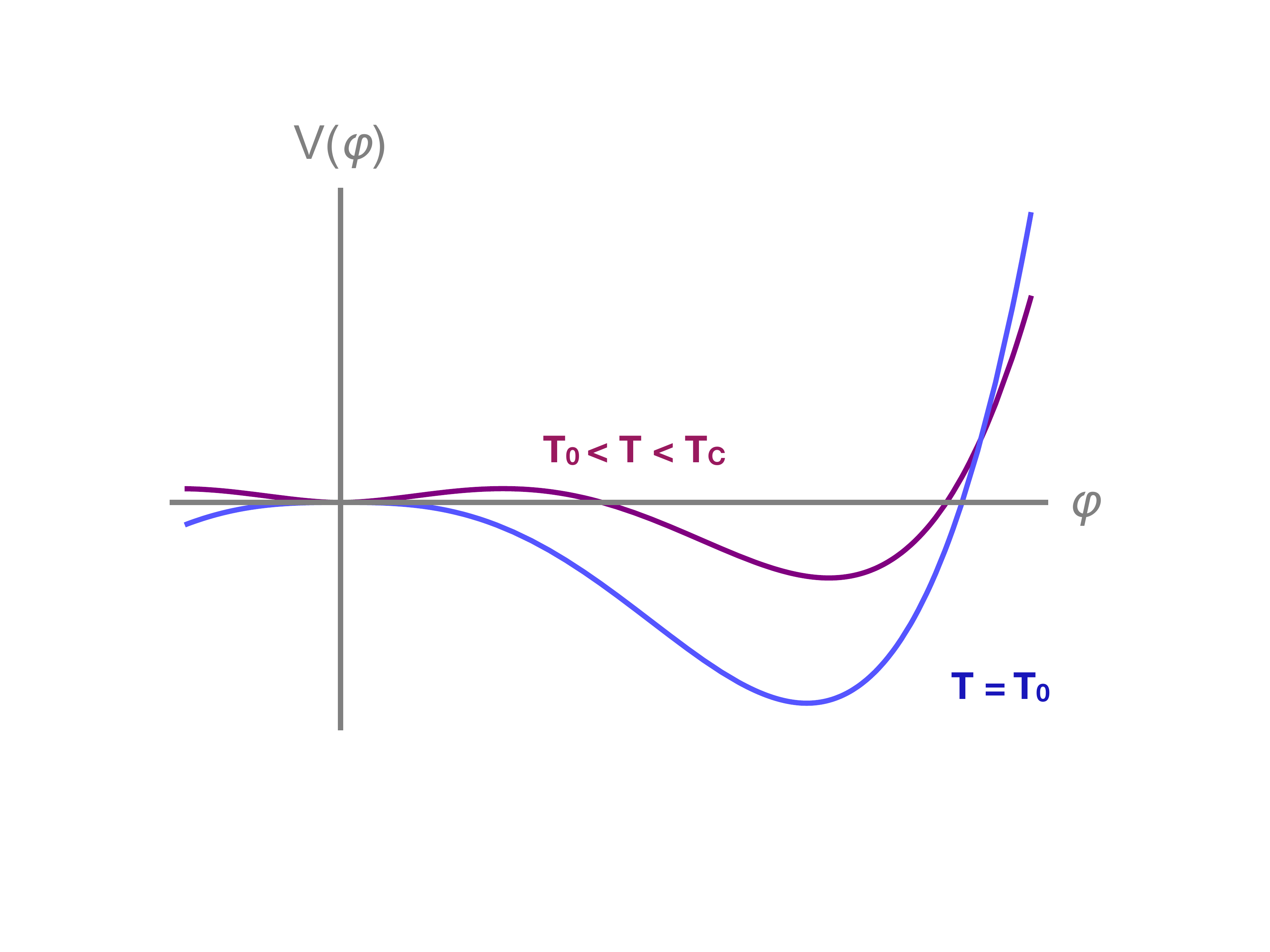}
    \caption{The evolution of the potential studied in this work is shown as a function of temperature. \emph{Left panel:} At very high temperatures, the potential has a unique minimum at zero. As the temperature lowers to $T_{1}$, a local inflection point appears away from zero. Below temperature $T_{1}$ the potential develops a new minimum at $\varphi = \nu_{+}$, sitting at higher energy than the origin. The dashed line shows that as the universe cools further, the $\nu_{+}$ minimum lowers. At the critical temperature $T_{C}$, the $\nu_{+}$ minimum is at the same level as the origin. At temperatures below $T_{C}$ a first order phase transition begins. \emph{Right panel:} Between temperatures $T_{C}$ and $T_{0}$ the potential has a minimum at the origin and another minimum at $\nu_+$. In that temperature range, phase transitions proceed through bubble nucleation. Below $T_{0}$, the potential has no minimum at the origin, and bubble nucleation ceases. For the models we study, the phase transition is complete before $T=T_0$.}
      \label{fig:sfo}
\end{figure}

In Figure \ref{fig:sfo}, we show the shape our potential takes during a first order phase transition. At very early times (high temperatures) the global minimum of the scalar field undergoing the phase transition is at a field value of zero; the universe is said to be in the symmetric phase. A local inflection point forms away from zero at a temperature $T_{1}$. As the temperature drops below $T_{1}$, a local maximum and local minimum form, given by the $\nu_{-,+}$ solution of~\eqref{eq:vevT} respectively. At this temperature the energy of the $\nu_+$ vacuum state is higher than the energy of the minimum at the origin - the $\nu_+$ minimum is energetically disfavoured. However, as the temperature decreases further the local minimum at $\nu_{+}$ descends until we reach a temperature $T_{C}$ at which it is energetically degenerate with the minimum at the origin. For temperatures $T < T_{C}$, the $\nu_+$ minimum is energetically favoured. 

By definition, the first order phase transition begins at $T < T_{C}$. Early on during the phase transition when $ T \lesssim T_{C}$, there is a minimum at $\nu_{+}$. At this point the universe is still in the vacuum state located at the origin. As the phase transition progresses, bubbles of the (now) lower energy $\nu_+$ phase nucleate and grow. These growing bubbles expand, percolate, and collide to create larger and larger regions in the $\nu_+$ phase. If we imagine for a moment that the universe is static, then the phase transition completes when the entire universe is in the $\nu_+$ phase - all the false vacuum regions have been converted to the true vacuum state. However in an expanding universe two important factors need to be considered: (1) the bubble nucleation and growth process needs to be fast compared to Hubble expansion and (2) an expanding universe cools down, which affects the nature of the phase transition. For the phase transition to be first order and proceed through bubble nucleation, the potential needs to have two minima: in this case at the origin and at $\nu_+$. However, at some temperature $T_{0}$, the minimum at the origin vanishes, and so the phase transition is no longer first order for temperatures below $T_{0}$. Hence for our purposes, it is useful to determine whether the phase transition completes quickly relative to the rate of Hubble expansion, before the universe reaches a temperature $T_{0}$. An important temperature in this regard is the nucleation temperature $T_{N}$, at which the bubble nucleation rate per unit volume per unit time is comparable to the Hubble rate of the universe. At $T_{N}$ the phase transition is efficient and the entire universe is converted to the $\nu_+$ phase. If $T_{N} > T_{0}$, it is safe to assume the entire transition is first order -- this criterion for first order phase transitions was first set out in \cite{Anderson:1991zb}. The nucleation temperature $T_N$ will also be the characteristic temperature for the production of gravitational waves via bubble collisions and associated acoustic waves generated in the primordial plasma.

We now begin a quantitative treatment of first order phase transitions caused by the scalar potential in Eq.~\eqref{eq:temppotentialhighT}. First, we analyze the potential at the origin ($\varphi = 0$) to validate some of our prior description of the potential for temperatures above and below $T_0$. Note that as depicted in Figure \ref{fig:sfo}, $T=T_0$ occurs well after any first order phase transition has completed. Because the extrema of the potential at various temperatures determine the characteristics of its phase transition, we consider the second derivative of Eq.~\eqref{eq:temppotential},
\begin{equation}\label{eq:D2V}
\frac{\partial^{2}V\left(\varphi,T\right)}{\partial \varphi^{2}} = 2D\left(T^{2} - T_{0}^2\right) - 6E T \varphi + 3\lambda\varphi^{2}.
\end{equation}
It can be seen from~\eqref{eq:D2V} that the extremum at $\varphi = 0$ is a minimum for temperatures $T > T_{0}$, a saddle point at $T = T_{0}$, and a maximum when $T < T_{0}$. At $T = T_{0}$, there is only one non-zero extremum, the minumum at $\nu_{+}\left(T_{0}\right) = \frac{3ET_{0}}{\lambda\left(T_{0}\right)} \equiv \nu_{T_{0}}$.

Now we turn to the dynamics of the potential around temperature $T_{1}$, which is defined by the formation of an inflection point at $\nu_+$. Until the universe reaches this temperature, there is only one minimum at the origin. At $T_1$ the term under the square root in~\eqref{eq:vevT} vanishes and the potential forms an extremum at $\varphi = \nu_+$.
The value of this extremum is found using~\eqref{eq:vevT}, $\nu_{T_{1}} \equiv \frac{3ET_{1}}{2\lambda\left(T_{1}\right)}$. Provided the temperature dependent quartic interaction term $\lambda\left(T_{1}\right)$ is positive, this is an inflection point that turns into a minimum at temperatures below $T_{1}$. In our case, $\lambda\left(T_{1}\right) < 0$ only occurs at extremely high temperatures above the Plank scale, well beyond the region of validity for our calculations. Initially, $V\left(\nu\left(T_{1}\right), T_{1}\right) > 0$ and the minimum at zero is energetically favoured since $V\left(0, T_{1}\right) = 0$. However, as the universe expands and cools, we reach some temperature $T = T_{C}$ where $V\left(\nu\left(T_{C}\right), T_{C}\right) = 0$. This condition simplifies to
\begin{equation}\label{eq:TC}
T^2\left[1 - \frac{E^{2}}{\lambda D}\right] \ \Bigg|_{T=T_{C}} = T_{0}^{2}.
\end{equation}
Bubbles of true vacuum form at $T_{C}$, since at this temperature a new true minimum appears at $\nu_+$. Below $T_C$, bubbles may nucleate. To compute the volumetric rate of bubble nucleation we use a Euclidean action treatment \cite{Coleman:1977py,Callan:1977pt,Kamionkowski:1993fg},
\begin{equation}\label{eq:bubblerate}
\Gamma = A \, \exp (-S_{E}),
\end{equation}
where $A$ is a dimensionful $\mathcal{O}\left(T^{4}\right)$ term and $S_{E}$ is the Euclidean action for tunneling from $\varphi = 0$ to the new minimum. Near the time of bubble nucleation $t_{N}$, which corresponds to a nucleation temperature $T_{N}$, the Euclidean action $S_{E}$ in~\eqref{eq:bubblerate} can be Taylor expanded around time $t_N$, 
\begin{equation}
\begin{split}
S_{E} & \simeq S_E(t_N) - \left(t - t_{N}\right) \beta,
\end{split}
\end{equation}
where $\beta = - \frac{dS_{E}}{dt} \Big|_{t = t_{N}}$ can be understood as the inverse duration of the phase transition. Furthermore, the Euclidean action can be factorized into a spatial and temporal piece, $S_{E} = \frac{S_{3}}{T}$, such that
\begin{equation}
S_{3} = 4\pi \int_0^{\infty} dr \ r^{2} \left[ \frac{1}{2}\left(\frac{d\varphi}{dr}\right)^{2} - V\left(\varphi, T\right) \right],
\end{equation}
is the three dimensional Euclidean action. By solving the Euclidean action for \eqref{eq:temppotentialhighT}, an expression can be found for $\frac{S_{3}}{T}$ of the form \cite{Dine:1992wr}
\begin{align}
\frac{S_{3}}{T} & \simeq \frac{4.85}{E^{2}} \frac{\left[2D\left(T^{2} -T_{0}^{2}\right)\right]^{\frac{3}{2}}}{T^{3}} f\left(x\right)
\label{eq:S3toT}\,,\\ \vspace{1pt}
f\left(x\right) & = 1 + \frac{x}{4}\left(1 + \frac{2.4}{1 - x} + \frac{0.26}{\left(1 - x\right)^{2}}\right) \label{eq:fx}\,, \\
x  & = \frac{\lambda D\left(T^{2} -T_{0}^{2}\right)}{E^{2}T^{2}}\,. \label{eq:x}
\end{align}
With these expressions, we can now give a more precise definition of the nucleation temperature $T_{N}$.  Because we are most interested in the onset of the phase transition, we want to find the temperature at which one bubble is nucleated on average per Hubble volume ($V_{H} \equiv H^{-3}$); this nucleation temperature $T_N$ is defined by
\begin{equation}
\label{eq:Tnuc}
\int_{t_{C}}^{t_{N}} \ dt \frac{\Gamma}{H^{3}} = \int_{T_{N}}^{T_{C}} \ \frac{dT}{T} \frac{\Gamma}{H^{4}} = 1,
\end{equation}
where for the first equality we have assumed that the universe is expanding adiabatically. To simplify this expression, we can use the fact that in a radiation dominated universe the Hubble parameter is given by $H^{2} = \frac{\rho}{3M_{Pl}^2}$, where $\rho$ is the energy density and $M_{Pl}$ is the reduced Plank mass. We assume that at the time of the phase transition, only standard model particles in the plasma are relativistic, yielding $\rho = \frac{\pi^{2}}{30} \times g_{SM}T^{4}$ where $g_{SM} \approx 106.75$ is the total number of Standard Model effective relativistic degrees of freedom. Putting everything together, the definition of the nucleation temperature becomes
\begin{equation}
\left(\frac{90}{\pi^2 g_{SM}}\right)^{2} \int_{T_{N}}^{T_{C}} \ dT \frac{M^{4}_{Pl}}{T^{5}}\exp\left[- \frac{S_{3}}{T}\right] = 1.
\label{eq:TNequation}
\end{equation}
Since we expect that for strong first order phase transitions this integral is peaked around $T_{N}$, for practical purposes the approximate expression
\begin{equation}
\label{eq:Tnucdefapprox}
\frac{S_{3}}{T} \Bigg|_{T = T_{N}} \sim - 4\log\left[\frac{T_{N}}{M_{Pl}}\right]
\end{equation} \\
is often used to evaluate $T_{N}$. We note that we solve for $T_N$ explicitly with Eq~\eqref{eq:TNequation} in our numerical implementation.

Given the strong first order phase transition predicted by our model, one of its most interesting consequences and phenomenological signatures is the generation of gravitational waves (GW) from the well studied mechanism of bubble nucleation and collisions. The study of such a cosmological SGWB requires complex numerical simulations that are beyond the scope of this work. Instead, as in much of the literature about GWs from first order phase transitions, we present the most salient features of this phenomenon. The interested reader is invited to consult \cite{Espinosa:2010hh,Hindmarsh:2015qta,Caprini:2015zlo,Guo:2020grp,Kamionkowski:1993fg} and references therein for a broader discussion. As discussed in these references, the formulae we quote rely on simulations showing that a certain fraction of the latent heat released from the first order phase transition is deposited into graviational waves during the propagation of bubble wavefronts. For the regime we consider, GWs are primarily sourced by sound waves formed in the plasma after bubble collisions, but there are other secondary contributions and subtleties to consider: the bubble collisions themselves and in medium magnetohydrodynamic turbulence can become important, especially in regimes where the bubble walls are very relativistic. Other factors that can alter GW production is whether the background energy density is radiation or matter dominated and the lifetime of acoustic waves \cite{Guo:2020grp}. Contributions to produced GWs combine approximately linearly \cite{Caprini:2015zlo}:
\begin{equation}
    \Omega_{GW}h^{2} \simeq \Omega_{B}h^{2} + \Omega_{S}h^{2} + \Omega_{T}h^{2}\,,
\end{equation}
where $\Omega_{S}$ is the contribution from sound waves as the colliding bubbles dump their energy into the plasma, $\Omega_{B}$ the contribution from bubble collisions, and $\Omega_{T}$ the contribution from magnetohydrodynamic turbulence effects caused by the colliding bubbles. In this work we will focus mainly on $\Omega_{S}$, since this is expected to produce the dominant contribution to gravitational waves for the model we consider. 

In the treatment above, we have assumed that in a thermal background, the expansion of the universe is adiabatic, $\frac{dT}{dt} = - TH$, where $H$ is the Hubble parameter. This gives us an important parameter that determines the properties of gravitational waves emitted from first order phase transitions,
\begin{equation}\label{eq:gwparameter1}
\frac{\beta}{H} \Big|_{H = H_{N}} =  T  \frac{d}{dT}\left(\frac{S_{3}}{T}\right) \Big|_{T = T_{N}} \,,
\end{equation}
where $H_{N}$ is the Hubble parameter at nucleation.
In order to compute the resulting energy density of gravitational waves today, we must first compute the latent heat, or free energy density difference between the true and false vacuum at the time of the phase transition,
\begin{equation}
\epsilon = - V \left(\nu_{+},T_{N} \right) + T\frac{\partial V\left( \nu_{+},T\right)}{\partial T} \Bigg|_{T = T_{N}}\,.
\label{eq:epsilon}
\end{equation}
Another important parameter for gravitational wave generation is the ratio of the latent heat released during the phase transition to the energy density in the high energy phase
\begin{equation}
\label{eq:gwparam2}
\alpha = \frac{\epsilon\left(T\right)}{\rho_{vac}\left(T\right)} \Bigg|_{T = T_{N}}\,.
\end{equation}
Using $\alpha$, the redshifted value of the gravitational wave energy density produced by acoustic waves has been shown
\cite{Caprini:2015zlo,Hindmarsh:2015qta} to follow a power law spectrum
\begin{equation}
    \Omega_{S}h^{2} (f) = 2.6\times 10^{6}\,\kappa^{2}\,v_b \left(\frac{\alpha}{1 + \alpha}\right)^{2} \left(\frac{100}{g_{*}}\right)^{\frac{1}{3}} \left(\frac{H}{\beta}\right) S_{ac}(f)
    \label{eq:gwfrequency}
\end{equation}
where $\kappa$ is the efficiency with which GW are generated by acoustic waves and $S_{ac}\left(f\right)$ is a power law frequency spectrum obtained numerically,
\begin{equation}
    S_{ac}\left(f\right) = \left(\frac{f}{f_{ac}}\right)^{3}\left(\frac{7}{4+3\left(\frac{f}{f_{ac}}\right)^{2}}\right)^{7/2}.
\end{equation}
The peak frequency of gravitational wave production $f_{ac}$ and average bubble wall velocity $v_b$ will depend on a number of factors, including $\varphi$'s coupling to the plasma, the equation of state of the universe, and the relative velocity of bubbles as compared to the plasma's sound speed \cite{Espinosa:2010hh,Guo:2020grp}. Here we make some standard choices, that are supported by acoustic GW production simulations undertaken in \cite{Espinosa:2010hh,Hindmarsh:2015qta,Guo:2020grp}. We fix the average bubble velocity to $v_b =0.9$. For this velocity,
\begin{equation}
\label{eq:gwefficiency}
    \kappa \approx \frac{\alpha}{0.73 +0.083 \sqrt{\alpha}+\alpha}
\end{equation}
is the efficiency for transferring acoustic wave energy into gravitational energy \cite{Caprini:2015zlo,Espinosa:2010hh}, and the peak frequency is $f_{ac} \approx \frac{2 \beta}{\sqrt{3} v_b}$ which redshifts to 
\begin{equation}
    f_{ac}=1.9\times 10^{-5}\mathrm{Hz}\dfrac{1}{v_b}\left(\dfrac{\beta}{H}\right)\left(\dfrac{T_N}{100\mathrm{GeV}}\right)\left(\dfrac{g_*}{100}\right)^{\frac{1}{6}}\,.
\end{equation}

\begin{figure}[t!]
    \centering
  \begin{subfigure}[b]{0.48\textwidth}
    \includegraphics[width=\textwidth]	{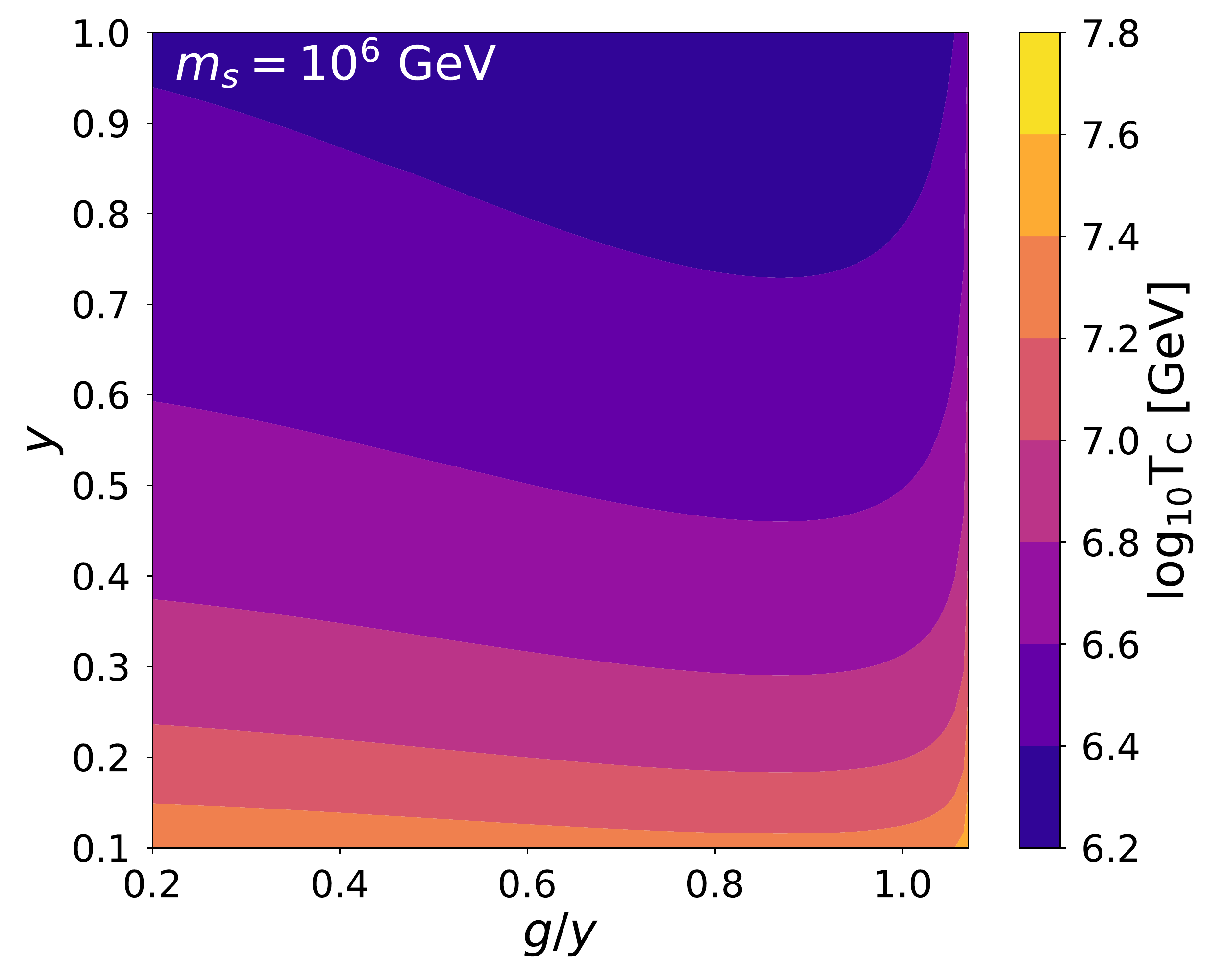}
    \label{fig:TCms1e6}
    \end{subfigure}
  \begin{subfigure}[b]{0.48\textwidth}
    \includegraphics[width=\textwidth]	{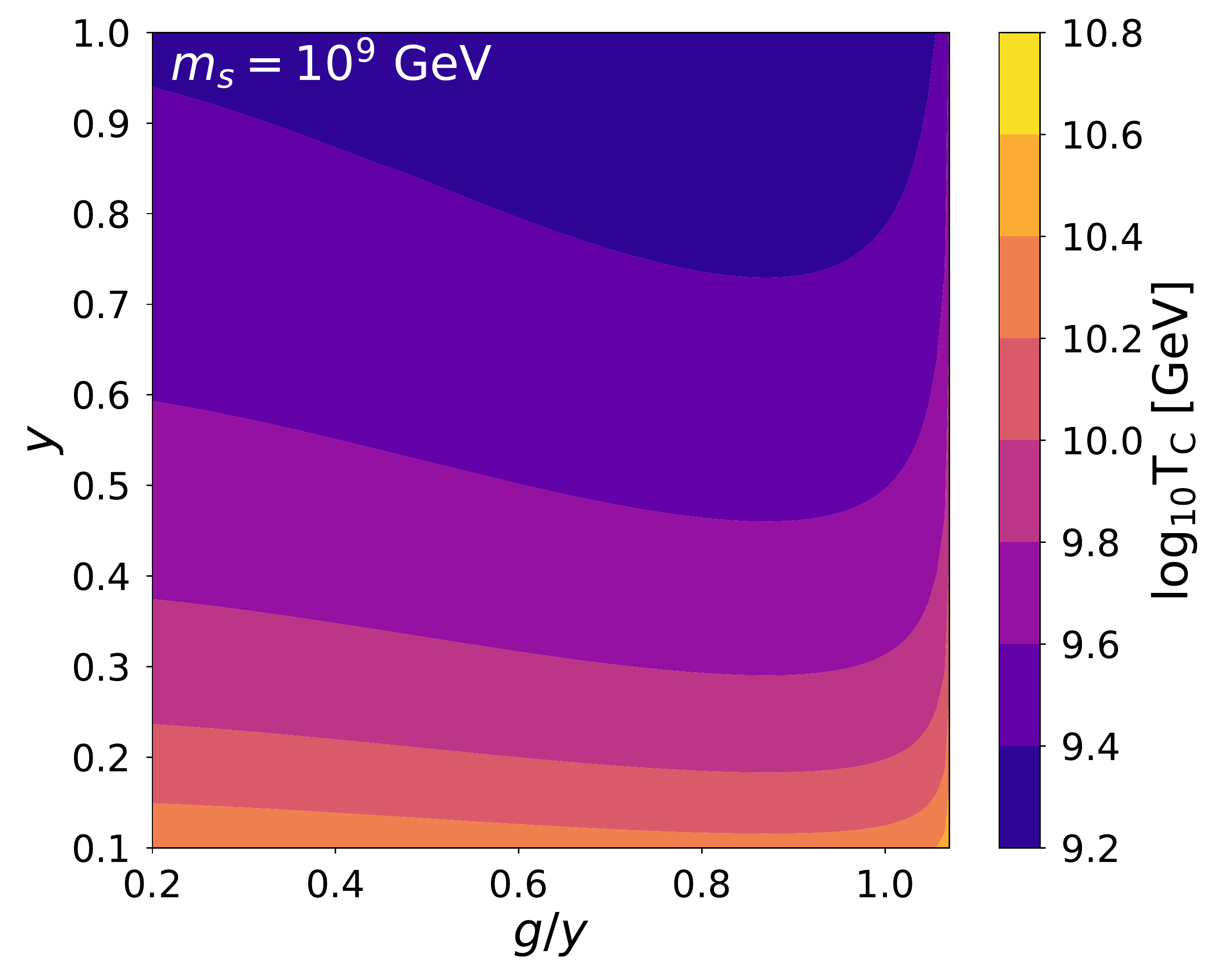}
    \label{fig:TCms1e9}
    \end{subfigure}
  \begin{subfigure}[b]{0.48\textwidth}
    \includegraphics[width=\textwidth]	{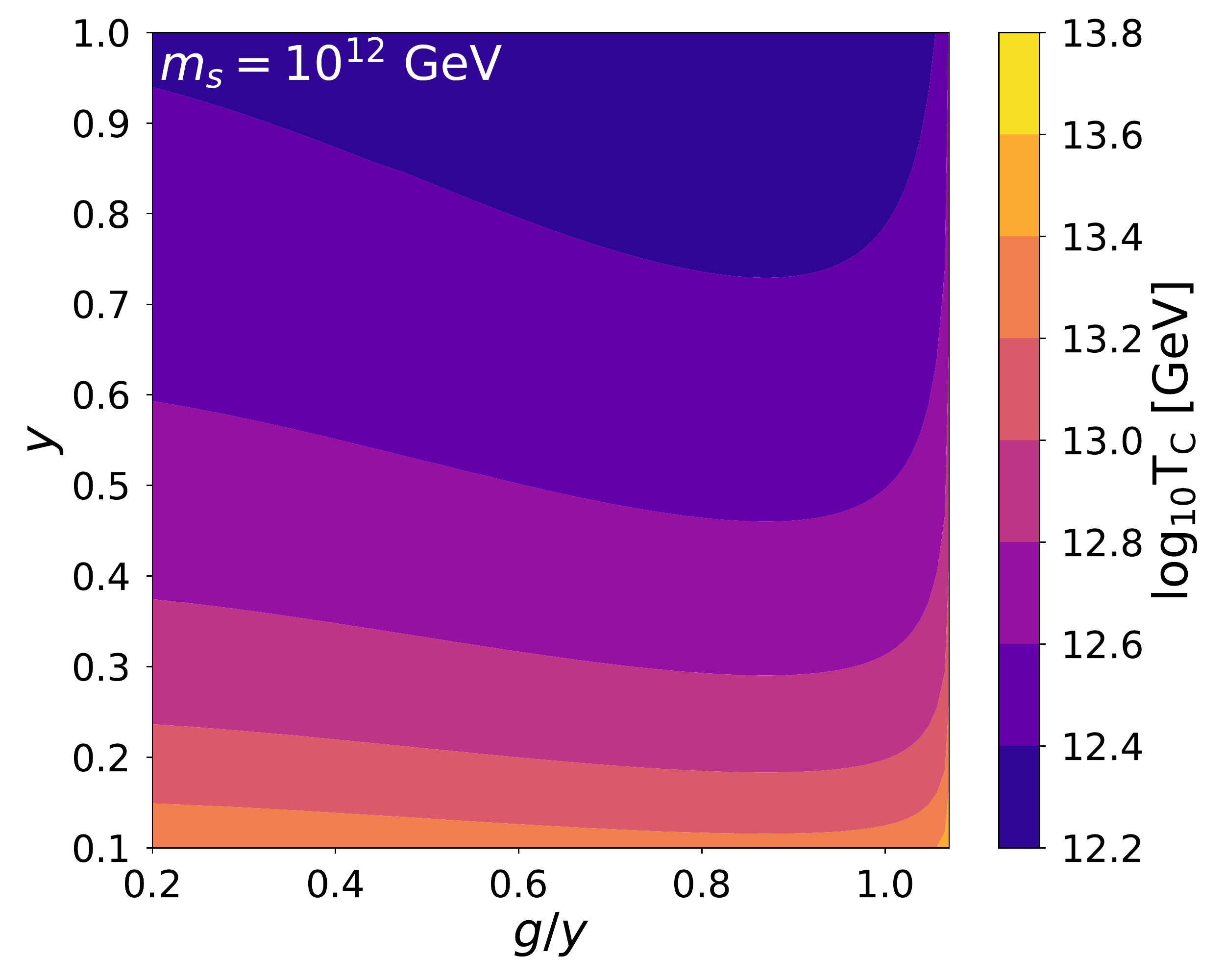}
    \label{fig:TCms1e12}
    \end{subfigure}
  \begin{subfigure}[b]{0.48\textwidth}
    \includegraphics[width=\textwidth]	{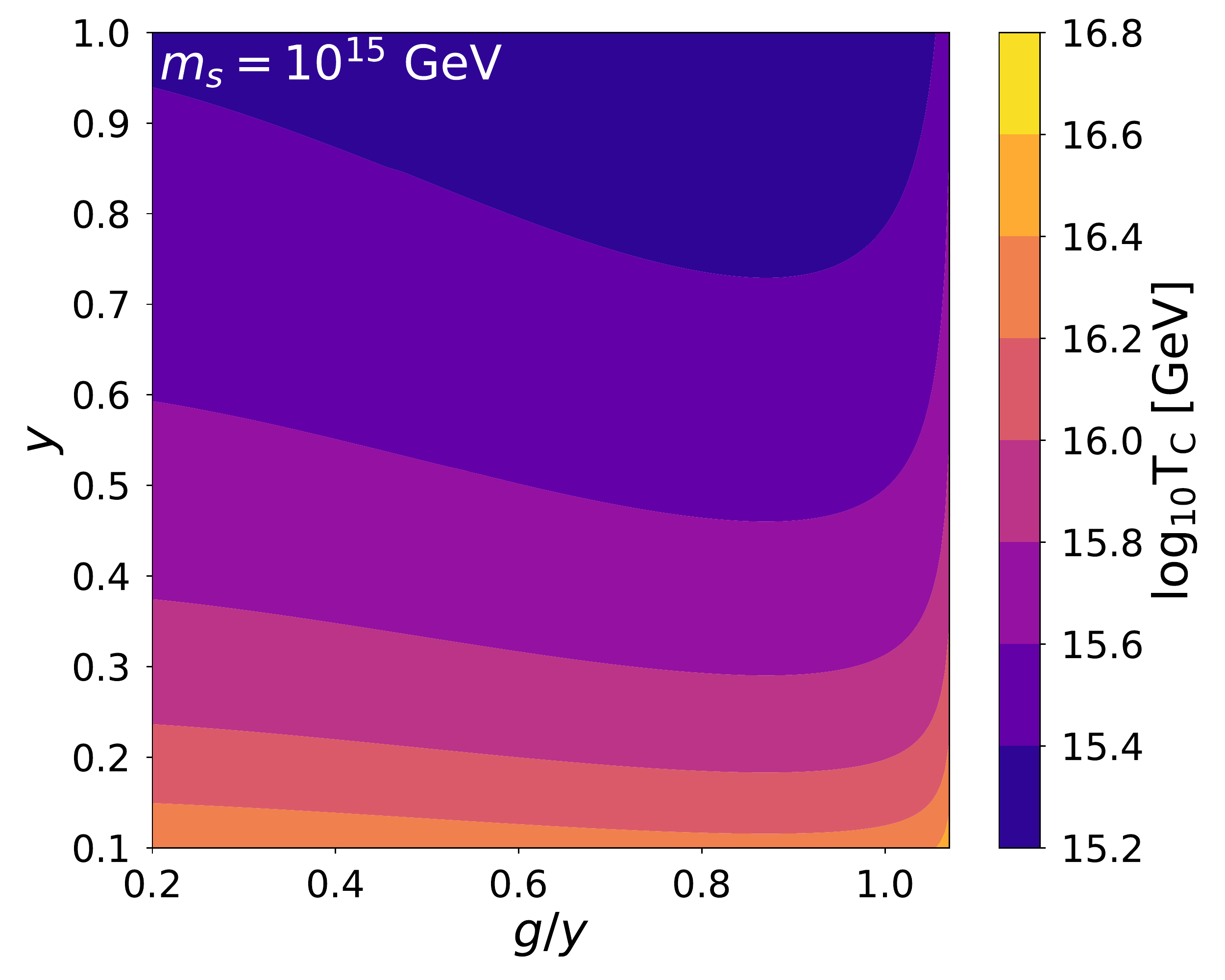}
    \label{fig:TCms1e15}
    \end{subfigure}
    \caption{The critical temperature $T_C$ in first order phase transitions for the potential given in Eq.~\eqref{eq:potential}, for scalar mass $m_S$, Yukawa coupling $y$, and gauge coupling $g$. The upper left, upper right, lower left and lower right panels correspond to the bare scalar mass $m_S=10^6~{\rm GeV},~10^9~{\rm GeV},~10^{12}~{\rm GeV}$ and $10^{15}$~GeV as indicated. The x-axis is the ratio of gauge to Yukawa coupling, which  varies between 0.2 and 1.07. Note that this and the following plots are only accurate to one-loop order, two loop corrections may be sizable for the region where $g/y \gtrsim 1.07$ (not plotted), as discussed in Section \ref{sec:Yukawa}.}
    \label{fig:TC}
\end{figure}

\begin{figure}[t!]
    \centering
  \begin{subfigure}[b]{0.48\textwidth}
    \includegraphics[width=\textwidth]	{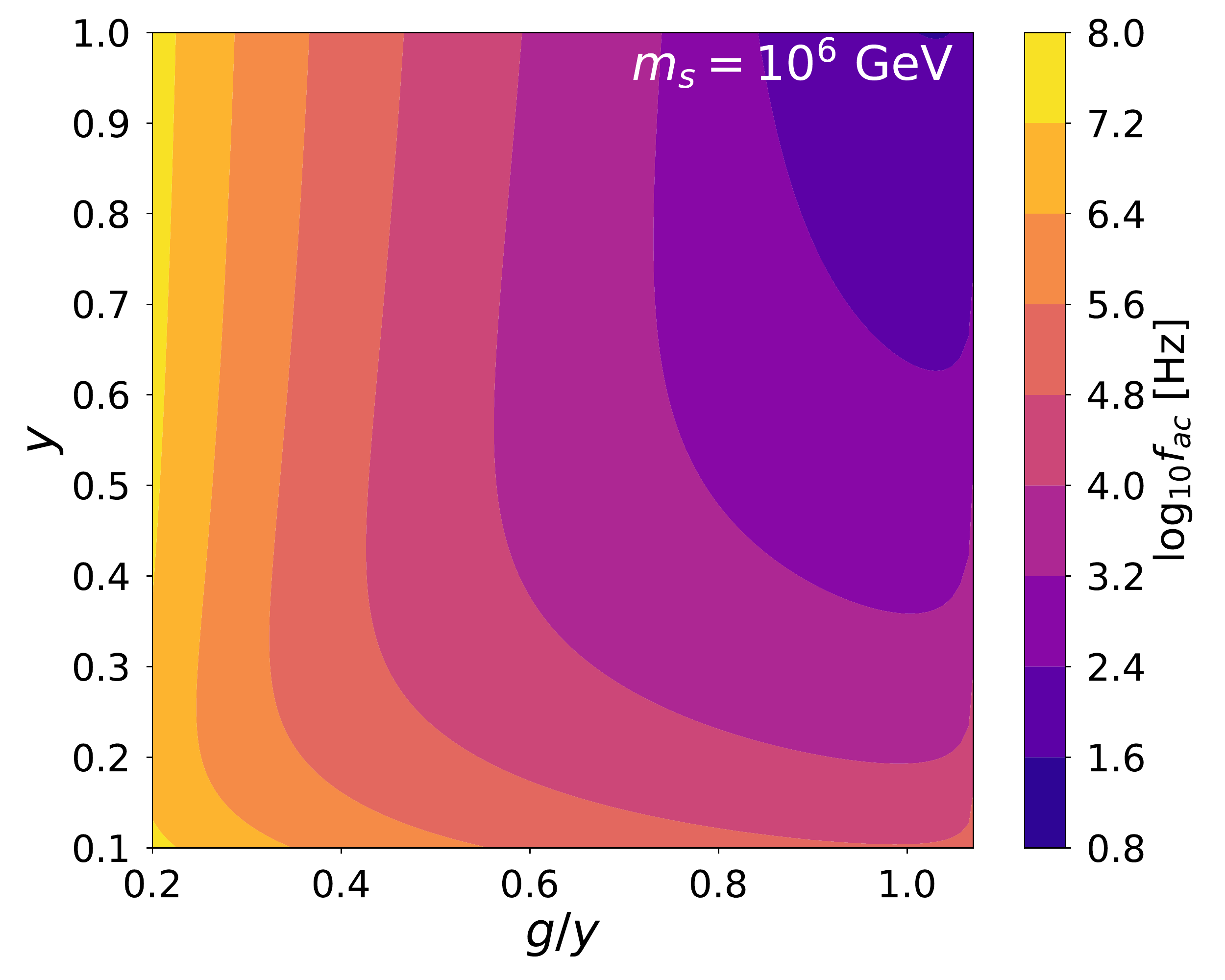}
    \label{fig:facms1e6}
    \end{subfigure}
  \begin{subfigure}[b]{0.48\textwidth}
    \includegraphics[width=\textwidth]	{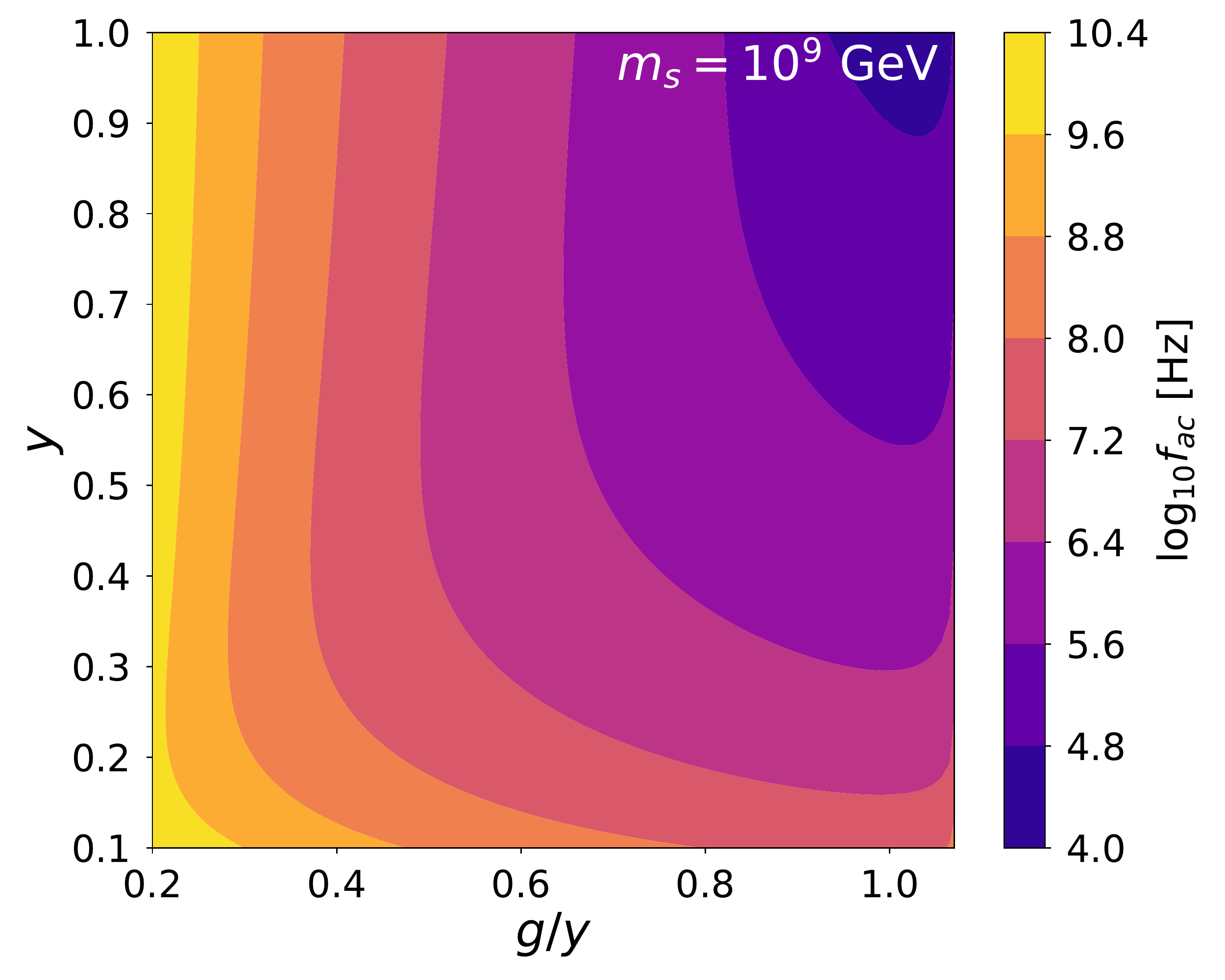}
    \label{fig:facms1e9}
    \end{subfigure}
  \begin{subfigure}[b]{0.48\textwidth}
    \includegraphics[width=\textwidth]	{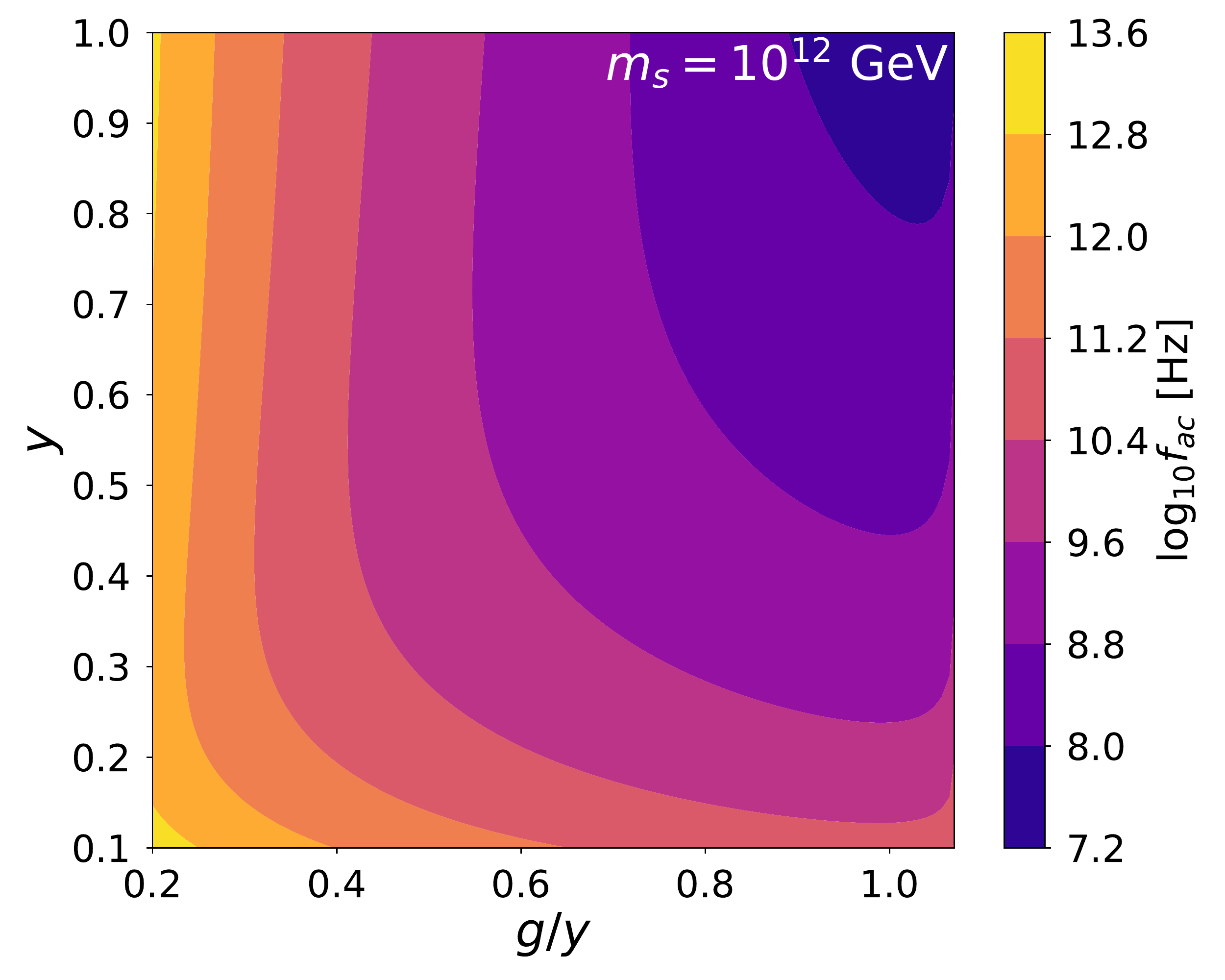}
    \label{fig:facms1e12}
    \end{subfigure}
  \begin{subfigure}[b]{0.48\textwidth}
    \includegraphics[width=\textwidth]	{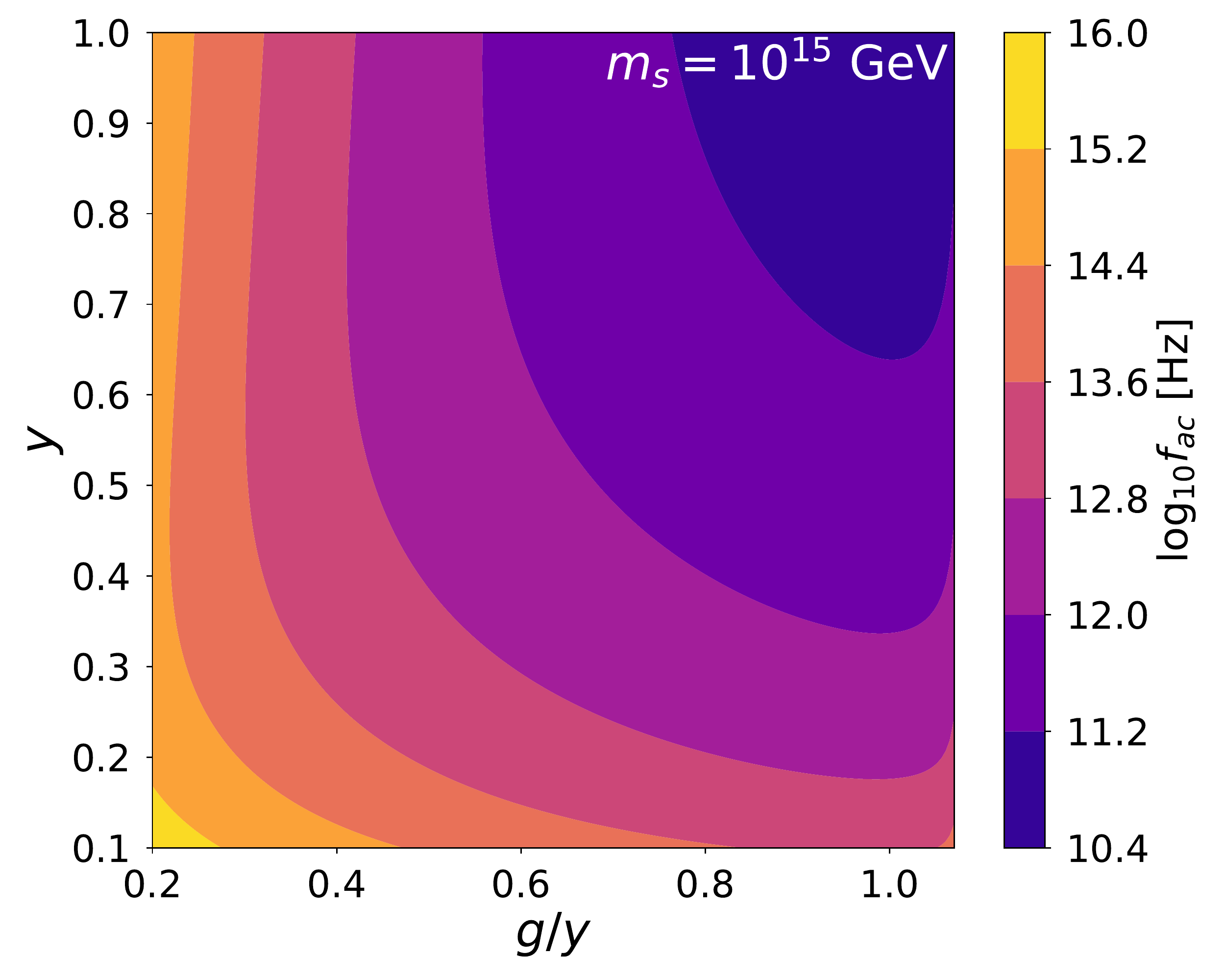}
    \label{fig:facms1e15}
    \end{subfigure}
    \caption{The peak frequencies of gravitational waves produced in a first order phase transition $f_{ac}$, for the potential given in Eq.~\eqref{eq:potential}, for scalar mass $m_S$, Yukawa coupling $y$, and gauge coupling $g$. The panels and axes are the same as in Figure~\ref{fig:TC}.}
    \label{fig:fac}
\end{figure}
\begin{figure}[t!]
    \centering
  \begin{subfigure}[b]{0.48\textwidth}
    \includegraphics[width=\textwidth]	{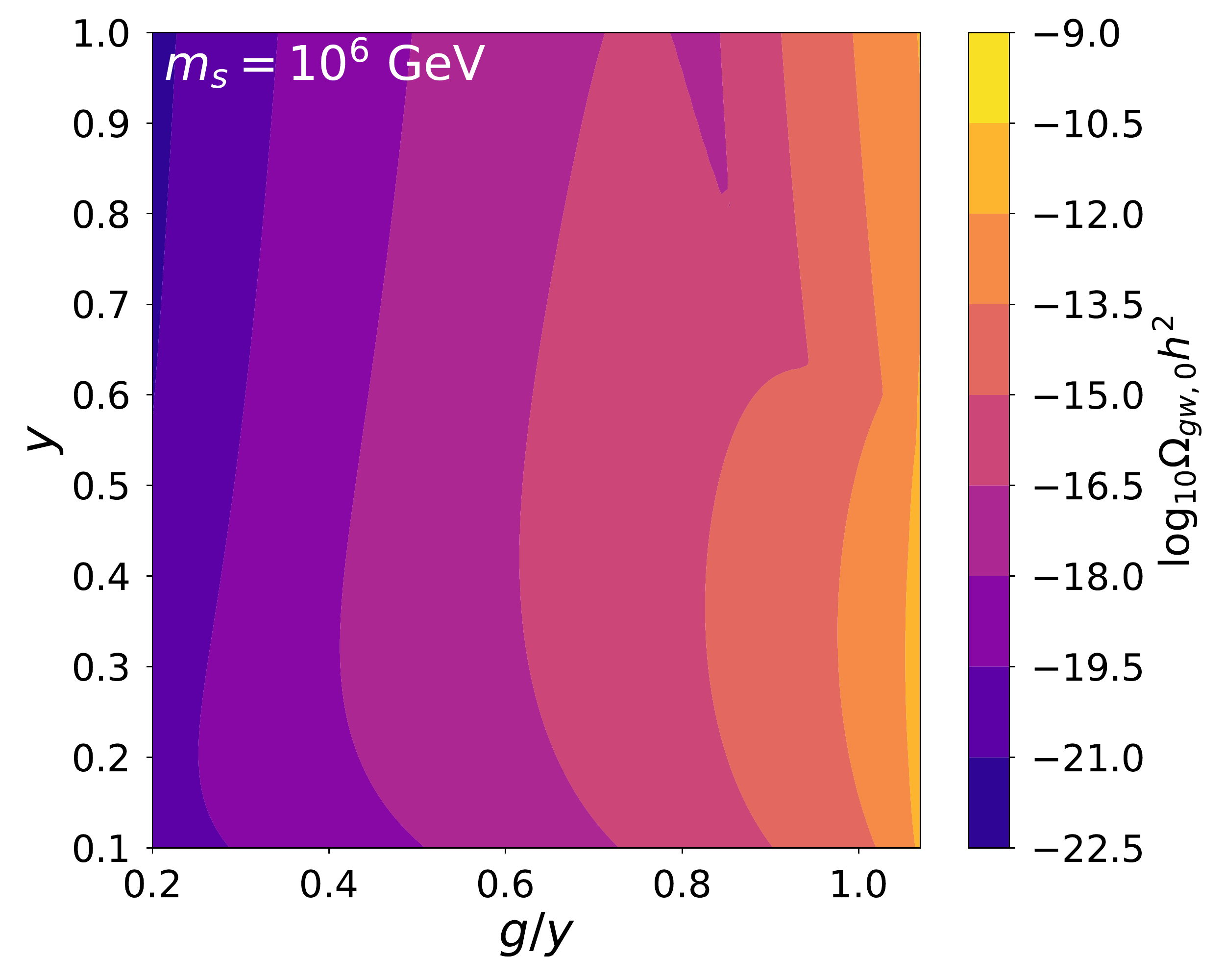}
    \label{fig:omegah2ms1e6}
    \end{subfigure}
  \begin{subfigure}[b]{0.48\textwidth}
    \includegraphics[width=\textwidth]	{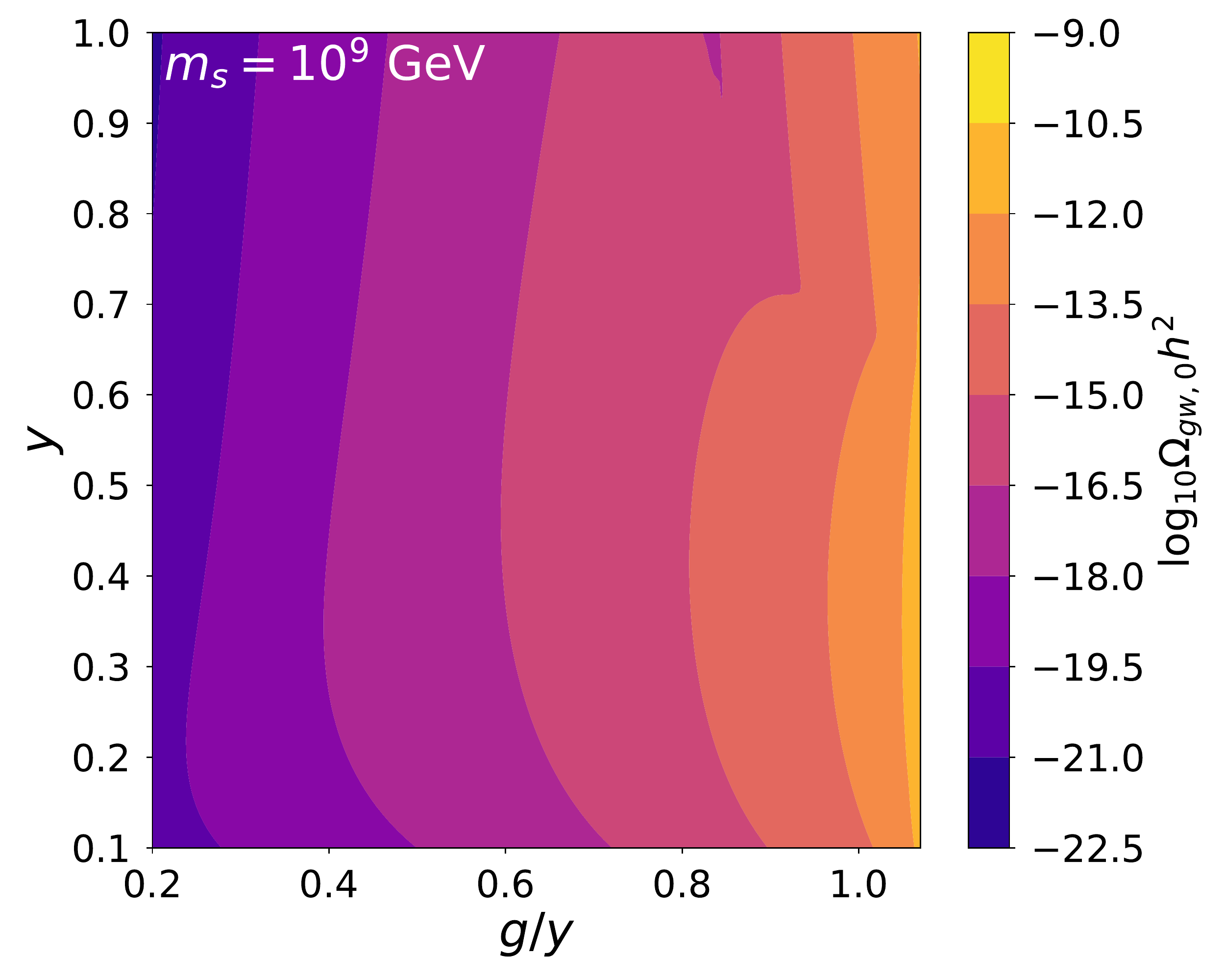}
    \label{fig:omegah2ms1e9}
    \end{subfigure}
  \begin{subfigure}[b]{0.48\textwidth}
    \includegraphics[width=\textwidth]	{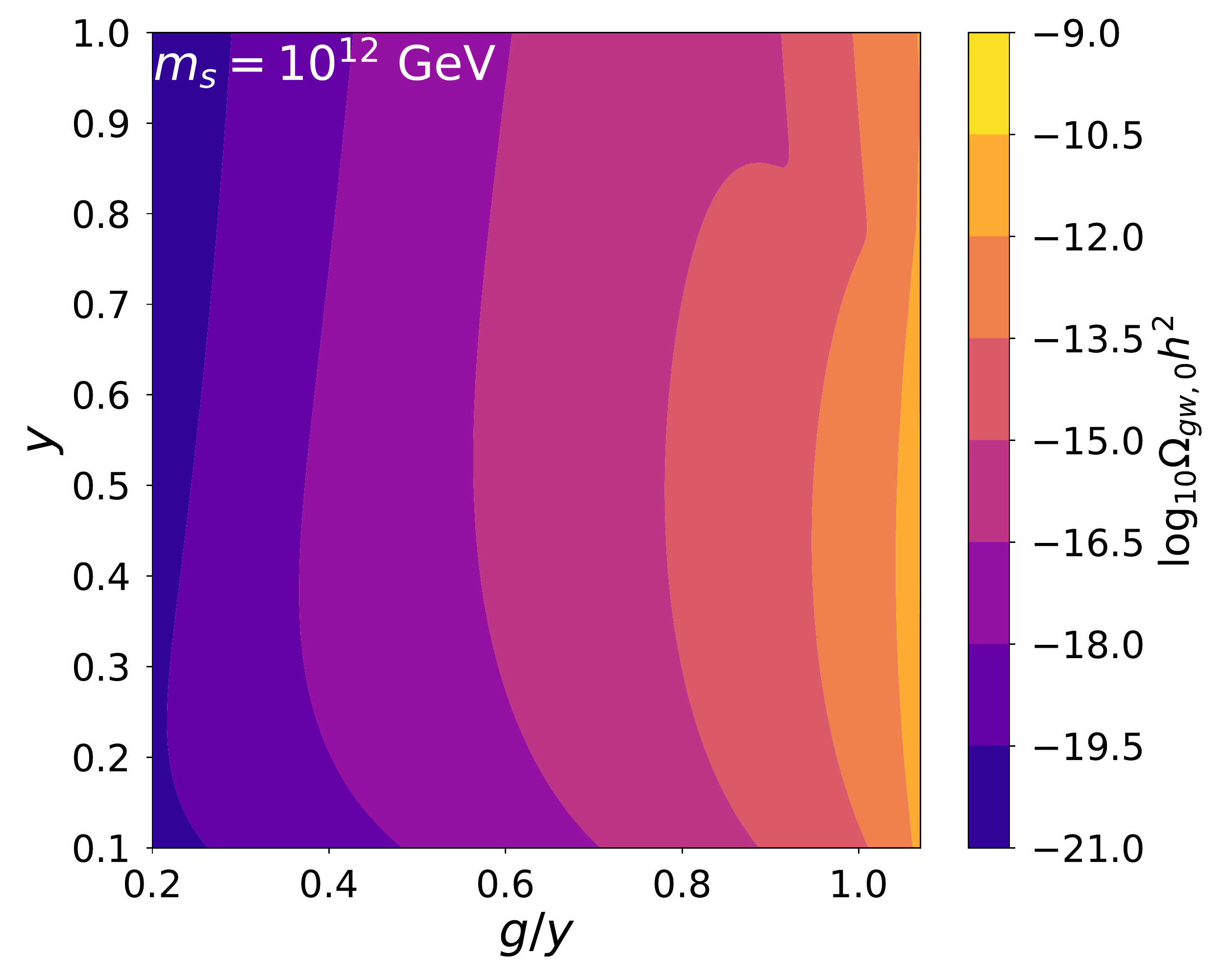}
    \label{fig:omegah2ms1e12}
    \end{subfigure}
  \begin{subfigure}[b]{0.48\textwidth}
    \includegraphics[width=\textwidth]	{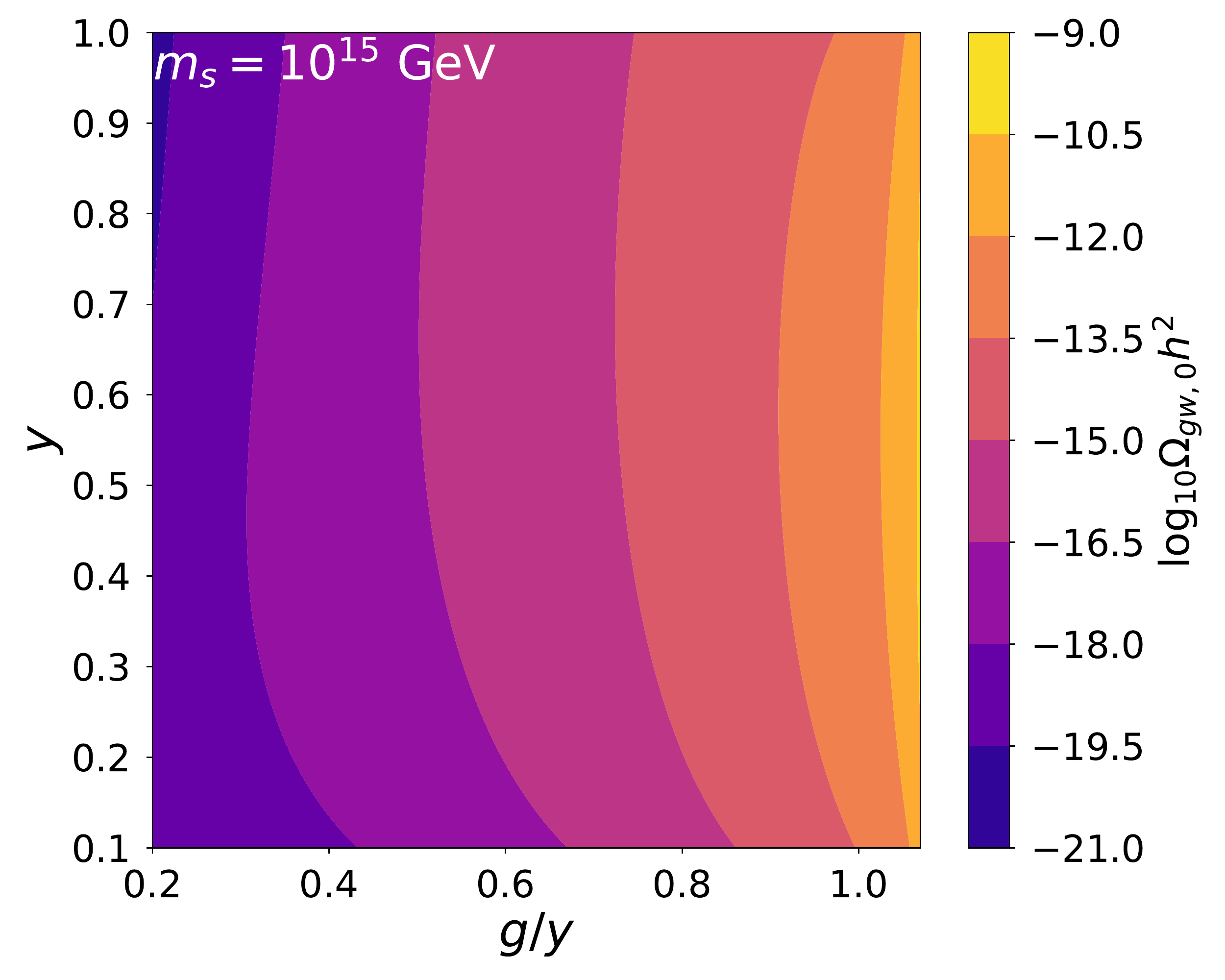}
    \label{fig:omegah2ms1e15}
    \end{subfigure}
    \caption{The energy density of gravitational waves today $\Omega_{gw} h^2$ at peak frequencies $f_{ac}$ generated from first-order phase transition, for the potential given in Eq.~\eqref{eq:potential}. The panels and axes are the same as in Figure~\ref{fig:TC}.}
    \label{fig:omegah2}
\end{figure}
With our model and gravitational wave observables defined, we are ready to investigate the generation of gravitational waves from first order phase transitions in the early universe. In Figures \ref{fig:TC} - \ref{fig:omegah2} we have plotted the critical temperature $T_C$, peak gravitational wave frequency $f_{ac}$, and gravitational wave energy density $\Omega_{gw} h^2$, for our scalar potential, in terms of its three model parameters, the Yukawa coupling $y$, gauge coupling $g$, and scalar mass $m_S$. In these plots we have detailed phase transition signatures for a range of scalar masses $m_S$, from $10^6$~GeV to $10^{15}$~GeV. In producing these plots, we do not use the high temperature approximation, but rather solve numerially for the full thermal potential given in Eq.~\eqref{eq:temppotential}. It is worth noting that higher order corrections and renormalization group improvements to the thermal potential, e.g. daisy corrections and taking into account the scale dependence of the effective potential do not have a significant impact on our results, as we discuss in Appendices~\eqref{sec:thermappendix} and~\eqref{sec:RGEImprovement}. 

\begin{figure}[t!]
    \centering
  \begin{subfigure}[b]{0.48\textwidth}
    \includegraphics[width=\textwidth]	{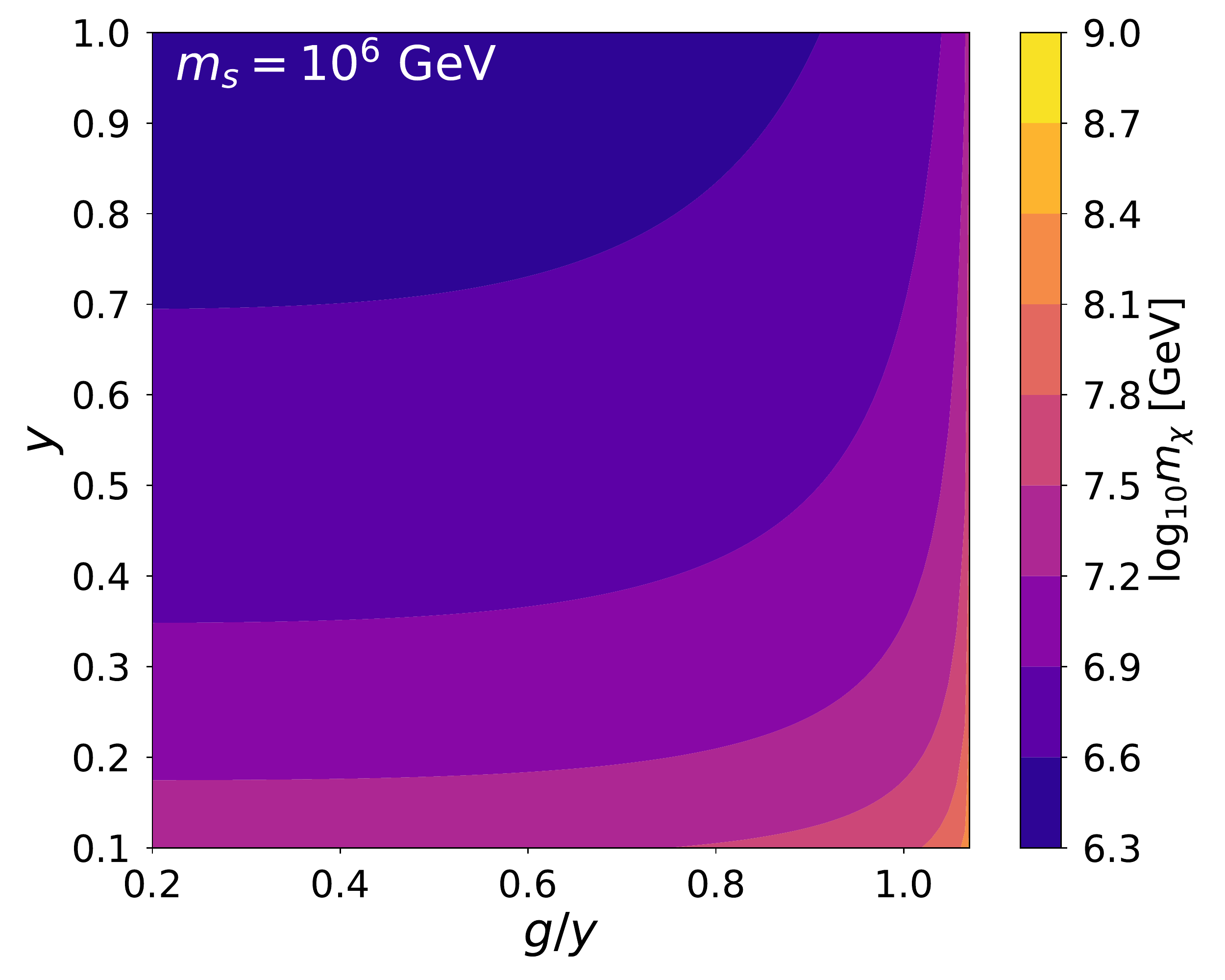}
    \label{fig:mfms1e6}
    \end{subfigure}
  \begin{subfigure}[b]{0.48\textwidth}
    \includegraphics[width=\textwidth]	{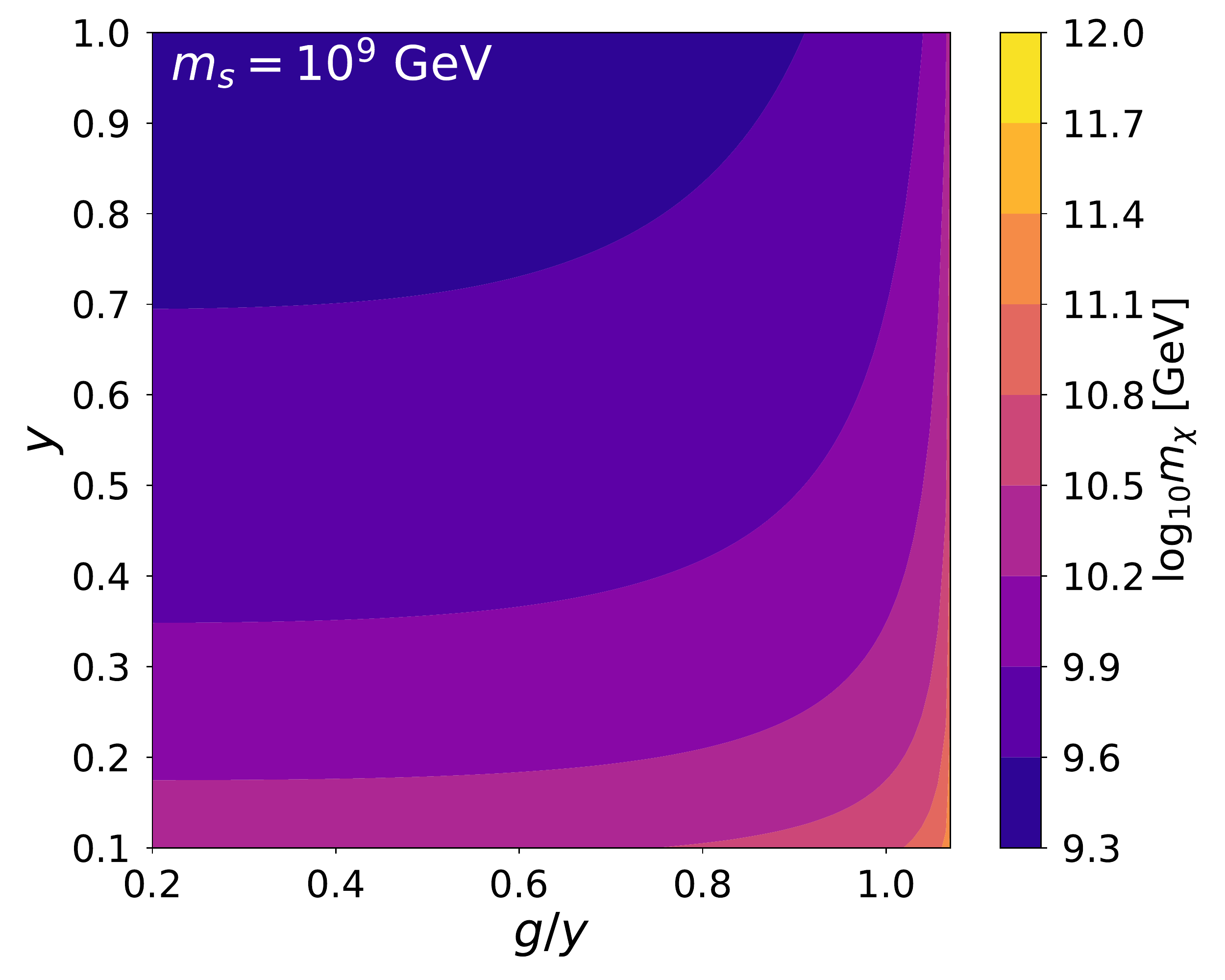}
    \label{fig:mfms1e9}
    \end{subfigure}
  \begin{subfigure}[b]{0.48\textwidth}
    \includegraphics[width=\textwidth]	{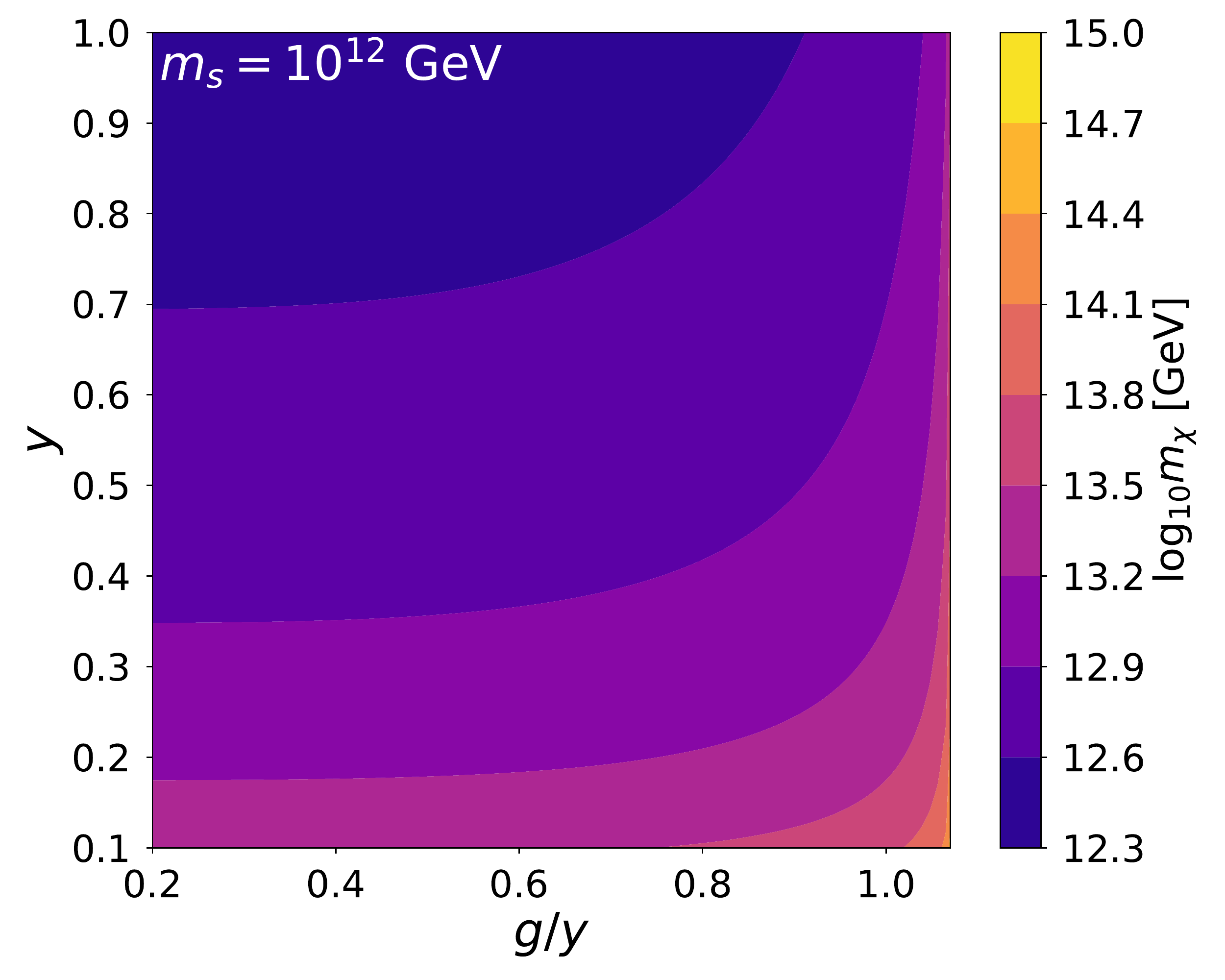}
    \label{fig:mfms1e12}
    \end{subfigure}
  \begin{subfigure}[b]{0.48\textwidth}
    \includegraphics[width=\textwidth]	{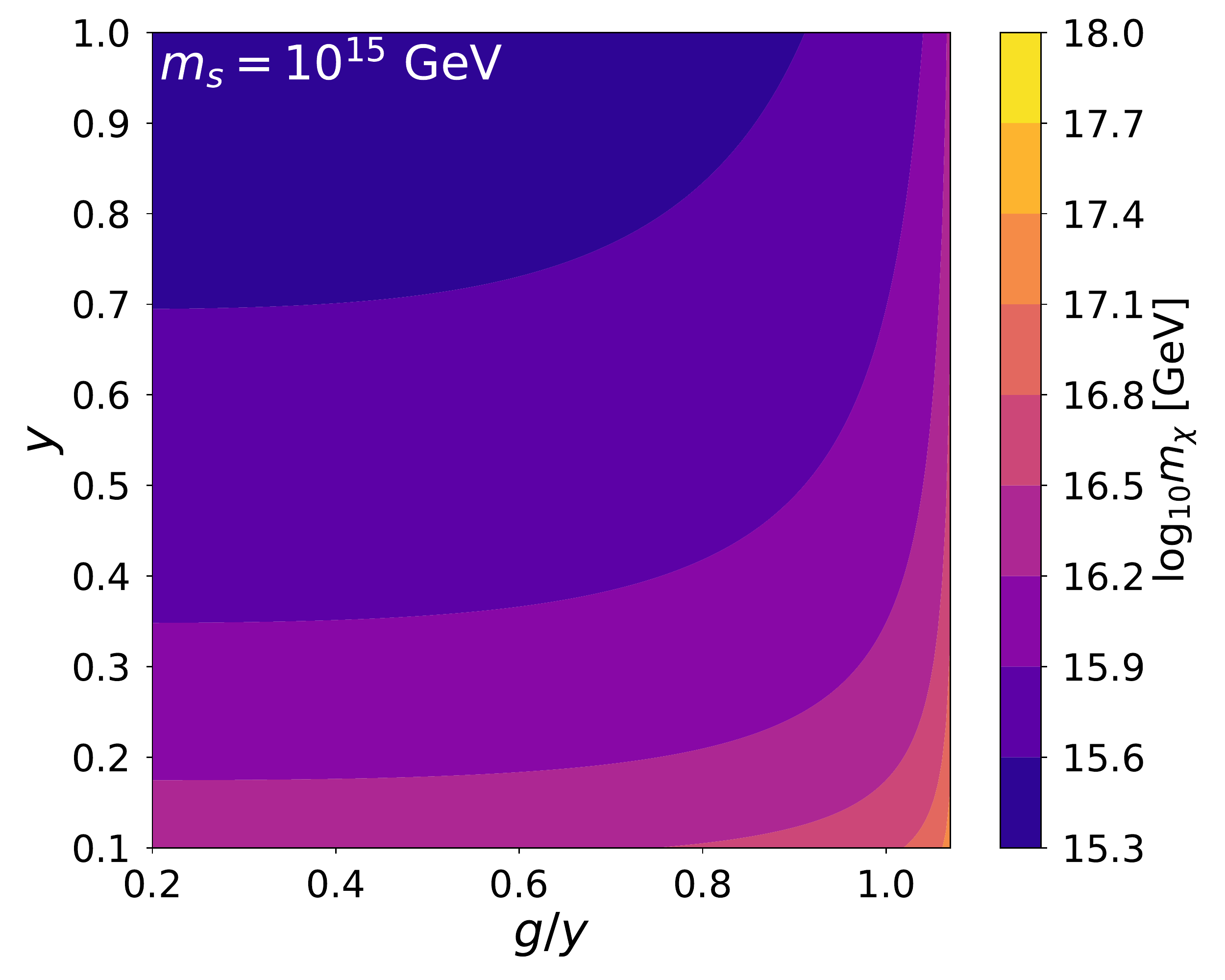}
    \label{fig:mfms1e15}
    \end{subfigure}
    \caption{The fermion mass $m_\chi$, as determined by $\chi$'s Yukawa coupling to scalar field $\varphi$, after a first-order phase transition. The panels and axes are the same as in Figure~\ref{fig:TC}. Note that this plot is accurate to one-loop order in $g$ and $y$; two-loop corrections may be substantial for $g/y \gtrsim 1.07$ (not plotted), see particularly discussion after Eq.~\eqref{eq:massdiscussion}.}
    \label{fig:mf}
\end{figure}

To be specific, we compute the critical temperature $T_C$ numerically by solving
\begin{equation}
    V(\nu_+,T_C)=0\,,
\end{equation}
where the minimum $\nu_+$ is defined by the value of the potential where $\frac{dV}{d\varphi}=0$ and $\frac{d^2V}{d\varphi^2}>0$. The nucleation temperature $T_N$ is obtained by solving Eq.~\eqref{eq:TNequation}, where $S_3/T$ is taken from Eq.~\eqref{eq:S3toT}. Note that at temperatures close to $T_C$, $x$ in Eq.~\eqref{eq:S3toT} may exceed one, in which case $f(x)$ in Eq.~\eqref{eq:fx} is not defined. However, this does impede our computations since at $T\sim T_C$, the energetic difference between the true and false vacuum is very small, and bubble nucleation is not yet efficient. We therefore set $\exp(-S_3/T)=0$ for $x>1$ in solving Eq.~\eqref{eq:TNequation}. The latent heat $\epsilon$ is attained numerically using Eq.~\eqref{eq:epsilon}, which is then combined with Eq.~\eqref{eq:gwfrequency} to find the energy density sources as a gravitational wave background.

In Figure \ref{fig:mf}, we show the final mass of the fermion field, $m_{\chi}$, at the end of the phase transition. The parameter space where $\chi$ recieves a large mass from $\varphi$, which occurs at $y \sim1, g/y \sim 1$, can be compared to the critical temperature, peak frequency, and SGWB densities for the same parameter space shown in Figures~\ref{fig:TC}-\ref{fig:omegah2}. Generally a larger Yukawa coupling indicates a lower phase transition temperature, as can be understood from Eq.~\eqref{eq:TC} since both $1/D$ and $T_0$ decrease with a larger Yukawa coupling. As $g/y$ approaches a particular value $3g^4 = 4y^4$, the critical temperature grows drastically due to an exceptionally small effective scalar quartic, $\lambda(T) \ll 1$. In Figure \ref{fig:fac}, we show the frequency at which the produced gravitational wave spectrum is peaked. As $m_S$ varies from $10^{15}$~GeV to $10^6$~GeV, the peak frequency drops from more than a GHz to less than 100 Hz, which falls within the frequency reach of LIGO. Compared with $T_C$ and the peak frequency $f_{ac}$, and the final fermion mass $m_\chi$, the gravitational wave energy density today at peak frequencies is most sensitive to the gauge-Yukawa ratio $g/y$ as can be seen from Figure~\ref{fig:omegah2}. Intriguingly, several features, including a low gravitational wave peak frequency, high GW energy density, and a large fermion mass (which can be dark matter) are all attained when $3g^4\sim 4y^4$.

\section{Gravitational Waves from Inflation}
\label{sec:inflat}

We now turn to a different potential source of gravitational waves: scalar field dynamics at the end of inflation. The inflaton models we will consider here attain large vacuum expectation values after inflation, much like the simplified thermal scalar potential studied in the previous section. Such potentials could therefore lead to mass boosted dark matter, as will be discussed in Section \ref{sec:massboost}.

If an inflaton field $\phi$ oscillates in its potential after inflation, it can self-interact in a way that causes large inhomogeneities to develop in its field density. These field configuration are sometimes called ``oscillons'' \cite{Copeland:1995fq,Broadhead:2005hn,Amin:2011hj,Lozanov:2017hjm}. The formation of these inhomogeneities in the inflaton field density will source a gravitational wave background, which may be detected at future gravitational wave observatories. The possibility that inflatons could be detected through post-inflation gravitational wave production has been studied previously in \cite{Zhou:2013tsa,Antusch:2016con,Antusch:2017flz,Amin:2018xfe,Lozanov:2019ylm}. This phenomenon is also interesting from the standpoint of detecting generic non-inflationary superheavy dark sectors, because this source of gravitational waves does not require a special thermal potential that results in a strong first order phase transition, like the one we investigated in Section \ref{sec:phasetransition}. Compared to prior work on inflatons generating gravitational waves at the end of inflation, we will be considering both E- and T-model potentials, whereas previously only T-model potentials were studied in \cite{Lozanov:2019ylm}. We will obtain a broad range of results for both E- and T-Model features, and find that E-model potentials result in different gravitational wave signatures than T-model potentials. 

Here we will be interested in computing gravitational wave background spectra produced by two classes of inflationary potential, the E-Model
\begin{equation}\label{eq:Emodel}
    V\big(\phi\big) = \Lambda^4\bigg(1 - e^{-\sqrt{\frac{2}{3\alpha}}\frac{\phi}{M_{Pl}}}\bigg)^{2}
\end{equation}
and T-Model potentials
\begin{equation}\label{eq:Tmodel}
    V\big(\phi\big) = \Lambda^4\tanh^{2}{\bigg(\frac{\phi}{\sqrt{6\alpha}M_{Pl}}\bigg)},
\end{equation}
where $\Lambda$ has dimensions of energy, $\phi$ is the inflaton field, and $\alpha$ is a constant that determines the shape of the potential. These models were proposed in \cite{Carrasco:2015rva,Galante:2014ifa,Kallosh:2013hoa}, where they were shown to arise from Kahler and superpotentials terms \cite{Carrasco:2015rva}. They are useful as generic inflation models, since for certain $\alpha$ values, E-Model and T-Model inflation are identical to a range of inflationary models which match present CMB observations, including Starobinsky \cite{Starobinsky:1980te} and Higgs inflation \cite{Bezrukov:2007ep}. 
Previously, gravitational wave background spectra from post-inflatonary field oscillations were computed for a T-Model inflationary potential in \cite{Lozanov:2019ylm}. Here we extend this treatment to E-Model inflation, and compute results for a broader range of T-Model potentials.

Another useful feature of the E-Model and T-Model classes of inflationary potentials, is that the free parameters of these potentials ($\Lambda$, $\alpha$) are fixed by certain inflationary observables, including the primordial scalar power spectrum and the primordial tensor power spectrum. Specifically, the ratio of these primordial fluctuations, $i.e.$ the tensor-to-scalar ratio ($r$) will have a fixed relationship to ($\Lambda$, $\alpha$).  Here we define the tensor power spectrum and the scalar power spectrum as is customary,
\begin{equation}\label{eq: tensor power spec}
    {P}_{T}(k) = {A}_{T}(k_*)\bigg(\frac{k}{k_*}\bigg)^{n_T}\,,
\end{equation}
\begin{equation}\label{eq: scalar power spec}
    {P}_{s}(k) = {A}_{s}(k_*)\bigg(\frac{k}{k_*}\bigg)^{n_s - 1}
\end{equation}
where $k$ is the wavenumber, $k_*$ is the pivot tensor/scalar, ${A}_T(k_*),{A}_s(k_*)$ are the amplitudes of the tensor and scalar power spectra which depend on the choice of $k_*$, and $n_T,n_s$ are the tensor and scalar spectral indices. We have defined these using the usual conventions given in \cite{Aghanim:2018eyx}; for a review on inflationary models see $e.g.$ \cite{Baumann:2009ds}. For slow-roll inflation, $r$, $n_s$, and $A_s$ are expressible in terms of the slow-roll parameters $\epsilon$ and $\eta$ \cite{Liddle:1992COBE, Lyth:2009text},
\begin{equation} \label{eq: epsandeta}
    \epsilon = \frac{1}{2}\bigg(\frac{M_{Pl}V^{\prime}}{V}\bigg)^2, \qquad  \qquad \eta = \frac{M_{{Pl}}^2V^{\prime\prime}}{V},
\end{equation}
\begin{equation}\label{eq: randns}
    r = 16\epsilon, \qquad \qquad n_s = 1 - 6\epsilon + 2\eta, \qquad  \qquad A_s = \frac{V}{24\pi^2M_{Pl}^4\epsilon},
\end{equation}
where $M_{Pl}$ is the reduced Planck mass, $V$ is the inflaton potential, and $V^{\prime}$ and $V^{\prime\prime}$ are the first and second derivative of the inflation with respect to $\phi$. With these quantities defined, it can be shown that for E-Model and T-Model potentials, $r$, $n_s$, and $A_s$ take the following form 
\begin{equation}\label{eq: randnsforalpha}
    r = \frac{12\alpha}{N_e^2}, \qquad n_s = -\frac{2}{N_e} + 1, \qquad A_s  = \frac{\Lambda^4N_e^2}{ 18 \pi^2 \alpha M_{Pl}^4 },
\end{equation}
where these expressions are accurate so long as $r \lesssim 10^{-3}$, and we have defined \mbox{$ N_e \equiv \int_{t_0}^{t_e} H dt $} as the number of e-folds of inflation lasting from time $t_0$ to $t_e$.

In our study of gravitational waves from oscillating fields after inflation, we fix the parameters of our inflaton potentials using current results from Planck's 2018 data release \cite{Aghanim:2018eyx}. Currently, Planck has observed $n_s = 0.9649 \pm 0.0042$ at the $68\%$ CL \cite{Akrami:2018odb} and $A_s = (2.100 \pm 0.030) \times 10^{-9}$ at the $68\%$ CL \cite{Aghanim:2018eyx}. For simplicity, we fix the number of inflationary e-folds to $N_e = 60$. The $r$ values explored in this paper are $10^{-4}, 10^{-6}, 10^{-8}, 10^{-10}$ which are well below the current upper bound $r < 0.061$, set by combining Planck 2018 and BICEP2/Keck Array data \cite{Akrami:2018odb}. Upcoming experiments such as LiteBIRD \cite{2020updatedLiteBIRD, LiteBIRDwhitepaper}, Simons Observatory \cite{SO:Ade:2018sbj, SOwhitepaper}, and CMB-S4 \cite{CMBS4sciencebook, CMBS4whitepaper} will be able to resolve an $r$ value on the order of $10^{-3}$. It is therefore interesting to note that this work explores $r$ values much lower than those discoverable with any planned CMB experiments. The SGWBs we compute are given in terms of the fraction of the universe's energy density they take up,
\begin{equation}\label{eq:omegagw}
    \Omega_{\text{gw}}(f) \equiv \frac{1}{\rho_{\text{crit}}}\frac{d\rho_{\text{gw}}}{d\ln{f}},
\end{equation}
where $\rho_{\text{crit}}$ is the critical energy density and $\rho_{\text{gw}}$
the energy density of gravitational waves.

\begin{figure}[t!]
    \centering
  \begin{subfigure}[b]{0.48\textwidth}
    \includegraphics[width=\textwidth]	{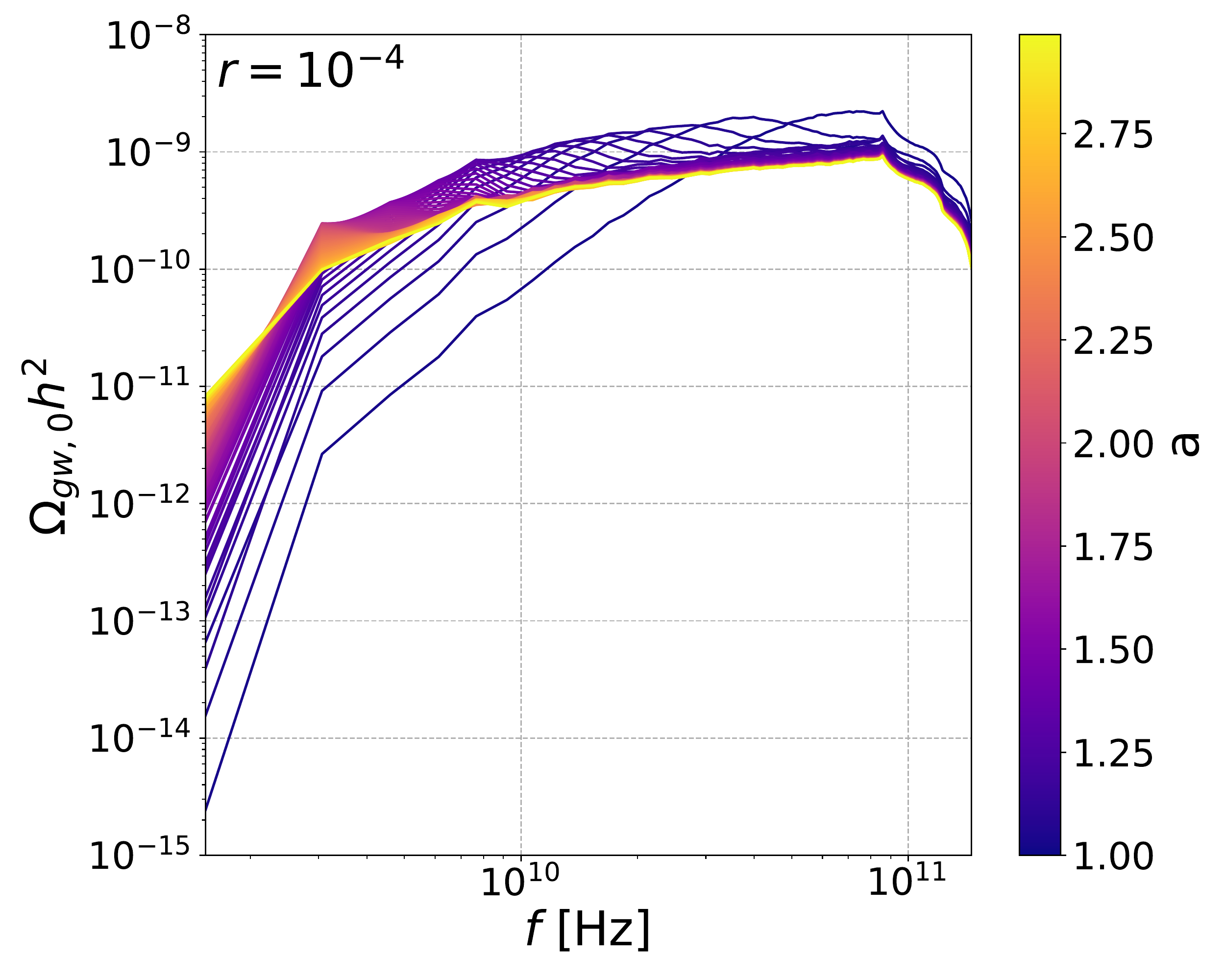}
    \label{fig:E-r10e-4}
	\end{subfigure}
 \begin{subfigure}[b]{0.48\textwidth}   \includegraphics[width=\textwidth]	{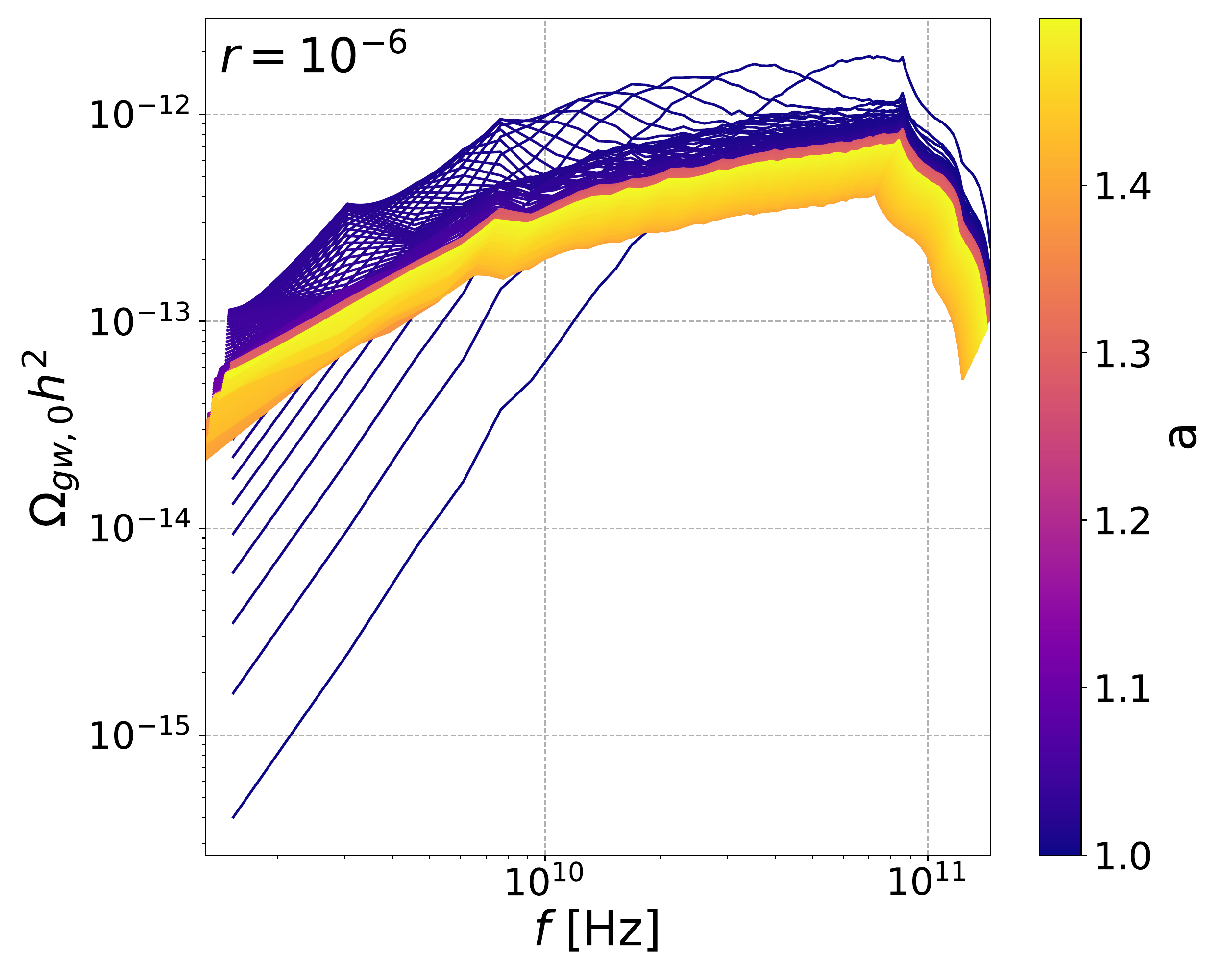}
    \label{fig:E-r10e-6}
    \end{subfigure}
    \caption{Stochastic background gravitational wave spectra produced after inflation from oscillations of the inflaton in a E-Model inflationary potential, for tensor-to-scalar ratio values $r$, as indicated. Colour coding indicates the evolution of the SGWB as the scale factor of the universe evolves from $a = 1$ to 3 for $r = 10^{-4}$ and from $a = 1$ to 1.5 for $r = 10^{-6}$ after inflation. The spectra does not evolve significantly for the $r = 10^{-6}$ after $~a = 1.3$ which is why the results are only shown up to $a = 1.5$. These spectra were created using $n = 256$,  and $L = 0.003H^{-1}$ for both $r$ values, and the initial conditions listed in table \ref{tab:simparams}.}
    \label{fig:E-Models}
\end{figure}

\begin{figure}[t!]
    \centering
  \begin{subfigure}[b]{0.48\textwidth}
    \includegraphics[width=\textwidth]	{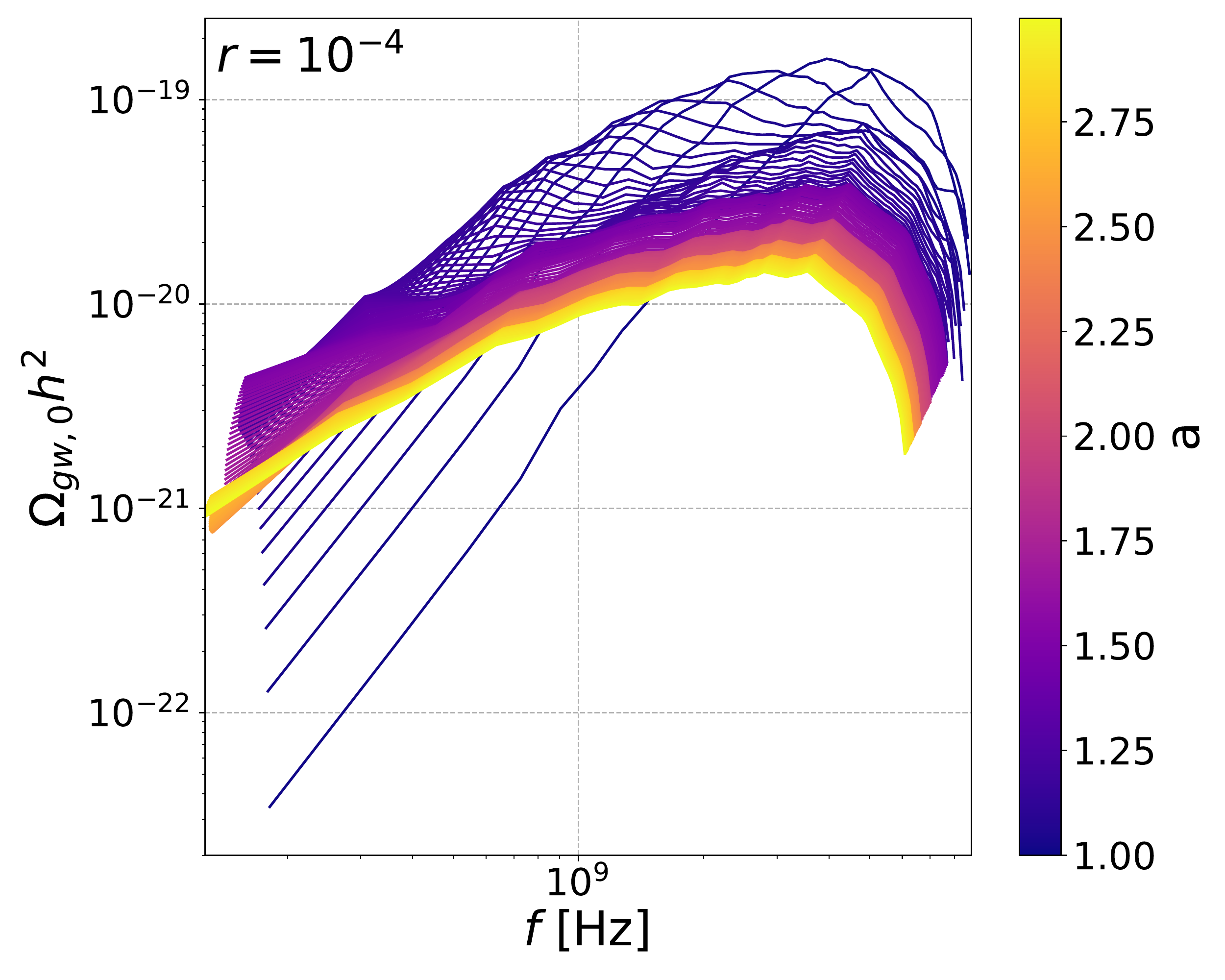}
    \label{fig:T-r10e-4}
	\end{subfigure}
  \begin{subfigure}[b]{0.48\textwidth}
    \includegraphics[width=\textwidth]	{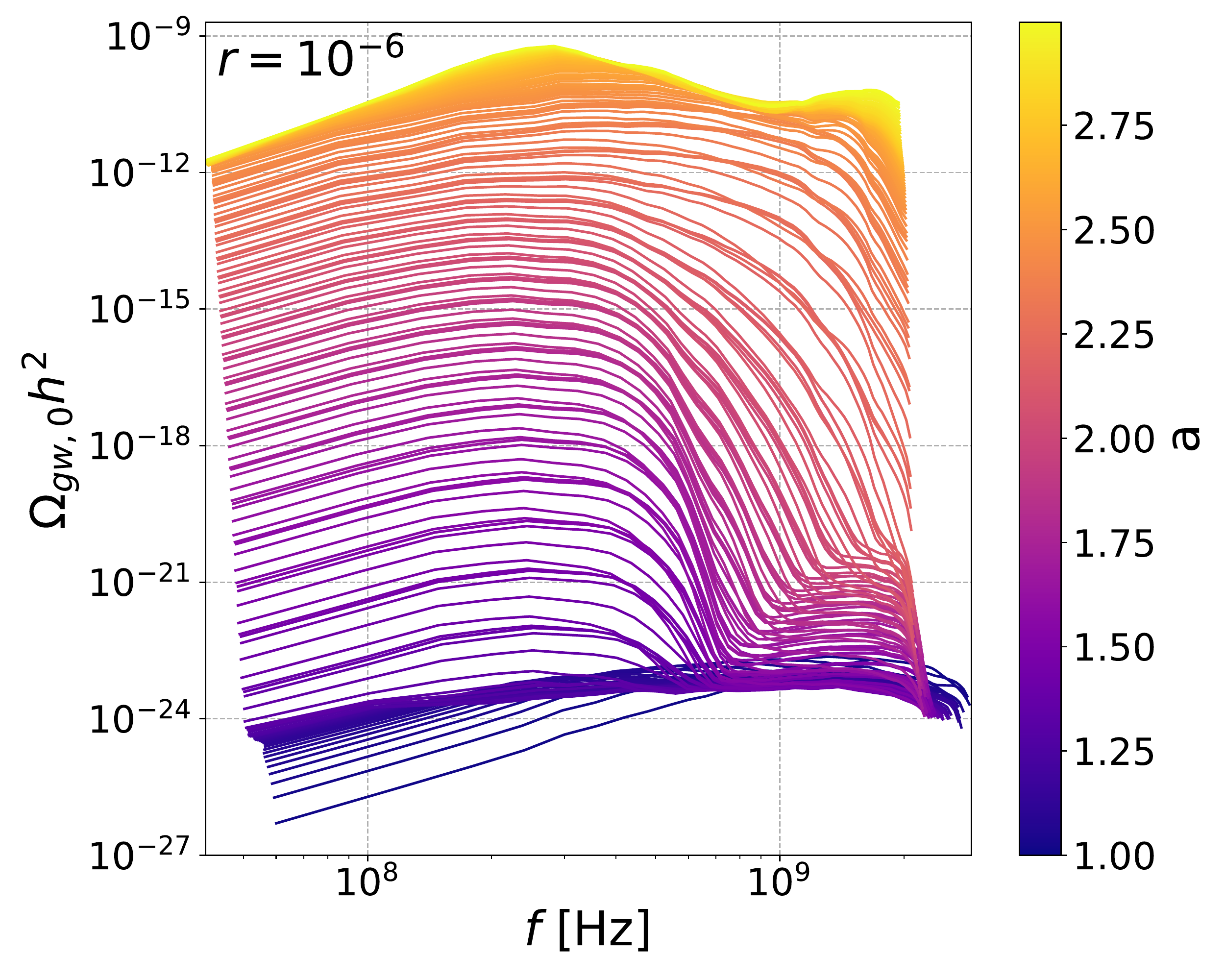}
    \label{fig:T-r10e-6}
	\end{subfigure}
 \begin{subfigure}[b]{0.48\textwidth}   \includegraphics[width=\textwidth]	{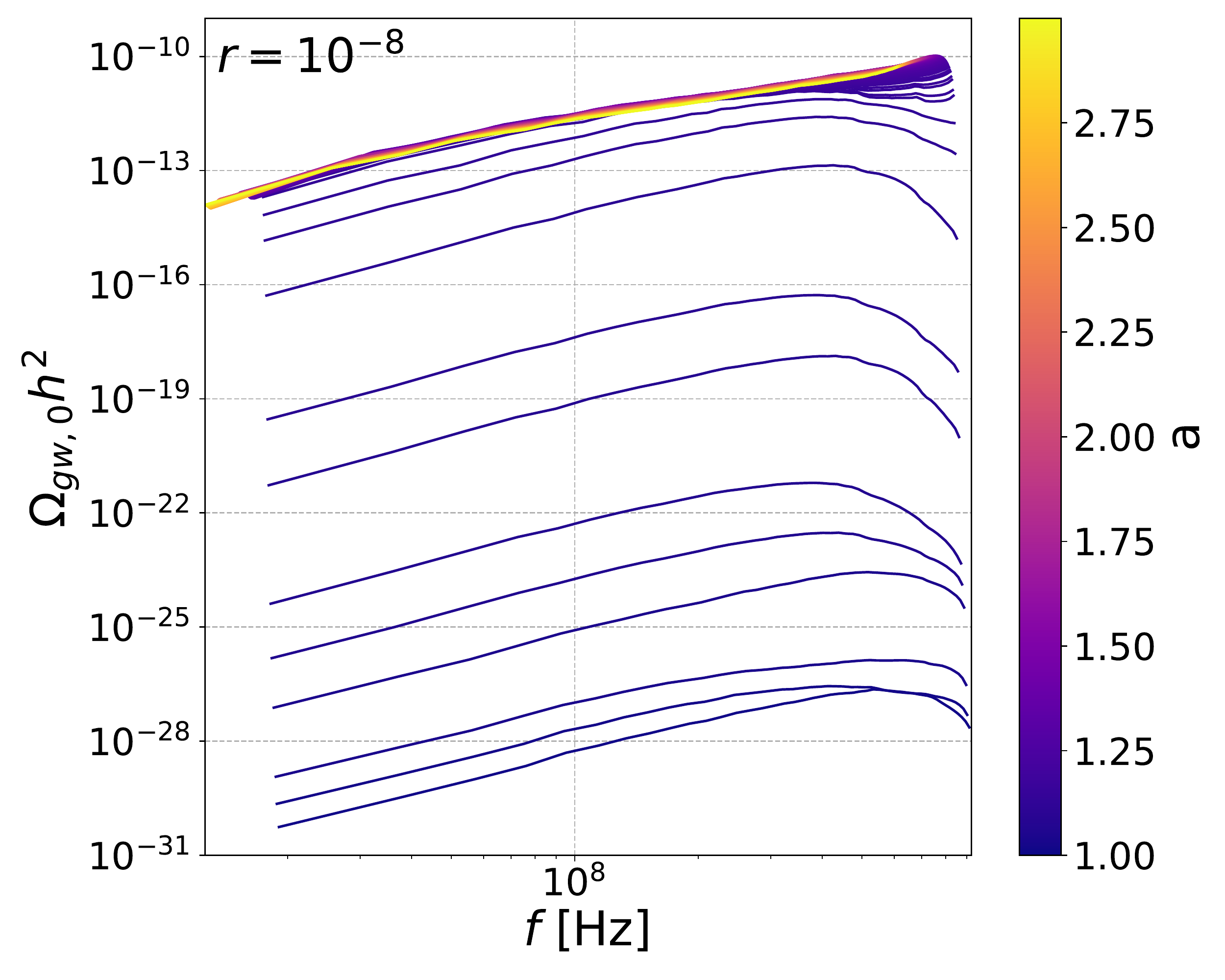}
    \label{fig:T-r10e-8}
    \end{subfigure}
  \begin{subfigure}[b]{0.48\textwidth}
    \includegraphics[width=\textwidth]	{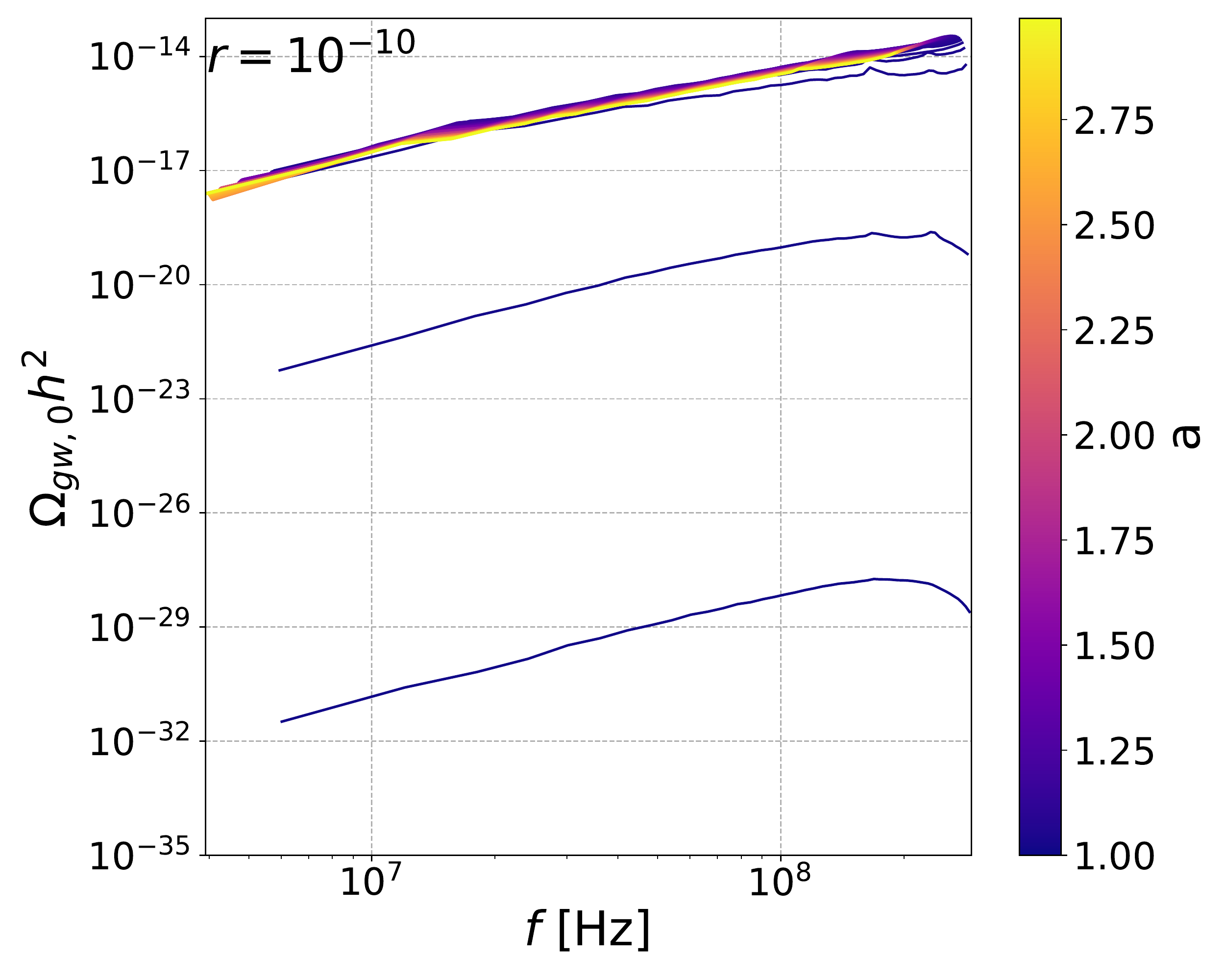}
    \label{fig:T-r10e-10}
	\end{subfigure}
    \caption{Stochastic background gravitational wave spectra produced after inflation from oscillations of the inflaton in a T-Model inflationary potential, for tensor-to-scalar ratio values $r$, as indicated. Colour coding indicates the evolution of the SGWB as the scale factor of the universe evolves from $a = 1$ to 3 after inflation. These spectra were created using $n = 128$, $L = 1.0H^{-1}$, and the initial conditions listed in table \ref{tab:simparams}.}
    \label{fig:T-Models}
\end{figure}

In order to determine the SGWB from E-Model and T-Model inflatons oscillating in their potentials at the end of inflation, we simulated the dynamics and self-interactions of these inflatons using the lattice field theory code HLattice \cite{Huang:2011gfHLattice}. Before the lattice simulation starts, HLattice solves for the inflaton field value at the end of inflation by evolving the field according to the Klein-Gordon equation, $ \ddot{\phi} + 3H\dot{\phi} + V' = 0, $ until the following condition is met: $ \dot{\phi} + H\phi <0.$ Because we are studying low scale inflation models, we replaced this HLattice condition with $\epsilon \geq 1$, where $\epsilon$ is the slow-roll parameter given in Eq.~\eqref{eq: epsandeta}. For low scale inflation models, this resulted in significantly smaller initial field values for $\phi$, compared to the original HLattice condition.

We ran simulations that lasted for just over an e-fold after the end of inflation, which corresponds to the scale factor of the universe, $a$, expanding by a factor of up to three for the T-Models. For the E-Model cases, the simulations lasted for just over an e-fold after inflation for $r = 10^{-4}$, with $a$ expanding by a factor of three and less than an e-fold after the end of inflation, corresponding to the scale factor expanding by approximately a factor of a half for $r =10^{-6}$. The $r = 10^{-6}$ case was only run to this value of $a$, because it did not evolve significantly past this value and was therefore not shown. Given technical limits and finite computational resources, the $r = 10^{-8}, 10^{-10}$ cases were not computed for the E-Model. For all cases shown, we found that the gravitational wave spectra converged on a fixed result well before the end of the simulation, as can be verified in Figures \ref{fig:E-Models} and \ref{fig:T-Models}. Note that the spectra given in these figures for each $a$ value, represent the energy density of gravitational waves at that $a$ value redshifted to present day. The lattice box size ($L$) and box resolution ($n$) were chosen in order to ensure the simulation smoothly resolved SGWB spectral features. We checked that the results were robust to minor variations in the lattice spacing and simulation box size. For the E-Models, $L = 0.003H^{-1}$ and $n = 256$ was used for both $r$ values. For the T-Models, $L = 1.0H^{-1}$ and $n = 128$ were used for all $r$ values. The frequencies that the simulation probed ranged up to $n\frac{\pi}{L}$ with a spacing of $\frac{\pi}{L}$.

\begin{table}[t!]
\centering
E-Model\\
\vspace{.2cm}
\begin{tabular}{|c|c|c|c|c|}
\hline
\rule{0pt}{3ex} \boldmath{$r$} & \boldmath{$\alpha$} & \boldmath{$\phi_0$ $[M_{Pl}]$} & \boldmath{$\phi_\text{e}$ $[M_{Pl}]$} & \boldmath{$\dot{\phi}_0$ $[{M_{Pl}}^2]$} \\ \hline
\rule{0pt}{3ex} $10^{-4}$ & \num{3e-2} & \num{1.67} & \num{5e-1} & \num{-9.5e-17}  \\
\rule{0pt}{3ex} $10^{-6}$ & \num{3e-4} & \num{2.65e-1} & \num{1e-1} & \num{-9.5e-17}  \\ \hline
\end{tabular}
\vspace{.5 cm}
\\
T-Model\\
\vspace{.2cm}
\begin{tabular}{|c|c|c|c|c|}
\hline
\rule{0pt}{3ex} \boldmath{$r$} & \boldmath{$\alpha$} & \boldmath{$\phi_0$ $[M_{Pl}]$} & \boldmath{$\phi_\text{e}$ $[M_{Pl}]$} & \boldmath{$\dot{\phi}_0$ $[{M_{Pl}}^2]$}  \\ \hline
\rule{0pt}{3ex} $10^{-4}$ & \num{3e-2} & \num{1.82} & \num{6e-1} & \num{-9.5e-17}  \\
\rule{0pt}{3ex} $10^{-6}$ & \num{3e-4} & \num{2.80e-1} & \num{1e-1} & \num{-9.5e-17}  \\
\rule{0pt}{3ex} $10^{-8}$ & \num{3e-6} & \num{3.77e-2} & \num{2e-2} & \num{-9.5e-17}  \\
\rule{0pt}{3ex} $10^{-10}$ & \num{3e-8} & \num{4.75e-3} & \num{2e-3} & \num{-9.5e-17}  \\\hline
\end{tabular}
\caption{Simulation and model parameters for the E-Model and T-Model inflationary potentials given in Equations~\eqref{eq:Emodel} and \eqref{eq:Tmodel}, where $\phi_0$ is the inflaton field value at the beginning of inflation, selected so that inflation lasts $N_e \simeq 60$ e-folds, $\phi_e$ is the field value at the end of inflation, and $\dot{\phi}_0$ is the initial kinetic field value set in HLattice. Stochastic background gravitational wave spectra for these models are shown in Figures \ref{fig:E-Models} and \ref{fig:T-Models}.}
\label{tab:simparams}
\end{table}

Table \ref{tab:simparams} gives simulation parameters used for the E-Model and T-Models respectively. Given a value of $\alpha$, which determines the flatness of the inflationary potential, $\phi_0$ was chosen such that inflation lasted for approximately 60 e-folds, $i.e.$ $N_e \simeq 60$. The initial kinetic energy of the inflaton $\dot{\phi}_0$ was chosen to be very small, and is the same for every case. (For the sake of being completely reproducible, we note that $\dot{\phi}_0 = -10^{-10}$ in HLattice code, where HLattice uses the convention ${M_{Pl}}=1024$.) For comparison, in Table \ref{tab:simparams} we have also given $\phi_\text{e}$, the field value where the slow-roll conditions are no longer met, which we computed using standard slow-roll formulae.

In Figure \ref{fig:E-Models}, the stochastic gravitational wave background spectra are shown for E-Model inflationary potentials for tensor-to-scalar ratios $r = 10^{-4}, 10^{-6}$, where as usual, smaller $r$ corresponds to a smaller energy density at the end of inflation. As a consequence of this decrease in energy density, the overall amplitude of the SGWB produced from inflation decreases with a decreasing $r$ value. Contrary to prior work \cite{Hasegawa:2017iay}, we do not observe growth in the SGWB consistent with the formation of oscillons for $\alpha \lesssim 10^{-3}$ for the E-Models. We believe this is because our HLattice simulations include the effect of metric fluctuations and associated gravitational dynamics, which can affect the formation of oscillons, since as we observe, there is a large amount of background gravitational waves generated very soon after the end of inflation. This initial large inhomogeneity in the E-Model potential, as compared to the T-Model, may be the result of the E-Model potential not being symmetric around its minimum. We see in Figure \ref{fig:E-Models} that the SGWB amplitude decreases by approximately three orders of magnitude. Based on preliminary investigation, we do not expect oscillon configurations to form, and these SGWB to increase at later times.

In Figure \ref{fig:T-Models}, the T-Model SGWB spectra were computed for $r = 10^{-4}, 10^{-6}, 10^{-8}, 10^{-10}$. Here the formation of oscillon field configurations is evident for the $r = 10^{-6}, 10^{-8}, 10^{-10}$ SGWB spectra, whose SGWB amplitudes increase markedly after $a\approx 2$, when oscillons form. This can be contrasted with $r=10^{-4}$, for which the SGWB amplitude is relatively static, because oscillons do not form. On the other hand, before the formation of these oscillons (at $a=1$), it can be seen that the initial amplitude of the stochastic gravitational background decreases with a decreasing $r$ value, as does the frequency of peak SGWB amplitude. In comparison with prior results \cite{Hasegawa:2017iay}, as expected we observe that T-Model inflatons only form oscillons for $\alpha \lesssim 10^{-4}$. Here we find that, more precisely, oscillon configurations can still form for $\alpha \leq \num{3e-4}$. It is interesting to note that the peak amplitude of SGWB spectra decreases by a few orders of magnitude between $r=10^{-6}$ and $r=10^{-8}$, and by four orders of magnitude between  $r=10^{-8}$ and $r=10^{-10}$.

\section{Heavy Nonthermal Dark Matter From a Mass Boost}
\label{sec:massboost}

In this section we explore how scalar fields detailed in Sections \ref{sec:Yukawa} and \ref{sec:inflat}, can give a mass boost to nonthermally produced dark matter, resulting in dark matter produced out of thermal equilibrium, with a final dark matter mass of up to $m_{\chi} \sim 10^{19}$ GeV. If dark matter is produced out-of-equilibrium at a temperature well above the temperature at which the universe reheats $T_{rh}$, then assuming that the universe remains radiation dominated from $T_{rh}$ until the era of matter-radiation equality in the early universe, the relic abundance of dark matter can be computed by comparing it to the presently observed radiation energy density,
\begin{equation} \label{eq:basicrelic}
\Omega_{DM} h^2 = \Omega_R h^2 \left( \frac{T_{rh}}{T_0} \right) \frac{\rho_{DM,e}^{i}}{\rho_e}.
\end{equation}
Here $\rho_e$ is the energy density of the universe at the end of inflation, $\rho_{DM,e}$ is the dark matter energy density at the end of inflation, $\Omega_{DM} h^2 \simeq 0.12$ is the physical dark matter density parameter,  $\Omega_{R} h^2 \simeq 4.3 \times 10^{-5}$ is the physical radiation density parameter, and $T_0 \simeq 2.34 \times 10^{-13}$ GeV is the present temperature of the universe. In an expanding universe, radiation energy density dilutes as $T^4$, while matter dilutes as $T^3$, so the factor of $\frac{T_{rh}}{T_0}$ accounts for shifts in the density of matter relative to radiation as the universe expands from $T_{rh}$ to $T_0$. Note that there is an implicit assumption in Eq.~\eqref{eq:basicrelic} that the ratio $\rho_{DM,e}/\rho_e$ does not shift significantly between the end of inflation and the time at which the inflaton decays -- in practice this may introduce a small correction to the final dark matter relic abundance.

The standard method to compute the abundance of gravitationally created particles at the end of inflation $\rho_{DM, e}$, is to integrate the particle's Bogoliubov operator to the end of inflation \cite{Chung:1998zb,Kuzmin:1998kk}. The precise abundance of particles created will depend on cosmological dynamics following inflation \cite{Kolb:2017jvz}, however, to good approximation the gravitationally produced relic abundance for a fermion with mass $m_{fi}$ is
\begin{equation}
\rho_{DM,e} \simeq 10^{-2} H_e^4 e^{-2m_{fi}/H_e},
\label{eq:fermabund}
\end{equation}
where this expression holds so long as $m_{fi} \gtrsim H_e$, where $H_e$ is the Hubble constant at the end of inflation.

Here we will be most interested in how the relic abundance is altered, if after a density $\rho_{DM,e}$ of dark matter particles is created at the end of inflation, these particles undergo a mass shift $m_{fi} \rightarrow m_{\chi}$ during a dark sector phase transition. This mass shift would boost the dark matter mass, and corresponding relic abundance, by a factor $m_{\chi}/m_{fi}$. For simplicity, we can assume that the universe remains radiation-dominated after reheating, so that there are not additional corrections to the dark matter relic abundance, from $e.g.$ a post-reheating period of matter-dominated expansion. Then substituting the Friedmann relation $\rho_e =3 M_{Pl}^2 H_e^2  $ and Eq.~\eqref{eq:fermabund} into \eqref{eq:basicrelic}, we arrive at the total relic abundance for dark matter that is gravitationally produced,
\begin{equation}
    \Omega_{DM} h^2 \simeq 0.003~ \Omega_R h^2 \left( \frac{T_{rh}}{T_0} \right) \frac{H_e^2}{M_{Pl}^2} e^{-2m_{fi}/H_e} \left( \frac{m_{\chi}}{m_{fi}} \right).
    \label{eq:finalrelic}
\end{equation}
The final term in this expression accounts for dark matter undergoing a mass boost from $m_{fi} \rightarrow m_{\chi}$, at some time after the end of inflation. Note that this expression for the relic abundance is only correct so long as $m_{fi} \gtrsim H_e$, which is required by the cosmology detailed above. Finally, we note also that Eq.~\eqref{eq:finalrelic} does not assume instantaneous reheating, allowing for a Hubble constant at reheating that is much smaller than the Hubble constant at the end of inflation.  This would occur, for example, in models of inflation where the decay of the inflaton is slow compared to the Hubble rate at the end of inflation.  In particular, in Eq.~\eqref{eq:finalrelic} $T_{rh}$ has been fixed to achieve the correct relic abundance of dark matter.

\begin{figure}[t!]
    \centering
    \includegraphics[width=0.49\textwidth]{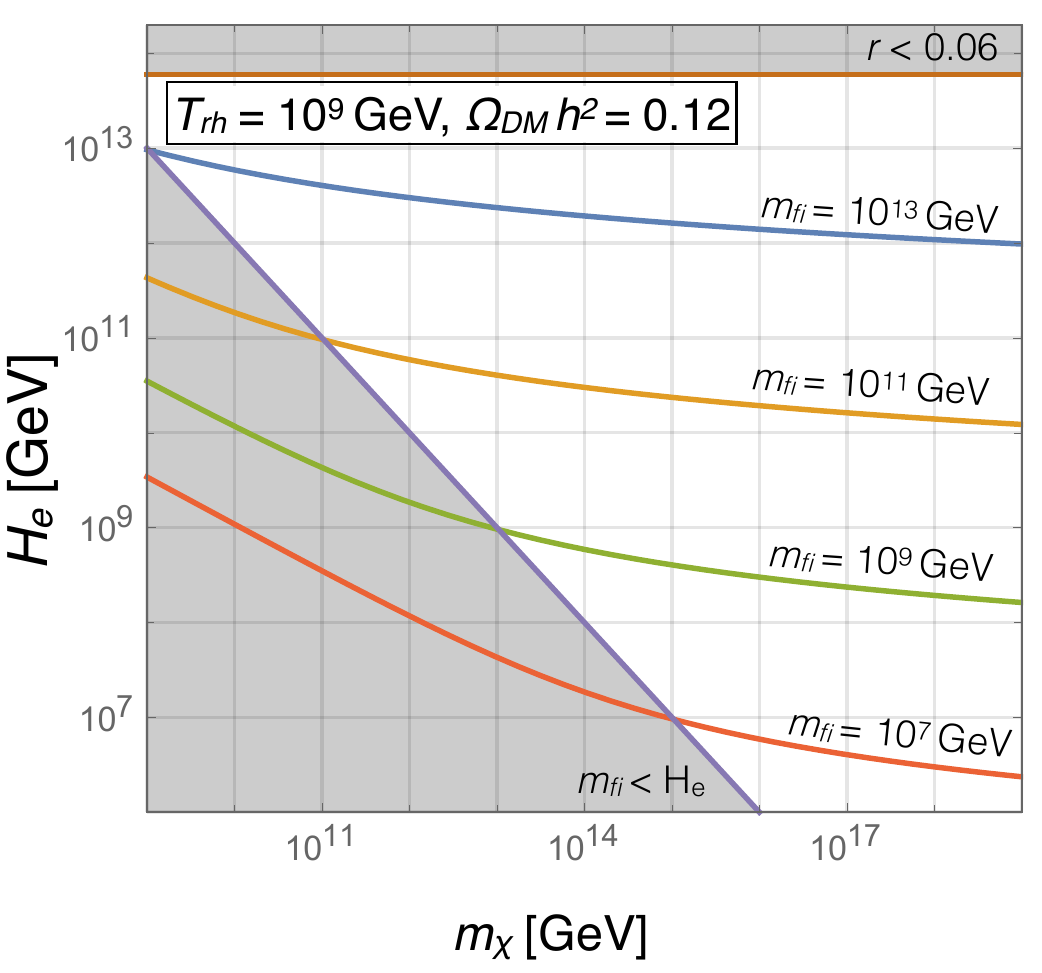}
     \includegraphics[width=0.49\textwidth]{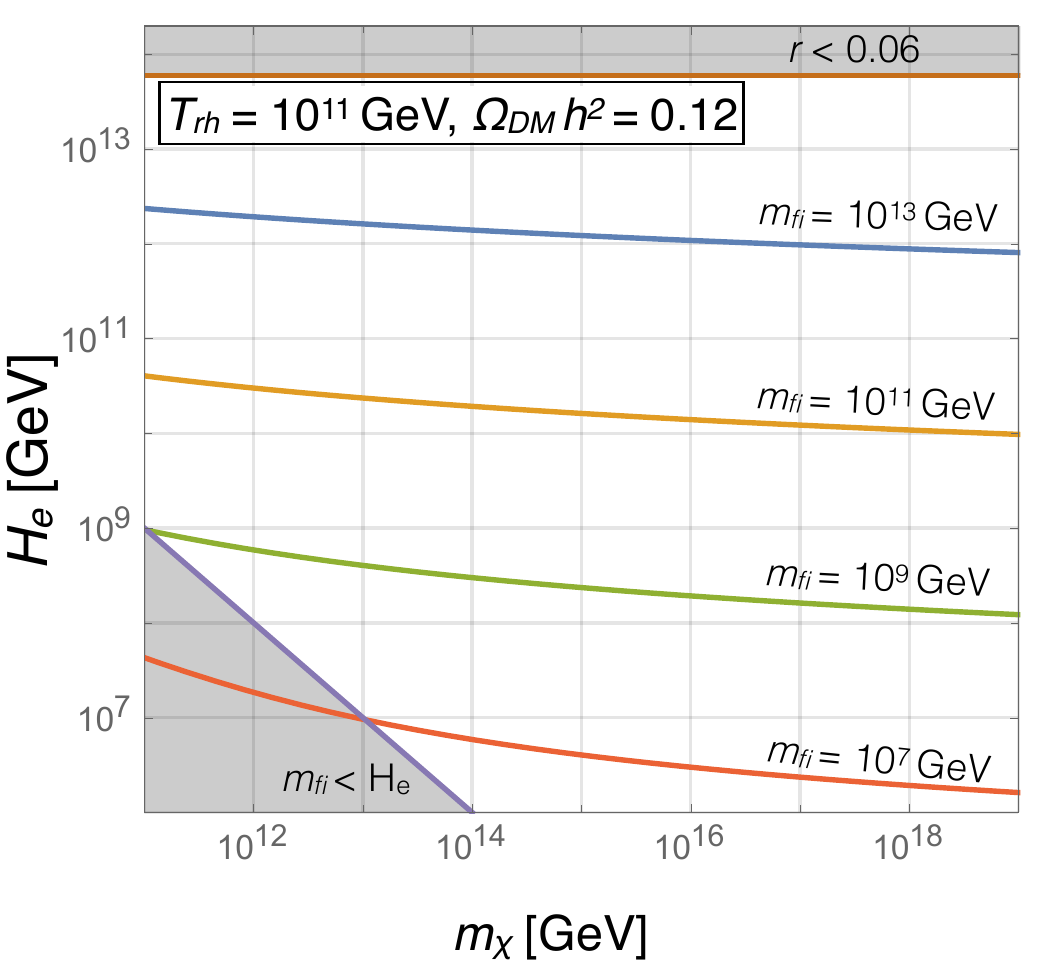}
    \caption{Dark matter relic abundance parameter space is shown for heavy dark matter produced gravitationally and nonthermally at the end of inflation, in terms of the initial mass $m_{fi}$, which is later mass-boosted from $m_{fi} \rightarrow m_\chi$ by a dark sector phase transition. For simplicity, we have required that initial dark matter mass is larger than the Hubble constant at the end of inflation $(m_{fi}> H_e)$, and have selected the range of the final dark matter mass $m_\chi$, to exceed the reheating temperature $T_{rh}$. For details about specific cosmologies, which can place further restrictions on this parameter space, please see text following Eq.~\eqref{eq:finalrelic}. The Planck 2018  and BICEP2/Keck Array BK15 bound on the tensor-to-scalar ratio ($r <0.061$) place an upper bound on $H_e$ \cite{Akrami:2018odb}.}
    \label{fig:mbrelic}
\end{figure}

In Figure \ref{fig:mbrelic} we show parameter space for mass-boosted dark matter, in terms of the Hubble constant at the end of inflation $H_e$, the temperature of the universe at reheating $T_{rh}$, the initial dark matter mass $m_{fi}$, and the final dark matter mass $m_\chi$. The present dark matter relic density has been fixed to match observations for a number of initial dark matter masses $m_{fi}$. For simplicity, we have restricted the parameter space to the regime of validity of Eq.~\eqref{eq:basicrelic}, so that the initial mass of the dark matter $m_{fi} > H_e$ exceeds the Hubble constant at the end of inflation. In Figure \ref{fig:mbrelic} it must also be required for consistency that the reheat temperature is low enough relative to the energy density associated with $H_e$. Thus, we have truncated $H_e$ y-axis value so that $H_e \gg H_{rh}$ throughout this plot for consistency.o that $H_e \gg H_{rh}$ throughout this plot for consistency.

We now provide a discussion of the relic abundance of dark matter, as pertains to Eq.~\eqref{eq:finalrelic}. There are a number of conditions which can be specified to ensure that the relic abundance of dark matter given by Eq.~\eqref{eq:finalrelic} represents the total relic abundance of dark matter, particularly so that additional dark matter is not produced by other processes, for example dark matter coming into thermal equilibrium with the radiation bath after the universe reheats at $T_{rh}$. These conditions will depend on the particulars of the mass-boosting cosmology after the end of inflation. 

First we discuss when a mass boost will occur for the simple scalar potentials we will consider. In order to provide a compact treatment of mass-boosting scalars in this section, we will be considering simplified versions of the scalar potentials detailed in Section \ref{sec:Yukawa} and \ref{sec:inflat}. In particular, we will focus on the fact that these scalar potentials $V \left( \varphi \right)$ are sufficiently flat in the early universe, that $V'' \left( \varphi_{i} \right) \ll H^{2}$. We assume that these scalars attained initial field values close to $\varphi = 0 $ prior to inflation. During inflation, such a scalar will obtain a vacuum expectation value from Hubble fluctuations of order $\varphi \approx H_{e}$. As previously stated, for all cases we consider $H_e \ll m_{fi}$, so this Hubble-induced vacuum expectation value $\varphi \sim H_{e}$ will not alter the mass of $\chi$ during inflation, since from inflationary fluctuations the induced mass will be $y H_{e} \lesssim H_{e}$. In this scenario, the scalar acts just like a curvaton field \cite{Lyth:2001nq}, or like any scalar/
axion/modulus with a flat potential in the early universe, and the initial field value $\varphi \sim H_e$ remains fixed by Hubble friction so long as $V'' \left( \varphi_{i} \right) \lesssim H^{2}$, see \textit{e.g.} \cite{Lyth:2001nq,Bramante:2016yju,Davoudiasl:2019xeb}. Therefore we can safely assume that during and after inflation $\varphi$ maintains its initial field value because of Hubble friction, until the Hubble constant drops to a smaller value than the scalar mass, which is when $H^2 \lesssim V''(\varphi) \equiv m_{eff}^2$. This provides the first restriction on the dark matter mass-boosting scalar considered here: the effective scalar mass should be smaller than the Hubble constant at the end of inflation $V''(\varphi) \equiv m_{eff}^2 < H_e^2$. With this simple scalar framework laid out, we are now ready to address the additional conditions required to ensure Eq.~\eqref{eq:finalrelic} is the majority contribution to the abundance of dark matter. This will depend on whether the mass boost occurs before or after reheating.

\begin{enumerate}
    \item[(A)] \emph{$m_{\chi} > T_{rh} $}.  First we consider the case that a dark matter mass boost occurs after the end of inflation, but before reheating. For the scalar fields we consider in this section, this can be enforced by the requirement that the effective mass of the scalar field $m_{eff}$ follows $H_e > m_{eff} > H_{rh}$, so that the scalar will roll down its potential between the end of inflation and $T_{rh}$. In this case, it is safe to assume that dark matter will not be produced in the thermal bath after reheating if $m_\chi \gg T_{rh}$ .
    
    \item[(B)] \emph{$m_{fi} > T_{rh} $}. Next we consider that case that a mass boost occurs only after reheating, but the initial dark matter mass $m_{fi}$ exceeds the reheating temperature. In this case again, it is usually safe to assume a negligible abundance $\chi$ is produced after reheating. In this case, the parameter space below the $m_{fi} = 10^9$ and $10^{11}$ GeV lines in the left and right panels of Figure~\ref{fig:mbrelic} should be excluded.
    
    \item[(C)] \emph{$m_{fi} < T_{rh} $}. Lastly for completeness and because it has some bearing on (A) and (B) we consider the possibility that a mass boost occurs after reheating, with the dark matter fermion mass smaller than the reheating temperature. This can be accommodated so long as $\chi$ and $\phi$ are weakly coupled to the radiation bath (because $\chi$ is gravitationally produced, no non-gravitational coupling need be assumed). In this case the requirement that $\chi$ is not substantially produced depends on the coupling to Standard Model particles in the radiation bath, for a complete treatment of ``freeze-in'' production and how small a coupling is required, we defer to \cite{Hall:2009bx}. In the following, for simplicity we will only consider cases (A) and (B).
\end{enumerate}

So far our treatment has been general, since we have not specified what scalar field provides the mass boost indicated by the ratio $m_\chi/m_{fi}$. We now consider two possibilities: this scalar field may either be a scalar field that goes through a phase transition, much like that detailed in Section \ref{sec:Yukawa}, but here different because we will set the gauge coupling to zero for simplicity, or a scalar field that rolls to a large field value after inflation, as explored in Section \ref{sec:inflat}. 

First we consider a mass-boosting scalar field, with a quartic self-coupling generated at loop-level from its Yukawa coupling to a fermion as laid out in Section \ref{sec:Yukawa}. Setting the gauge coupling to zero ($g=0$), and omitting the logarithmic term which will yield a small correction when $\varphi$ has settled at its minimum, Equation \eqref{eq:potential} becomes
\begin{equation}
    V ( \varphi ) = -\frac{m_S^2}{2}\varphi^2 + \frac{17y^4}{48 \pi^2} \varphi^4,
    \label{eq:mbscalar}
\end{equation}
where we remind ourselves that $y$ is also the Yukawa coupling between the scalar and the dark matter field $\chi$. 

 After the mass-boosting scalar rolls to its minimum, the dark matter mass will be boosted through its Yukawa coupling to the scalar. In the case that the Yukawa mass is much larger than the dark matter's initial mass, the final mass of the dark matter will be $m_\chi \approx y \nu $, where $\nu$ is the final vacuum expectation value of $\varphi$. Then assuming that after $\varphi$ has settled at its minimum, the Yukawa term $i y \bar \chi \chi \varphi$ provides the majority of the dark matter mass, the dark matter mass will be
\begin{equation}
    m_\chi \approx y \nu = \sqrt{\frac{24}{31}} \frac{\pi m_S}{y},
    \label{eq:mxfinal}
\end{equation}
where here we have used the exact vacuum expectation value as given in Eq.~\eqref{eq:vev}. 

It will be instructive to consider some benchmark parameter using Eqs.~\eqref{eq:finalrelic} -- \eqref{eq:mxfinal}. 
\begin{enumerate}
\item[]Benchmark I: For $m_{fi} = 10^{14}$ GeV, $m_{S} = 10^{12}$ GeV, $H_e = 10^{13}$ GeV, and $T_{rh}=10^{11}$ GeV ($cf.$ the right panel of Figure \ref{fig:mbrelic}) the correct relic abundance is obtained for $m_{\chi} = 10^{15}$ GeV, corresponding to $y = 2.8 \times 10^{-2}$ and $\nu = 4 \times 10^{17}$ GeV. For this benchmark point, $\varphi$ will roll to the bottom of its potential and boost the mass of $\chi$ when $H \sim m_{S} \sim 5 \times 10^{12}$ GeV, well before reheating at $H_{rh} = 2300$ GeV.

\item[]Benchmark II: For $m_{fi} = 10^{11}$ GeV, $m_{S} = 10^{9}$ GeV, $H_e = 10^{13}$ GeV, and $T_{rh}=10^{9}$ GeV ($cf.$ the left panel of Figure \ref{fig:mbrelic}) the correct relic abundance is obtained for $m_{\chi} = 10^{13}$ GeV, corresponding to $y = 2.8 \times 10^{-3}$ and $\nu = 4 \times 10^{16}$ GeV. For Benchmark II, $\varphi$ will roll to the bottom of its potential and boost the mass of $\chi$ when $H \sim m_{S} \sim 10^{9}$ GeV, well before reheating at $H_{rh} = 0.2$ GeV. In this case since $m_S \sim T_{rh}$, $\varphi$ could undergo a first order phase transition as in Section \ref{sec:phasetransition}.

\end{enumerate}

In principle, arbitrarily large dark matter masses can be obtained in the limit that $y \rightarrow 0$ in Eq.~\eqref{eq:mxfinal}, because for the model detailed in Section \ref{sec:Yukawa}, the Yukawa coupling scales inversely with the scalar's vacuum expectation value, $y \propto \sqrt{m_S / \nu }$, a large-enough dark matter mass implies a super-Planckian vacuum expectation value for $\varphi$ (where here we reiterate that the gauge coupling $g=0$ for simplicity), 
\begin{equation}
    m_\chi \approx 4 \times 10^{16}~{\rm GeV} \left( \frac{m_S}{6 \times 10^{13} ~{\rm GeV}} \right) \left( \frac{\nu}{10^{19}~{\rm GeV}} \right)^{1/2}
\end{equation}
where here we normalized the scalar mass (which must satisfy $m_S \lesssim H_e$) to the maximum value allowed by the Planck 2018 bound \cite{Akrami:2018odb} on the Hubble constant at the end of inflation $ H_e^{\rm max} \lesssim 6 \times 10^{13} ~{\rm GeV}$. We note that this does not necessarily restrict the maximum mass of mass-boosted dark matter, since there are a number of models that accommodate super-Planckian scalar field vacuum expectation values \cite{Banks:2003sx,Kaloper:2008fb,McAllister:2008hb,delaFuente:2014aca}.  However, there is one additional considerations in this regard. Namely, we should consider whether a vacuum energy added to cancel the contribution of $V(\varphi)$ at its minimum, which would be $\sim - m_S^2 \nu^2$ in Eq.~\eqref{eq:mbscalar}, implies some departure from the cosmology we have already laid out. Because there exist mechanisms that dynamically adjust the cosmological constant (see $e.g.$ \cite{Bellazzini:2013fga,Coradeschi:2013gda,Bellazzini:2015wva}), such an addition is not necessarily required. However, if we assume this additional term is present, this would place an additional restriction that $\nu \lesssim M_{Pl}$, since this would ensure that when $\varphi$ begins rolling to the bottom of its potential after inflation at $H \approx m_S$, the vacuum energy term would be subdominant to the universe's energy density at this time, $i.e.$ $3 H^2 M_{Pl}^2$.

Lastly, we discuss under what conditions the inflationary potentials examined in Section \ref{sec:inflat} can provide a mass boost to gravitationally produced dark matter.  These potentials have fields that move over Planckian field ranges (see Table \ref{tab:simparams}), and so it is natural to consider whether they might provide a large mass boost to dark matter, which is gravitationally produced towards the end of inflation. Given that the inflaton field value itself is shifting appreciably during inflation, so too will the dark matter's mass, considerably complicating the resulting dynamical production of dark matter. The standard Bogoliubov coefficient evolution that yields Eq.~\eqref{eq:fermabund}, would need to be re-computed to incorporate a dynamical mass for $\chi$. We leave a full investigation of this to future work. 

Here we will consider the simpler case, that the scalar potentials given in Section \ref{sec:inflat} are not inflatons, but rather low-mass scalar fields, that stay fixed at some field value during and after inflation $\phi = \phi_i$, and then roll to their minima sometime after the end of inflation. For the case we consider, we will choose to set this initial field value to the ``$\phi_e$'' field values given in Table \ref{tab:simparams}. We choose these field as a starting field value, because for larger field values, these potentials would initiate a second period of inflation. (This is allowed \cite{Burgess:2005sb}, but beyond the scope of our consideration here). In this case, we note that the scalar would couple to the fermion $\chi$ with a Yukawa coupling $\mathcal{L}_{yuk} = - y \chi \bar \chi (\phi - \phi_e)$. The dynamics of these potentials are similar to curvaton fields \cite{Lyth:2001nq,Bramante:2016yju}, in that they will slowly roll to their final field value sometime after the end of inflation, although in this case we do not require that they necessarily produce the observed spectrum of primordial perturbations, as for a curvaton field. 

The dynamics of dark matter production in this scenario will be the same as above, but with the additional restriction that the effective mass of the late-time rolling scalars is less than the Hubble constant at the end of inflation
\begin{equation}
    m_{eff}^2 = V''(\phi_e) \ll H_e^2
\end{equation}
This condition ensures that the scalar will not begin oscillations until after the end of inflation, and this can be satisfied for all of the potentials studied in Section \ref{sec:inflat}. Finally, note that so long as the scalar field is free to oscillate in its potential (and is not for example damped by scattering with background fields) it should produce the same gravitational wave signatures detailed in Section \ref{sec:inflat}. For concreteness, let us briefly examine the E-Model for $\alpha = 3 \times 10^{-4}$. In this case, the effective mass $m_{eff} = \sqrt{ V''(\phi_e)} \approx 2 \times 10^{12}~{\rm GeV}$, well below the bound $H_e \lesssim 10^{14} ~{\rm GeV}$ indicated in Figure \ref{fig:mbrelic}. Looking at Table \ref{tab:simparams}, the vev of this scalar, and the corresponding mass of a coupled dark matter field ($m_\chi \approx y \phi_e$) will shift by up to $\sim 0.1 M_{Pl}$ at the end of the scalar's oscillations.

\section{Conclusions}
\label{sec:conc}

\begin{figure}[ht!]
\centering
\includegraphics[scale=.45]{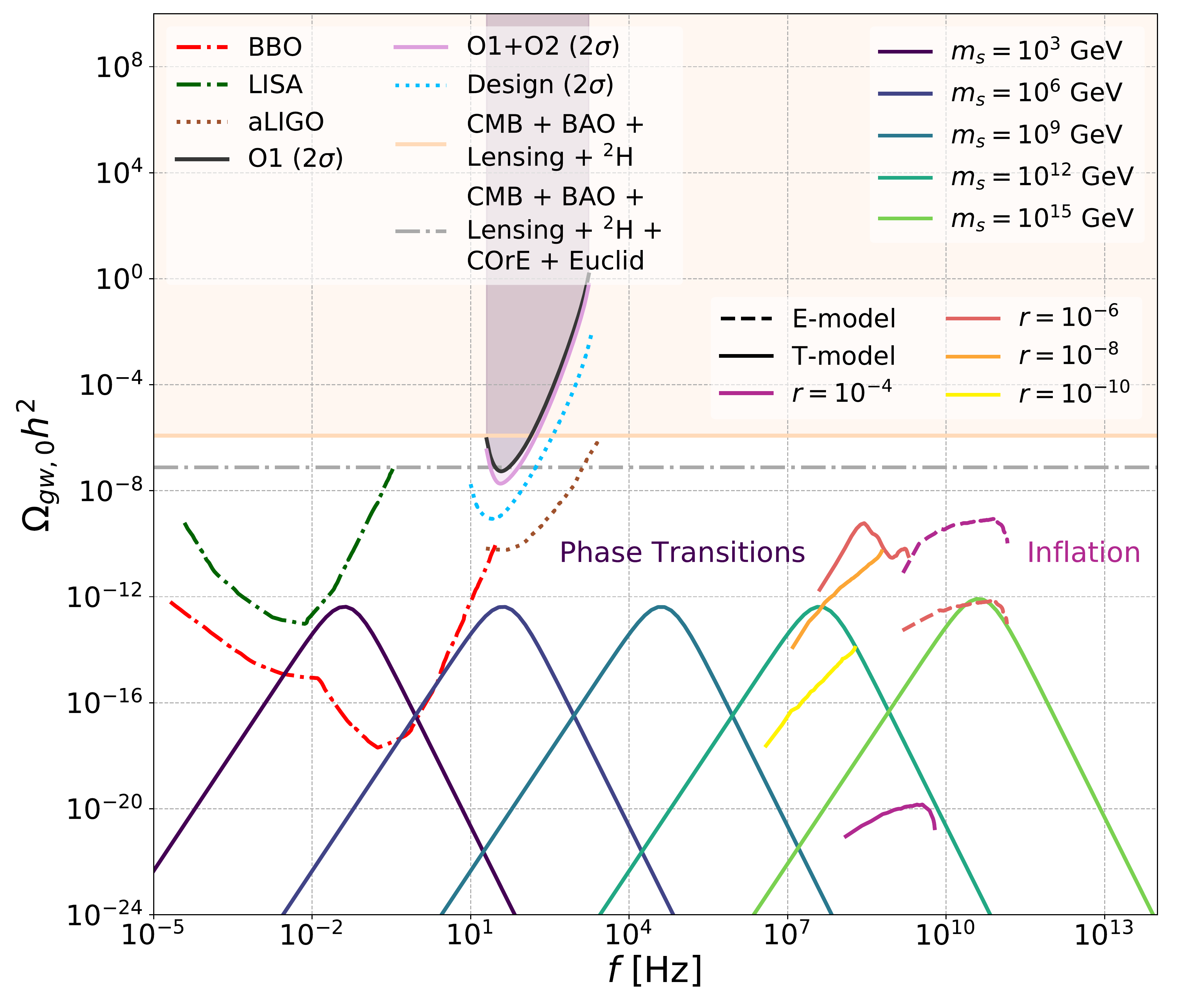}
\caption{Gravitational wave spectra for oscillating fields after inflation and dark sector phase transition scenarios. The T-Model SGWB spectra sourced by oscillating inflatons are given for $r = 10^{-4}, 10^{-6}, 10^{-8}, 10^{-10}$ as indicated. Note that the E-Models are only plotted for $r = 10^{-4}, 10^{-6}$. Phase transition spectra are shown for $y = 1$, $g/y = 1.05$, with scalar masses $m_s$ indicated. Acoustic GW production predominates and we use a bubble velocity $v_b =0.9$. The dashed-dotted sensitivity curves are projections from \cite{2007noiseCampeti:2020xwn} that include the contribution of the astrophysical foregrounds. Dotted curves show a projection that did not include astrophysical foregrounds. The solid sensitivity curves are the present LIGO bounds from \cite{O1O2LIGOScientific:2019vic} along with a bound on SGWB contributions to relativistic degrees of freedom, derived from a fit to CMB power, baryon acoustic oscillations, lensing, and primordial nuclear abundances \cite{Pagano:2015hma}. Prospects for an improvement of this cosmological SGWB bound, using projected constraints from the COrE and EUCLID satellites \cite{Pagano:2015hma} is shown with dash-dotted line.} \label{fig:2007noisePTandE-T-mod}
\end{figure}

Gravitational waves produced by scalar fields may soon reveal details from the first moments of our universe. We have found connections between gravitational waves sourced by scalars and dark matter, including a mechanism whereby scalar fields boost the mass of non-thermally produced dark matter. For many cases of mass boosted dark matter, there will be an associated SGWB, produced through a first order phase transition or oscillations of a scalar field after inflation. In our study of first order phase transitions, we particularly examined a scalar field whose self-interactions are generated at one-loop by its Yukawa coupling to a fermion and gauge coupling to a vector field. For oscillating fields after inflation, our lattice simulations have extended prior work on oscillating scalar field configurations formed by E-Model and T-Model inflationary potentials, and their SGWBs. Both of these scalar fields tend to produce larger amplitude gravitational wave backgrounds for larger dark matter masses.

Figure \ref{fig:2007noisePTandE-T-mod} shows stochastic gravitational wave spectra from inflation, phase transitions, and sensitivity curves for upcoming and current gravitational wave experiments. The E-Model and T-Model spectra shown correspond to the results from Section \ref{sec:inflat}, with T-Model results ranging from $r=10^{-4} - 10^{-10}$ and E-Model results ranging from $r=10^{-4}-10^{-6}$.  Gravitational wave spectra from phase transitions are shown for $g = 1.05$, $y = 1$, and $m_S = 10^{3}- 10^{15}$ GeV as indicated. Sensitivity curves are also shown, both for direct GW observatories, and indirect cosmological probes. Dashed-dotted projected sensitivity curves include the contribution of astrophysical foregrounds while dotted curves do not include astrophysical foregrounds. Solid curves represent bounds on SGWB. The sensitivity curves for BBO, LISA, and aLIGO were taken from \cite{2007noiseCampeti:2020xwn} and take into account the number of detectors, mission observation time, and the overlap reduction function of each detector pair. The BBO and LISA curves additionally take into account astrophysical foregrounds. The LIGO sensitivity curve for the first observing run (O1), the first and second observing run (O1+O2), and the design sensitivity are taken from the data supplement to \cite{O1O2LIGOScientific:2019vic}. A bound on SGWB contributions to relativistic degrees of freedom, derived from a fit to CMB power, baryon acoustic oscillations, lensing, and primordial nuclear abundances and prospects for an improvement of this cosmological SGWB bound, using projected constraints from the COrE and EUCLID satellites is taken from \cite{Pagano:2015hma}. 

There are a number of interesting conclusions that can be drawn from Figure \ref{fig:2007noisePTandE-T-mod}. For low mass scalar phase transitions, the peak of the SGWB could possibly be detected in upcoming space-based gravitational wave observatories such as the Big Bang Observatory (BBO). If a SGWB is detected, its provenance from either acoustic waves during a first order phase transition or oscillating field dynamics could be determined using the strikingly different SGWB spectral features apparent in these two cases. While BBO appears sufficient to find dark sector phase transitions for scalar masses $m_S \sim {\rm TeV}$, our work has reinforced the importance of MHz-GHz searches, since the peak amplitude of many SGWB signatures lie in the MHz-GHz frequency range, motivating the development of sensitivity in this regime. Experiments have so far obtained sensitivity to MHz-range SGWBs exceeding the current critical density, $e.g.$ $\Omega_{gw,0} h^2 \lesssim 10^{13}$ at MHz frequency \cite{Chou:2016hbb} and $\Omega_{gw,0} h^2 \lesssim 10^{26}$ at 0.1 GHz \cite{Akutsu:2008qv}. There are some laboratory proposals at GHz frequencies \cite{Ito:2019wcb}. However, even in the absence of laboratory tests, the cosmological impact of MHz-GHz gravitational waves, which can contribute to the number of relativistic degrees of freedom in the early universe \cite{Pagano:2015hma}, or convert into photons in the presence of a magnetic field, are both promising high-frequency SGWB detection methods \cite{Dolgov:2012be,Domcke:2020yzq}. 

In conclusion, we have shown that a range of gravitational waves background signatures could reveal the early universe dynamics of an inflaton or hidden sector phase transition, and might also be a signature of very heavy dark matter which received a mass boost in the early universe. In future work, it would be interesting to examine the extent to which gravitational spectra are produced, as the scale of inflaton potentials and corresponding tensor-to-scalar ratio $r$ are decreased by many orders of magnitude. It would also be interesting to revisit our computations at higher loop order, to allow for analysis of the simplified potential we have studied in a more strongly-coupled limit. In addition, while we have studied a very simple model demonstrating links between mass boosts for heavy dark matter, gravitational waves from phase transitions, and a simplified scalar potential, these connections could be re-examined for an enlarged model, particularly since we have found observable phase transitions that could be associated with Grand Unified Theories predicting many scalars with masses around $ 10^{15}$ GeV.

\section*{Acknowledgements}
The work of AB, JB, SN, NS is supported by the Natural
Sciences and Engineering Research Council of Canada (NSERC). We thank, Avi Friedlander, Alan Goodman, Andrew Long, Amalia Madden, and Pauline Perrin for helpful discussions and especially Mustafa Amin, Djuna Croon, Kaloian Lozanov, and Graham White for comments on the manuscript. Research at Perimeter Institute is supported in part by the Government of Canada through the Department of Innovation, Science and Economic Development Canada and by the Province of Ontario through the Ministry of Colleges and Universities.

\appendix

\section{Thermal Corrections and Higher Order Effects} 
\label{sec:thermappendix}
This appendix provides further details of calculations for the thermal potential discussed in Sections \ref{sec:Yukawa} and \ref{sec:phasetransition}.  The validity of the perturbative expansion is addressed using analytical arguments and numerical results which show that daisy contributions, which can dramatically affect the nature of phase transitions, are subdominant. The derivations are similar to those of \cite{Carrington:1991hz,Arnold:1992rz}, with a few key differences. 

We begin with the calculation of the finite temperature effective potential. It will be useful to go into some detail about the more formal aspects of this exercise to highlight how our model differs from scalar potentials with a tree level quartic term. As with the zero temperature case, we use the background field method. To compute the one loop effective potential at a finite temperature $T$, we will have to calculate an integral of the form \cite{Carrington:1991hz}:
\begin{equation}\label{eq:finiteTloopint}
    I\left(\varphi\right) = \frac{T}{2}\sum_{j} c_{j}\sum_{n}\int \frac{d^{3}\mathbf{k}}{\left(2\pi\right)^{3}} \ \log\left[\frac{k_{E}^{2} + M^{2}_{j}\left(\varphi\right)}{T^{2}}\right]
\end{equation}
Here the sum over $j$ is a sum over particle species and the expression $k_{E}^{2} + M^{2}_{j}\left(\varphi\right)$ is the inverse of the square of the propagator for that particle species (except for scalars where it is just the inverse of the propagator). We will return to this shortly, but for now let us proceed with evaluating this integral. $k_{E}$ is the Euclidean momentum at finite temperature, $k_{E}^{2} = \omega_{n}^{2} + k^{2}$, with $\omega_{n} = 2\pi T n, \ 2\pi T \left(n+1\right) $ the Matsubara frequencies for bosons and fermions respectively and $k$ the three momentum amplitude. To evaluate~\eqref{eq:finiteTloopint}, it is useful to define $E^{2} \equiv k^{2} + M^{2}\left(\varphi\right)$ and use the following: 

\begin{equation}
\begin{split}
    v_{B}\left(E\right) & = \sum_{n = -\infty}^{+\infty} \log\left[4\pi^{2}T^{2}n^{2} + E^{2}\right], \frac{\partial v_{B}}{\partial E} = \sum_{n = -\infty}^{+\infty} \frac{2E}{\left(4\pi^{2}T^{2}n^{2} + E^{2}\right)}   \end{split}
\end{equation}
for bosons and
\begin{equation}
    \begin{split}
    v_{F}\left(E\right) & = \sum_{n = -\infty}^{+\infty} \log\left[4\pi^{2}T^{2}\left(n+1\right)^{2} + E^{2}\right],  \frac{\partial v_{F}}{\partial E} = \sum_{n = -\infty}^{+\infty} \frac{2E}{\left(4\pi^{2}T^{2}\left(n+1\right)^{2} + E^{2}\right)} 
    \end{split}
\end{equation}
for fermions. The sums in $\frac{\partial v_{B,F}}{\partial E}$ can be evaluated exactly using 
\begin{equation}\label{eq:sumTint}
\sum_{n = 1}^{+\infty} \frac{y}{y^{2} + n^{2}} = -\frac{1}{2y} + \frac{\pi}{2}\coth\left(\pi y\right)  
\end{equation}
With \eqref{eq:sumTint}, $\frac{\partial v_{B,F}}{\partial E}$ can be written in closed form and integrated with respect to $E$ to obtain $v_{B,F}$. The first term of the series sum~\eqref{eq:sumTint} gives the zero temperature effective potential while the second one gives the thermal part. We have skipped a step by step derivation here but for further details the reader may consult in particular references \cite{Dolan:1973qd} and \cite{Quiros:1999jp}. The integral expressions in~\eqref{eq:temppotential} are easily obtained, bearing in mind that gauge boson contributions come with a factor of $c_{B} = 3$ and fermions a factor of $c_{F} = -4$ from the Dirac trace. Let us however maintain a more general presentation. The one loop effective potential is then written as:
\begin{equation}\label{eq:oneloopgeneralT}
    V_{1}\left(\varphi,T\right) = V_{0}\left(\varphi,0\right) + V_{1}\left(\varphi,0\right) + \frac{T^{4}}{2\pi^{2}}\left[c_{B}J_{B}\left(\varphi,T\right) + c_{F}J_{F}\left(\varphi,T\right)\right] 
\end{equation}
where $V_{0}\left(\varphi,0\right)$ and $V_{1}\left(\varphi,0\right)$ are the tree level and one-loop effective potentials at zero temperature respectively. The thermal part of the potential reads 
\begin{equation}\label{eq:oneloopgeneralT2}
    J_{F,B}\left(\varphi,T\right) = \int_{0}^{\infty} \ dx \ x^{2}\log\left[1 \pm \exp\left(\sqrt{x^{2} + \frac{M^{2}_{F,B}\left(\varphi\right)}{T^{2}}}\right)\right] 
\end{equation}
The high temperature expansion is obtained by using the well know expressions, equations (174) and (200) of \cite{Quiros:1999jp}, for the bosonic and fermionic contributions respectively. 

\subsection*{Daisy Resummation}
Let us now pay closer attention to 
the argument inside the logarithm for expression~\eqref{eq:finiteTloopint}. The $M^{2}$ term is just the tree level field-dependent mass of the boson/fermion that contributes to the effective potential. We know, however, that this mass changes due to radiative corrections, and if we could include the effect of this ``effective mass" into our potential we would have a more accurate result. This is not just a matter of accuracy. At zero temperature, we only require that the coupling constants be small to ensure the validity of perturbation theory but at finite temperature this is not sufficient. Loop corrections at finite temperature depend both on coupling constants and temperature, and a high temperature can compensate for a small coupling constant, making perturbation theory unreliable - higher order corrections to the one-loop effective potential become increasingly important. 

To assess the impact of loop corrections, we need to define a loop expansion parameter $l$ which is dimensionless and depends on the coupling constants, temperature, and cutoff scale of the theory. The contribution to the effective potential of a loop amplitude of superficial degree of divergence $d$ is of the order $l^{d}$. If $l^{d} \ll 1$ higher order corrections, which come in powers of $l^{d}$, are negligible. At fixed order in perturbation theory daisies, loops with $d=2$, provide the largest contribution. Therefore, if $l^{2}$ is $\mathcal{O}\left(1\right)$ we cannot reliably trust perturbative methods. Fortunately, there is a relatively straightforward procedure to include their contributions to all orders in perturbation theory. \textit{Daisy resummation} consists of calculating $M^{2}$ to some order in perturbation theory and substituting $M^{2} \rightarrow M^{2}_{eff}$ into the original expression for the effective potential. For reasons that will soon be apparent, let us - as is conventionally done in \cite{Curtin:2016urg} for example - write:
\begin{equation}
    M^{2} = M_{0}^{2} + \Pi^{2}\left(0\right)
\end{equation}
where $\Pi^{2}$ is called the \textit{polarization tensor} of the particle species in question, calculated to some order in perturbation theory. With this modification, the one-loop effective potential in~\eqref{eq:finiteTloopint} is modified to
\begin{equation}\label{eq:finiteTloopintDaisy}
    I\left(\varphi\right) = \frac{T}{2}\sum_{j} c_{j}\sum_{n}\int \frac{d^{3}\mathbf{k}}{\left(2\pi\right)^{3}} \ \log\left[\frac{k_{E}^{2} + M^{2}_{j} + \Pi_{j}^{2}\left(0\right)}{T^{2}}\right]
\end{equation}
It is useful to seperate out this expression so that we may write the final expression for the effective potential as: 
\begin{equation}
     V_{1}\left(\varphi,T\right) = V_{0}\left(\varphi,0\right) + V_{1}\left(\varphi,0\right) + \frac{T^{4}}{2\pi^{2}}\left[c_{B}J_{B}\left(\varphi,T\right) + c_{F}J_{F}\left(\varphi,T\right)\right] + V_{D}\left(\varphi,T\right).
\end{equation}
This potential is identical to~\eqref{eq:finiteTloopint} except for the daisy ``correction" $V_{D}\left(\varphi,T\right)$,
\begin{equation}\label{eq:daisypot}
    V_{D}\left(\varphi,T\right) = -\frac{T}{2}\sum_{j} c_{j}\sum_{n}\int \frac{d^{3}\mathbf{k}}{\left(2\pi\right)^{3}} \ \log\left[\frac{1}{T^{2}}\left(1 + \frac{\Pi_{j}^{2}\left(0\right)}{k_{E}^{2} + M^{2}_{j}}\right)\right]\delta^{0}_{n},
\end{equation}
where we have used the Kronecker delta $\delta^{0}_{n}$ to illustrate that in the Daisy resummation only the infrared (IR) divergent $n=0$ energy mode is important. This is equivalent to a high temperature expansion \cite{Curtin:2016urg}. The interested reader is invited to consult \cite{Carrington:1991hz} for an alternative but nonetheless in depth derivation of this effective potential. 

Proceeding, the resulting integral is
\begin{equation}\label{eq:daisypot2}
    V_{D}\left(\varphi,T\right) = -\frac{T}{2}\sum_{j} c_{j}\int \frac{d^{3}\mathbf{k}}{\left(2\pi\right)^{3}} \ \log\left[\frac{1}{T^{2}}\left(k^2 + M_{j}^{2}\left(\varphi\right) + \Pi_{j}^{2}\left(0\right)\right)\right] - \log\left[\frac{1}{T^{2}}\left(k^2 + M_{j}^{2}\left(\varphi\right)\right)\right],
\end{equation}
where $k$ is the three momentum integral. The integral can be evaluated using dimensional regularization,
\begin{equation}
    \int \frac{d^{D}k}{\left(2\pi\right)^{D}} \log\left(k^{2} + \Delta^{2}\right) = \frac{\Gamma\left(-\frac{D}{2}\right)}{\left(4\pi\right)^{\frac{D}{2}}}\left(\Delta^{2}\right)^{\frac{D}{2}},
\end{equation}
where $D = 3 + \epsilon$ and $\epsilon$ is set to zero at the end. The final result is
\begin{equation}\label{eq:DaisyPotFinal}
    V_{D}\left(\varphi,T\right) = -\frac{T}{12\pi}\sum_{j} c_{j} \left(M_{j}^{2}\left(\varphi\right) + \Pi_{j}^{2}\left(0\right)\right)^{\frac{3}{2}} - \left(M_{j}^{2}\right)^{\frac{3}{2}}
\end{equation}

This is the general form of the daisy potential. For our purposes, all we need now to calculate its exact form are polarization tensors for the scalar and vector, $\Pi^{2}_{S}\left(0\right)$ and $\Pi^{2}_{V}\left(0\right)$ respectively. We only consider bosonic contributions as fermionic ones are not IR divergent. Let us first consider the scalar contribution as it is the most important difference between our model and prior calculations in the literature. In the interest of generality, let us assume for the moment that we have a tree level term $\frac{\lambda}{4}\varphi^{4}$ in our lagrangian. We will set $\lambda$ to zero at the end of the calculation. The expression for $M^{2}_{S}$ to one loop order is,
\begin{equation}\label{eq:Scalrpol}
    M^{2}_{S} = -m_{S}^{2} + 3\lambda \varphi^{2}_{c} + \Pi^{2}_{S}\left(0\right) 
\end{equation}
The tree level contribution is $M^{2}_{0} = -m_{S}^{2} + 3\lambda\varphi^{2}_{c}$, which has a field dependent value due to the quartic term. 

Restricting ourselves only to first order contributions, we find that the following diagrams contribute at first order to the scalar polarization tensor.

The contribution due to the self interaction $\lambda$ 
\begin{equation}
    \Pi^{2}_{SS}\left(k\right) = 3\lambda T \sum^{\infty}_{n=-\infty} \int \frac{d^{3}k}{\left(2\pi\right)^{3}} \frac{1}{k^{2}+ m_{S}^{2} + \left(2\pi n\right)^2 T^{2}} \vert_{T\gg m_{S}}
\end{equation}
where the $SS$ subscript in $\Pi^{2}$ indicates that we are calculating the scalar contribution to the scalar polarization tensor. We are typically interested in the high temperature behaviour of the system, so it is customary to set $m_{S} = 0$ since its contribution is negligible at that temperature. Following the same steps as before, i.e., performing the sum first and then the three momentum integral, we find
\begin{equation}\label{eq:scalarpoltens}
    \Pi^{2}_{SS}\left(0\right) =\Pi^{2}_{SS}\left(k\right) \vert_{k=0} = \frac{\lambda T^{2}}{4}. 
\end{equation}
The other polarization tensors are calculated in the same manner. They are \cite{Carrington:1991hz}
\begin{equation}
    \begin{split}
        \Pi^{2}_{SF}\left(0\right) & = \frac{y^{2}}{2}T^{2} \\
        \Pi^{2}_{SV}\left(0\right) & = \frac{g^{2}}{2}T^{2} 
    \end{split}
\end{equation}
for the fermionic and vector contributions to the scalar polarization tensor respectively. This is everything we need to write the final form of the scalar contribution to the daisy potential 
\begin{equation}\label{eq:DaisyPotScalarContributionFinal}
    V_{DS}\left(\varphi,T\right) = -\frac{T}{12\pi}\left[ \left(-m_{S}^{2} + 3\lambda\varphi^{2}_{c} + \frac{\lambda T^{2}}{4} + \frac{y^{2}}{2}T^{2} + \frac{g^{2}}{2}T^{2} \right)^{\frac{3}{2}} - \left(-m_{S}^{2} + \frac{\lambda}{2}\varphi^{2}_{c}\right)^{\frac{3}{2}}\right].
\end{equation}
The difference between our model and effective potentials with a tree level quartic interaction is now much clearer: in our model, the scalar does not contribute to daisy contributions. This is evident by setting $\lambda$ to zero - the $\varphi$ dependence of $V_{DS}\left(\varphi,T\right)$ vanishes and $V_{DS}$ only contributes an irrelevant constant to the daisy resummed effective potential.  \\

\begin{figure}[t!]
    \centering
  \begin{subfigure}[b]{0.48\textwidth}
    \includegraphics[width=\textwidth]	{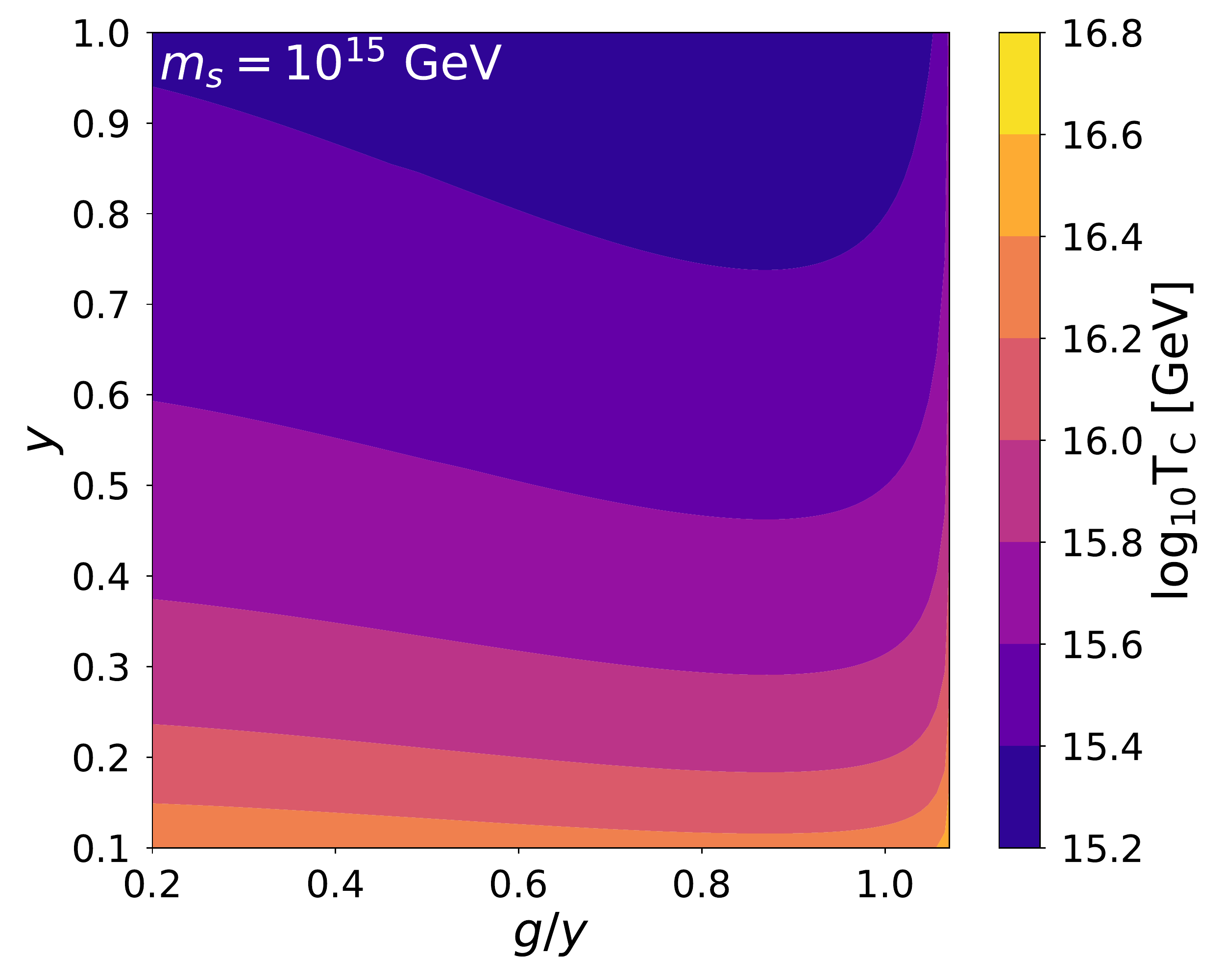}
    \label{fig:TCms1e6Second}
    \end{subfigure}
  \begin{subfigure}[b]{0.48\textwidth}
    \includegraphics[width=\textwidth]	{figures/TC_ms1e15_fig_cut.pdf}
    \label{fig:TCms1e15Second}
    \end{subfigure}
    \caption{Left: the critical temperatures $T_C$ obtained from first-order phase transition of a ring resummed potential. Right: the critical temperatures $T_C$ obtained from first-order phase transition without resummation.}
    \label{fig:TCCompareResum}
\end{figure}

For our model then, the only contribution to the daisy potential comes from the gauge bosons. Here the computation is slightly tricky since only the longitudinal part of the gauge boson field $A_{\mu}$ contributes. Let us start by defining transverse, $T_{\mu\nu}$, and longitudinal, $L_{\mu\nu}$, projection tensors
\begin{equation}
    \begin{split}
        T_{\mu\nu} & = \eta_{\mu\nu} - \frac{k_{\mu}k_{\nu}}{k^{2}} \\
        L_{\mu\nu} & = \frac{k_{\mu}k_{\nu}}{k^{2}},
    \end{split}
\end{equation}
where $\eta_{\mu\nu}$ is the Minkowski metric. It is easily verified that $T_{\mu\nu}T^{\mu\rho} = T_{\nu}^{\rho}$, $L_{\mu\nu}L^{\mu\rho} = L_{\nu}^{\rho}$, and $T_{\mu\nu}L^{\mu\nu} = L_{\mu\nu}T^{\mu\nu} = 0$. Then the daisy potential can be written as
\begin{equation}
    -\frac{T}{2}\sum c_{j}\int \frac{d^{3}\mathbf{k}}{\left(2\pi\right)^{3}} \ \log\left[\frac{1}{T^{2}}\left(k^2 + \left(g\varphi\right)^{2} + \left(T_{\mu\nu} + L_{\mu\nu}\right)\left(\Pi^{2}\right)^{\mu\nu}\left(0\right)\right)\right] - \log\left[\frac{1}{T^{2}}\left(k^2 + \left(g\varphi\right)^{2}\right)\right]
\end{equation}
To proceed, we will simply use a result obtained in \cite{Carrington:1991hz}, which the interested reader is invited to consult for a more in depth presentation:
\begin{equation}\label{eq:poltensgb1}
    \begin{split}
         \Pi^{2A}_{\mu\nu}\left(0\right) & = - L_{\mu\nu}\Pi^{2A}_{00}\left(0\right) \\
         \Pi^{2A}_{00}\left(0\right) & = \frac{\left(gT\right)^{2}}{3}
    \end{split}
\end{equation}
Here we see the purpose of introducing the polarization tensors: the key contribution to the daisy potential from gauge bosons come from their longitudinal polarizations. Putting everything together, the daisy potential for our model is therefore:
\begin{equation}
    V_{D}\left(\varphi,T\right) = - g^{3}\frac{T}{12\pi}\left[ \left(\varphi^{2} + \frac{T^{2}}{3}\right)^{\frac{3}{2}} - \varphi^{3} \right]
\end{equation}
Having computed the daisy resummed one loop effective potential, we now wish to assess the extent to which it affects our original calculations and preserves the nature of our phase transition. This is of course a difficult question requiring complex lattice simulation which are beyond the scope of this work. Here we will restrict ourselves to computing the critical temperature. If daisy corrections are significant they can change the nature of the phase transition from first to second order. The potential loses its valley - hill - valley shape at high temperature and there is no temperature at which the effective potential evaluated at its non-zero minimum is equal to zero, i.e., there is no solution for $T_{C}$. Therefore, determining regions of parameter space where $T_{C}$ exists for the daisy improved effective potential is a good indicator of a first order phase transition. We consider a $10^{15}$ GeV mass scalar and, as Fig.~\eqref{fig:TCCompareResum} shows, the critical temperature does not change drastically and daisy contributions are therefore subdominant.  

Retrospectively, let us examine why we should have expected this outcome from power counting arguments. Since Yukawa interactions do not contribute to the potential barrier between the two minima to leading order in perturbation theory, we neglect them. At high temperature the (non daisy resummed) effective potential when the temperature is close to $T_{0}$ can be written as:
\begin{equation}\label{eq:temppotentialhighTapprox}
\begin{split}
V\left(\varphi, T\right) = \nu_{eff}^{2}\varphi^{2} - E T \varphi^{3} + \frac{\lambda\left(T\right)}{4}\varphi^{4},
\end{split}
\end{equation}
where we have written the quadratic term in~\eqref{eq:temppotentialhighT} as $\nu^{2}_{eff}\varphi^{2}$, where $\nu_{eff}$ is temperature dependent. Near $T_{0}$ all three terms are roughly the same order of magnitude. This potential has a minimum $\varphi \sim \frac{g^{3}}{\lambda_{T_0}}$ and an effective mass $\nu^{2}_{eff} \sim \frac{g^{6}}{\lambda_{T_{0}}}T^{2}$ obtained by equating the second and third terms and first and third terms respectively. The loop expansion parameter is given by $l \simeq \frac{g^{2}T}{g\varphi} \sim \frac{\lambda_{T_{0}}}{g^{2}}$ \cite{Arnold:1992rz}. In our case, it can be verified that for the parameter space we are interested in $\lambda_{T_{0}} \ll g^{2}$ so higher order loop corrections to the effective potential are negligible.

\section{RG Improvement of the Effective Potential} 
\label{sec:RGEImprovement}
Here we discuss how the effective potential depends on the scale $\mu$ at which it is evaluated, $i.e.$ the renormalization group (RG) improvement of the model. We will find that for our model, all couplings run very weakly with this scale. In our setup for the zero temperature one-loop effective potential, we define the effective potential at a fixed scale $\mu = \nu$, where $\nu$ is given by~\eqref{eq:vev}. However, it is known that both the field value $\varphi$ and couplings $g\left(\mu\right)$ and $y\left(\mu\right)$ depend on $\mu$ and the values we used in our calculations are really just the field and coupling constant values at scale $\nu$. Our effective potential at another scale $\mu \neq \nu$ will be different from the one we have used for our calculations in this work and these are to be trusted only if that difference is small. In this Appendix we show that this is indeed the case.

\begin{figure}[ht!]
    \centering
  \begin{subfigure}[b]{0.48\textwidth}
    \includegraphics[width=\textwidth]	{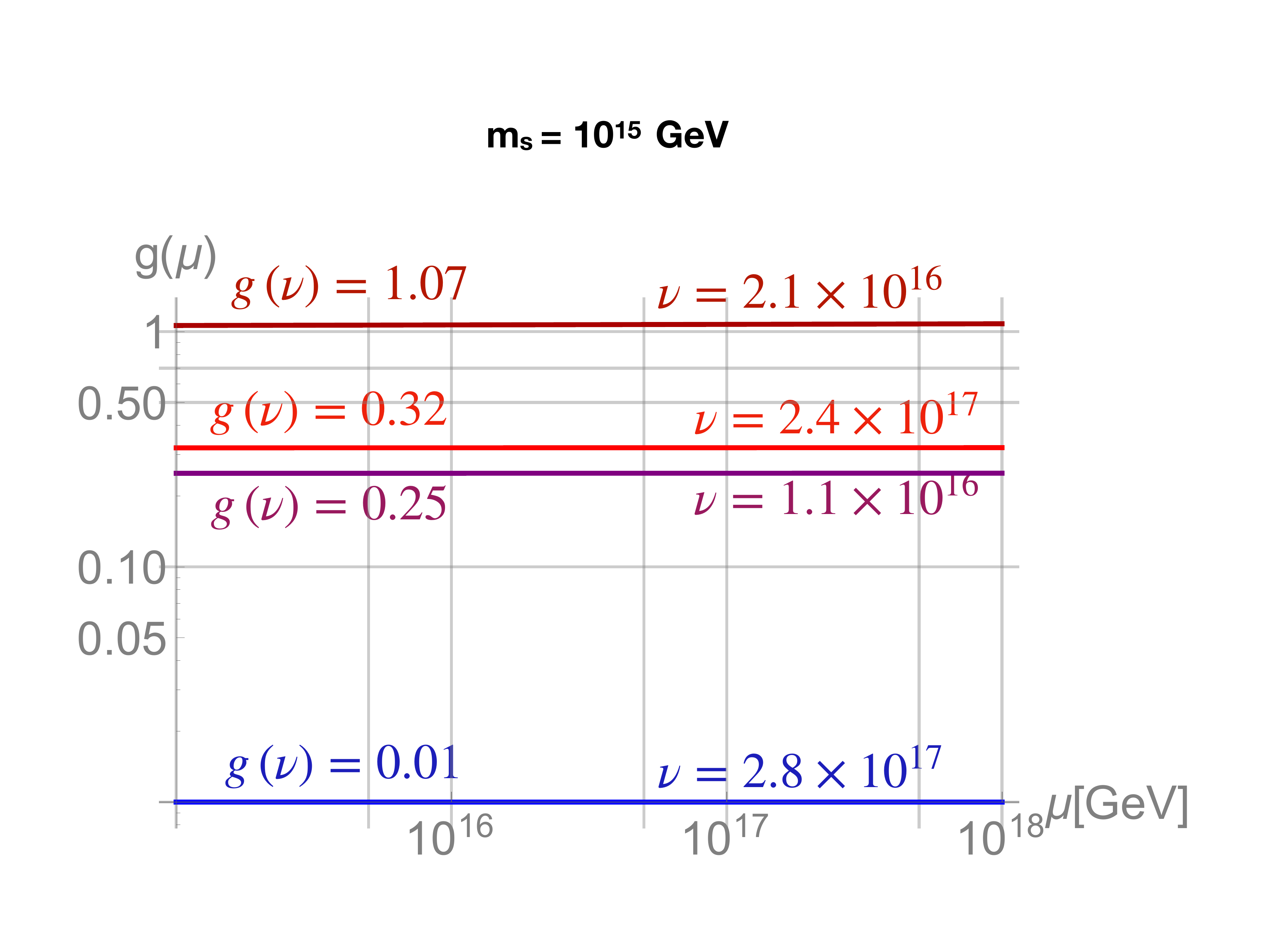}
    \label{fig:RGEGauge}
    \end{subfigure}
  \begin{subfigure}[b]{0.48\textwidth}
    \includegraphics[width=\textwidth]	{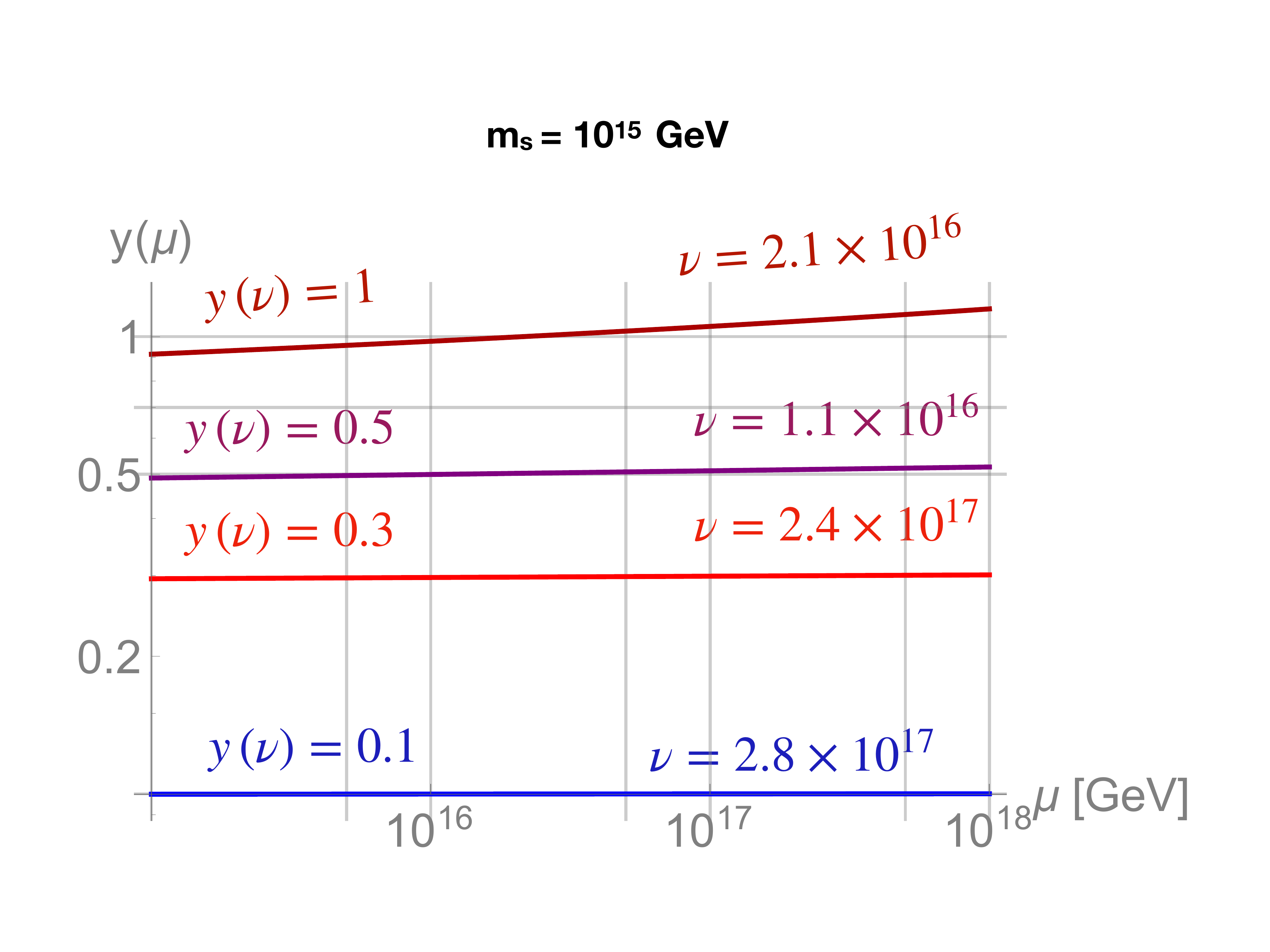}
    \label{fig:RGEYukawa}
    \end{subfigure}
    \caption{Plot of the variations in couplings with scale for $g_{\nu} = 1.07, 0.32, 0.25, 0.01$ and $y_{\nu} = 1, 0.5, 0.3,0.1$. The scalar mass parameter is fixed to $m_{S} = 10^{15}$~GeV. Left: Gauge coupling RG evolution for several initial values $g_{\nu}$. Right: Yukawa coupling RG evolution for several initial values $y_{\nu}$. As the plot shows, the variation in coupling constant value is small over the range of importance for our calculations. While the Yukawa couplings show more variation, their impact on the phase transition is not significant since, as discussed previously, this depends primarily on the gauge coupling.}
    \label{fig:RGECouplings}
\end{figure}

Let us start with the field value $\varphi$. A shift in scale $\mu \rightarrow \mu^{\prime}$ shifts the field value according to \cite{Ford:1992mv}
\begin{equation}\label{eq:phiRG}
    \begin{split}
        \varphi & \rightarrow \varphi\xi\left(t\right) \\
        \xi\left(t\right) & = \exp\left[-\int_{0}^{t}dt^{\prime}\gamma\left(t^{\prime}\right) \right] 
    \end{split}
\end{equation}
where $t\left(\mu\right) = \log\left(\frac{\mu}{\nu}\right)$ has been introduced for convenience. $\gamma\left(t^{\prime}\right)$ is the anomalous dimension of the scalar field. To one loop order it is given by
\begin{equation}\label{eq:anomalousdim}
    16\pi^{2}\gamma\left(t\right) = 6y^{2}\left(t\right) - 3g^{2}\left(t\right).
\end{equation}
\\
Equations~\eqref{eq:phiRG} and~\eqref{eq:anomalousdim} show that as long as the evolution of the coupling constants is small for the $\mu$ range under consideration, the effective potential does not change significantly. The couplings evolve according to their RG equations
\begin{equation}\label{eq:RGECouplings}
    \begin{split}
        16\pi^{2}\frac{d g}{dt} & = \frac{g^{3}}{3}, \ \ \ \ \
        g\left(t\right) = \frac{g_{\nu}}{1 - \frac{g_{\nu}^{2}}{24\pi^{2}}t} \\
        16\pi^{2}\frac{d y}{dt} & = 5y^{3}, \ \ \ \ \
        y\left(t\right) = \frac{y_{\nu}}{1 - \frac{5y_{\nu}^{2}}{8\pi^{2}}t},
    \end{split}
\end{equation}
where we again remind ourselves that $g_{\nu},y_{\nu}$ are the values of the gauge and Yukawa couplings at $\mu = \nu$ respectively. Since $t = \log\left(\frac{\mu}{\nu}\right)$, there is a logarithmic dependence on scale for both couplings.

In our case we started by defining the coupling values at a scale $\mu = \nu$ and then considered thermal effects leading to a first order phase transition, for which the scale range is naturally the range of temperatures $T_{0} \lesssim T \lesssim T_{C}$. To show that our calculations can be trusted, it is sufficient to prove that two conditions are met: (1) that the couplings do not vary significantly in the range $T_{0} \lesssim T \lesssim T_{C}$ and (2) that the coupling values in that range are not too different from the values at $\mu = \nu$. 

We find that typically $T_{C} \ \simeq \mathcal{O}\left(\nu\right)$, with $\nu$ higher by $\mathcal{O}\left(1\right)$ factors. Only in the special cases mentioned previously where $y \simeq g/1.05$, does $\nu$ differ from $T_{C}$ by a few orders of magnitude. Therefore, both conditions mentioned above are met if we can show that the couplings do not vary significantly in the range $T_{0} \leq \mu \leq \nu$, which we plot in Figure \ref{fig:RGECouplings}.
As we can see clearly, both the gauge and Yukawa couplings meet the two conditions established above. This can also be understood from the fact that the couplings depend logarithmically on scale, their variation over the range of interest is small even if the latter spans multiple orders of magnitude - a fractional amount for a scale range spanning a few orders of magnitude, for example. Therefore, our calculations appear to be trustworthy since the effective potential does not vary significantly over the range of interest to us.

\bibliographystyle{JHEP.bst}

\bibliography{ghdm.bib}

\end{document}